\documentclass[showpacs,aps,pra,floatfix,reprint,superscriptaddress,footinbib,citeautoscript,obeyspaces]{revtex4-2}
\usepackage{amssymb,amsfonts,amsmath}
\usepackage{graphicx}
\usepackage{verbatim}
\usepackage[T1]{fontenc}
\usepackage{soul,xcolor,colortbl}
\usepackage{multirow}
\usepackage{bm}
\usepackage{tabularx}
\usepackage{array}
\usepackage{color}
\usepackage{hyperref} \hypersetup{colorlinks=true,citecolor=blue,linkcolor=blue,urlcolor=blue}
\usepackage{mdframed}
\usepackage{longtable}
\usepackage{booktabs}
\usepackage{enumitem}
\usepackage{lmodern}
\usepackage{seqsplit}
\usepackage{enumitem}
\usepackage{listings}

\definecolor{codegreen}{rgb}{0,0.6,0}
\definecolor{codegray}{rgb}{0.5,0.5,0.5}
\definecolor{codepurple}{rgb}{0.58,0,0.82}
\definecolor{codebgcolorCommand}{rgb}{0.95,0.95,0.92}
\definecolor{codegreen}{rgb}{0,0.6,0}
\definecolor{codegray}{rgb}{0.5,0.5,0.5}
\definecolor{codepurple}{rgb}{0.58,0,0.82}
\definecolor{codebgcolorCommand}{rgb}{0.95,0.95,0.92}
\definecolor{codebgcolorFile}{rgb}{0.92,0.92,0.95}
\definecolor{binarycolor}{rgb}{0,0,1}
\lstdefinelanguage{aflux}{
basicstyle=\ttfamily\footnotesize,
frame=single,
breaklines=true,
backgroundcolor=\color{codebgcolorCommand},
escapechar=\%,
keywordstyle=\color{binarycolor},
commentstyle=\it\color{codegreen},
moredelim=[is][\it]{/*}{*/},
postbreak=\raisebox{0ex}[0ex][0ex]{\ensuremath{\color{red}\hookrightarrow\space}},
}
\setlist[itemize]{noitemsep, topsep=0pt}
\newlist{myitemize}{itemize}{3}
\setlist[myitemize,1]{label=\textbullet,leftmargin=1em}
\setlist[myitemize,2]{label=--,leftmargin=1em}
\setlist[myitemize,3]{label=$\diamond$,leftmargin=1em}
\setlist[myitemize]{noitemsep, topsep=0pt}

\setlist[itemize]{noitemsep, topsep=0pt}
\newlist{myitemize_nobullet}{itemize}{3}
\setlist[myitemize_nobullet,1]{label=,leftmargin=1em}
\setlist[myitemize_nobullet,2]{label=,leftmargin=1em}
\setlist[myitemize_nobullet,3]{label=,leftmargin=1em}
\setlist[myitemize_nobullet]{noitemsep, topsep=0pt}

\definecolor{pranab_green}{rgb}{0.31,0.53,0.10}
\definecolor{pranab_red}{rgb}{0.85,0.23,0.11}

\newcolumntype{L}[1]{>{\raggedright\arraybackslash}p{#1} }
\newcolumntype{C}[1]{>{\centering  \arraybackslash}p{#1} }
\newcolumntype{R}[1]{>{\raggedleft \arraybackslash}p{#1} }

\def\AFLOW{{\small AFLOW}}

\def\AFLOWorg{{\sf {a}{f}{l}{o}{w}{.}{o}{r}{g}}}
\def\AFLOWPOCC{{\small AFLOW-POCC}}
\def\AFLOWSYM{{\small AFLOW-SYM}}
\def\AFLOWCHULL{{\small AFLOW-CHULL}}
\def\AFLOWCCE{{\small AFLOW-CCE}}
\def\AFLOWAEL{{\small AFLOW-AEL}}

\def\AFLOWML{{\small AFLOW-ML}}

\def\RESTAPI{{\small REST-API}}
\def\API{{\small API}}
\def\AFLUX{{\small AFLUX}}

\def\AFLOWXTALFINDER{{{\small AFLOW}-XtalFinder}}

\def\JSON{{\small JSON}}
\def\DFT{{\small DFT}}
\def\DOS{{\small DOS}}

\def\VASP{{\small VASP}}
\def\POSCAR{{\small POSCAR}}

\def\WYCCAR{{\small WYCCAR}}
\def\CIF{{\small CIF}}
\def\QUANTUMESPRESSO{\textsc{Quantum {\small ESPRESSO}}}
\def\FHIAIMS{{\small FHI-AIMS}}
\def\ABINIT{{\small ABINIT}}
\def\ELK{{\small ELK}}

\def\GPa{{\,GPa}}

\def\AUID{{\small AUID}}
\def\AUIDs{{\small AUIDs}}

\def\FAIR{{\small FAIR}}
\def\AFLOWONLINE{{\small AFLOW Online}}
\def\OUTCAR{{\small OUTCAR}}

\def\URI{{\small URI}}
\def\URL{{\small URL}}
\def\AFLOWOPTIMADE{{\small AFLOW-OPTIMADE}}
\def\OPTIMADE{{\small OPTIMADE}}
\def\AFLOWSYM{{\small AFLOW-SYM}}
\def\POSCAR{{\small POSCAR}}
\def\PARTCAR{{\small PARTCAR}}
\def\KPOINTS{{\small KPOINTS}}
\def\LDA{{\small LDA}}
\def\PBE{{\small PBE}}
\def\SCAN{{\small SCAN}}
\def\AFLOWCHULLONLINE{{\small AFLOW-CHULL Online}}
\def\AFLOWMLONLINE{{\small AFLOW-ML Online}}

\setcitestyle{square}

\makeatletter \renewcommand\frontmatter@abstractwidth{\dimexpr\textwidth\relax} \makeatother

\begin{document}
\title{\AFLOWorg: a web ecosystem of databases, software and tools}

\author{Marco~Esters}
\affiliation{Department of Mechanical Engineering and Materials Science and Center for Autonomous Materials Design, Duke University, Durham, NC 27708, USA}
\author{Corey~Oses}
\affiliation{Department of Mechanical Engineering and Materials Science and Center for Autonomous Materials Design, Duke University, Durham, NC 27708, USA}
\author{Simon~Divilov}
\affiliation{Department of Mechanical Engineering and Materials Science and Center for Autonomous Materials Design, Duke University, Durham, NC 27708, USA}
\author{Hagen~Eckert}
\affiliation{Department of Mechanical Engineering and Materials Science and Center for Autonomous Materials Design, Duke University, Durham, NC 27708, USA}
\author{Rico~Friedrich}
\affiliation{Department of Mechanical Engineering and Materials Science and Center for Autonomous Materials Design, Duke University, Durham, NC 27708, USA}
\affiliation{Institute of Ion Beam Physics and Materials Research, Helmholtz-Zentrum Dresden-Rossendorf, 01328 Dresden, Germany}
\affiliation{Theoretical Chemistry, Technische Universit\"{a}t Dresden, 01062 Dresden, Germany}
\author{David~Hicks}
\affiliation{Department of Mechanical Engineering and Materials Science and Center for Autonomous Materials Design, Duke University, Durham, NC 27708, USA}
\affiliation{LIFT, American Lightweight Materials Manufacturing Innovation Institute, Detroit, MI 48216, USA}
\author{Michael~J.~Mehl}
\affiliation{Department of Mechanical Engineering and Materials Science and Center for Autonomous Materials Design, Duke University, Durham, NC 27708, USA}
\author{Frisco~Rose}
\affiliation{Department of Mechanical Engineering and Materials Science and Center for Autonomous Materials Design, Duke University, Durham, NC 27708, USA}
\author{Andriy~Smolyanyuk}
\affiliation{Department of Mechanical Engineering and Materials Science and Center for Autonomous Materials Design, Duke University, Durham, NC 27708, USA}
\affiliation{Institute of Solid State Physics, Technische Universit\"{a}t Wien, A-1040 Wien, Austria}
\author{Arrigo~Calzolari}
\affiliation{CNR-NANO Research Center S3, Via Campi 213/a, 41125 Modena, Italy}
\author{Xiomara~Campilongo}
\affiliation{Department of Mechanical Engineering and Materials Science and Center for Autonomous Materials Design, Duke University, Durham, NC 27708, USA}
\author{Cormac~Toher}
\affiliation{Department of Mechanical Engineering and Materials Science and Center for Autonomous Materials Design, Duke University, Durham, NC 27708, USA}
\affiliation{Department of Materials Science and Engineering and Department of Chemistry and Biochemistry, University of Texas at Dallas, Richardson, Texas 75080, USA}
\author{Stefano~Curtarolo}
\email[]{stefano@duke.edu}
\affiliation{Department of Mechanical Engineering and Materials Science and Center for Autonomous Materials Design, Duke University, Durham, NC 27708, USA}

\date{\today}

\begin{abstract}
  \noindent
  To enable materials databases supporting computational and
  experimental research, it is critical to develop platforms that both facilitate access to the data
  and provide the tools used to generate/analyze it --- all while considering the diversity of users'
  experience levels and usage needs. The recently formulated \FAIR\ principles (Findable, Accessible,
  Interoperable, and Reusable) establish a common framework to aid these efforts.
  This article describes \AFLOWorg, a web ecosystem developed to provide \FAIR-compliant access to the
  \AFLOW\ databases. Graphical and programmatic retrieval methods are offered, ensuring
  accessibility for all experience levels and data needs. \AFLOWorg\ goes beyond data-access by
  providing applications to important features of the \AFLOW\ software, assisting users in their own
  calculations without the need to install the entire high-throughput framework.
  Outreach commitments to provide \AFLOW\ tutorials and materials science education to a
  global and diverse audiences will also be presented.
\end{abstract}

\maketitle
\onecolumngrid
\tableofcontents
\newpage
\twocolumngrid

\section{Introduction}
\noindent
The advantage of big data has long been recognized in computational materials science, resulting in
large databases of experimental and hypothetical materials. The \underline{A}utomatic \underline{FLOW}
Framework for Materials Discovery~(\AFLOW) is the largest database for computationally investigated materials,
with over 3.5~million materials entries~\cite{curtarolo:art75} and over 1,100 crystallographic
prototypes to classify crystals and generate input structures.

The data should not just be publicly available, but also easily accessible to users with varying
usage requirements, from casual single-entry access to high-throughput retrieval. To provide a
common framework for such an infrastructure, the \FAIR\ principles for data stewardship were
developed, where \FAIR\ stands for ``Findable'', ``Accessible'', ``Interoperable'', and
``Reusable''~\cite{Wilkinson_FAIR_SciData_2016}.

This article introduces the \AFLOW\ web ecosystem, which consists of the \AFLOW\ databases and
web interfaces to access features of the \AFLOW\ software. We will start with the \AFLOWorg\ repository,
its data structure, and how it conforms with the \FAIR\ data principles. Next, it will be shown how to
access and retrieve that data using both graphical, low-throughput and programmatic, high-throughput
methods. Additionally, the Prototype Encyclopedia --- a database over 1,100 crystallographic
prototypes --- will be discussed along with its tools to create structures for \textit{ab-initio}
calculations.

To create the \AFLOW\ database, a variety of tools has been developed and bundled into the
open-source \AFLOW\ software suite. These features include structure manipulation and
creation~\cite{curtarolo:art58,curtarolo:art65,curtarolo:art110},
symmetry and structure analysis~\cite{curtarolo:art135,curtarolo:art170},
and post-processing tools~\cite{curtarolo:art150,curtarolo:art172}.
While packaging these tools into a monolithic codebase is ideal for generating standardized data,
the individual tools are also useful for many applications outside high-throughput calculations. For
these purposes, a single large codebase is detrimental: users may only require a subset (thus not
using the majority of features), there is a steep learning curve to find and properly use the desired
feature, and the lack of a graphical user interface requires that users be comfortable using the
command line. To alleviate these issues, \AFLOW\ provides web applications that interface to a large
variety of \AFLOW\ features~(\AFLOWONLINE), create convex hulls for thermodynamic
analysis~(\AFLOWCHULL), and use \AFLOW\ machine learning applications~(\AFLOWML). These applications
will be introduced after the discussion about the \AFLOW\ database.

Lastly, we will present the \AFLOW\ School and seminar series. Making software usable and
accessible goes beyond just making it open source. Tutorials are an invaluable way to reduce the
barrier to entry for new users. The Schools and seminars represent \AFLOW's commitment to providing
free education about its software and materials science to a global and diverse audience.
\section{The \AFLOWorg\ Repository}
\subsection{Database Content and Organization}
\label{database_organization}

\begin{figure*}
  \includegraphics[width=\textwidth]{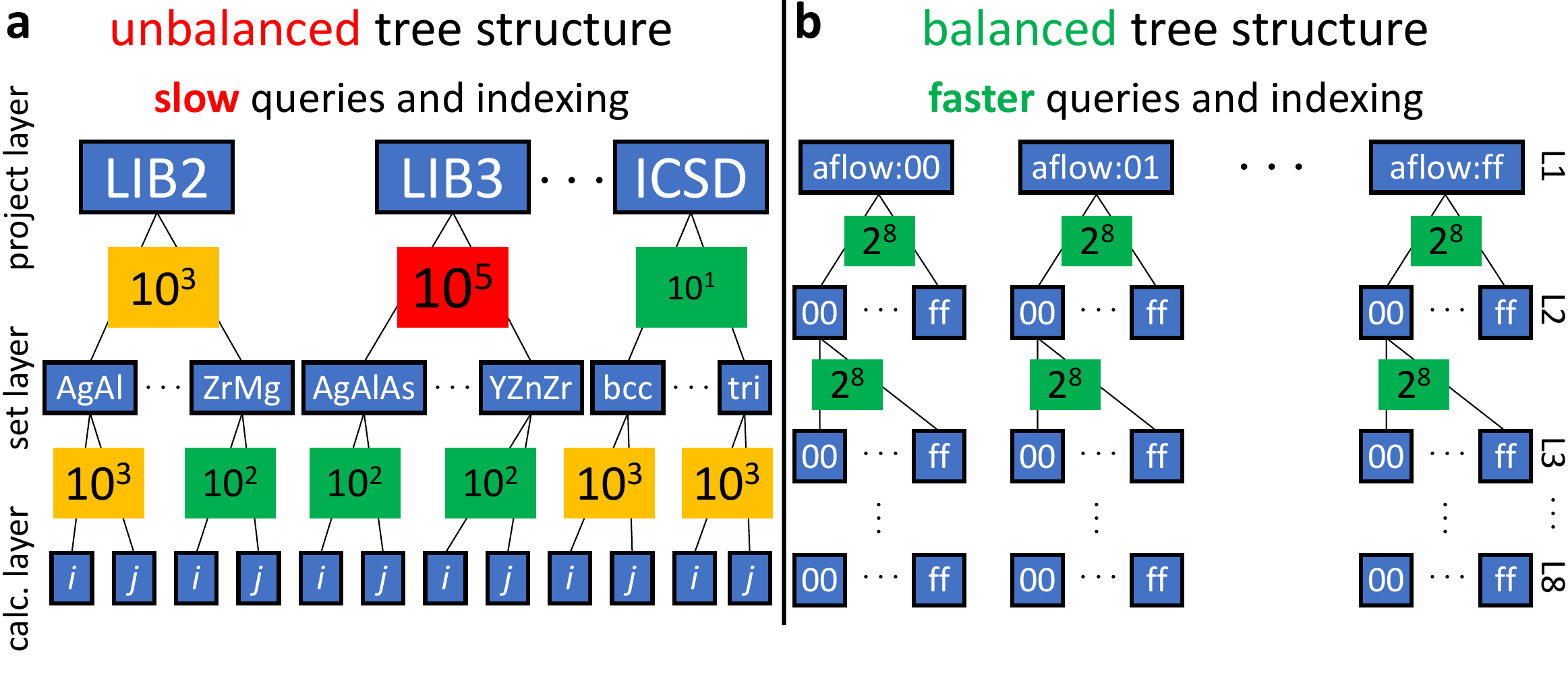}
  \caption{\textbf{\AFLOW\ data structure.} Illustration of the \AFLOW\ data structure as
  (\textbf{a}) a human-readable, but unbalanced search tree and (\textbf{b}) a machine-readable,
  balanced search tree.}
  \label{fig:database_tree}
\end{figure*}

\noindent
The full data set generated by the high-throughput \AFLOW\ process~\cite{curtarolo:art75} is currently over 40 TB.
Stored data includes input criteria and other calculation details
(to facilitate reproducibility and fulfill the ``Reusable'' criterion of the \FAIR\ principles),
calculated results, and derivative output.
To ensure the robust, long-term fulfillment of the \FAIR\ criteria of accessibility and findability for such a large data store,
the \AFLOW\ repository~\cite{curtarolo:art75} is organized in a human-navigable hierarchical directory structure,
allowing specific entries to be accessed independently of the database engine.
Specifically, the repository is organized into project, set and calculation
layers, as illustrated in Figure~\ref{fig:database_tree}.
At the project layer, the calculations are divided into different catalogs
based on the origin and composition of the entries~\cite{curtarolo:art92, curtarolo:art128}.
Within each catalog, entries are grouped into sets based on shared lattice type or alloy system.
The entries within each set contain the results of \underline{d}ensity \underline{f}unctional \underline{t}heory~(\DFT)
calculated properties for particular structures.

The \AFLOW-ICSD catalog contains the \DFT-calculated properties for over 60,000 experimentally observed materials listed in the
\underline{I}norganic \underline{C}rystal \underline{S}tructure \underline{D}atabase (ICSD)~\cite{ICSD, ICSD1}.
Internally, this catalog is organized by lattice type, and then by individual materials entry.
The entries in this catalog are grouped by Bravais lattice type into 14 sets:
\texttt{BCC},
\texttt{BCT},
\texttt{CUB},
\texttt{FCC},
\texttt{HEX},
\texttt{MCL},
\texttt{MCLC},
\texttt{ORC},
\texttt{ORCC},
\texttt{ORCF},
\texttt{ORCI},
\texttt{RHL},
\texttt{TET}, and
\texttt{TRI}.
The name of each materials entry is generated using the format
\texttt{<composition>\_ICSD\_<ICSD number>}.
To aid reproducibility, the species labels in the \texttt{composition} part of the entry name are
the full \VASP\ pseudopotential (POTCAR) names used for that calculation.
Since the materials in this catalog are already known to exist, the primary interest is in accurately calculating
electronic structure and thermo-mechanical properties.
Therefore, calculations for this catalog are generally performed using the Hubbard $U$ correction to the \DFT\
exchange-correlation functional~\cite{Liechtenstein1995, Dudarev1998} where appropriate, using a set of standardized $U$ values~\cite{curtarolo:art104}.

The entries in the other catalogs, such as \texttt{LIB1}, \texttt{LIB2}, and \texttt{LIB3},
are generated by decorating crystal structure prototypes to predict new hypothetical compounds, and contain unary, binary, and ternary materials, respectively.
Within each catalog, the entries are grouped by element and exchange-correlation functional in the case of \texttt{LIB1},
and by alloy system in the cases of \texttt{LIB2} and \texttt{LIB3}.
\texttt{LIB1} contains a total of 4k entries, while \texttt{LIB2} currently has 364k entries and \texttt{LIB3} has about 2.6 million.
Within each alloy system, the individual materials entries are named according to the relevant crystal prototype.
For example, the directory path for the composition AlCu$_2$Mg in the Heusler structure is
\texttt{LIB3\_WEB/AlCu\_pvMg\_pv/T0001.AB2C},
where \texttt{LIB3\_WEB} is the project layer corresponding to \texttt{LIB3}, \texttt{AlCu\_pvMg\_pv} is the set layer corresponding to the Al-Cu-Mg alloy system
(using the \texttt{Al}, \texttt{Cu\_pv} and \texttt{Mg\_pv} pseudopotential POTCARs), and \texttt{T0001.A2BC} is the \texttt{A2BC} decoration of the Heusler prototype \texttt{T0001}.
For these catalogs, the emphasis is on the discovery of new thermodynamically stable or metastable materials,
and on their use to generate the thermodynamic density of states for the prediction of the formation of disordered
materials such as metallic glasses~\cite{curtarolo:art112} or high entropy alloys~\cite{curtarolo:art139}.
Therefore, calculations in these catalogs are performed using the GGA-PBE exchange-correlation functional~\cite{PBE} without Hubbard $U$ corrections~\cite{curtarolo:art104}
so as to produce consistent energy differences, enabling the calculation of accurate  formation enthalpies.

Categorizing data based on the number of species, composition, and crystal prototype is an
intuitive, human-readable choice. However, this data structure introduces inefficiencies that make
it unsuitable for machine reading. The number of possible element combinations and crystal
prototypes increases with the number of species. This results in an unbalanced search
tree as illustrated in Figure~\ref{fig:database_tree}(a). For instance, the catalog of
binary~(\texttt{LIB2}) and ternary entries~(\texttt{LIB3}) may have $10^3$ and $10^5$ entries,
respectively. In the alloy layer, the number of entries may be unevenly distributed as well. This
data structure is inefficient to query and index: walking and searching through \texttt{LIB3} is
significantly slower than through \texttt{LIB2}, resulting in uneven performance and preventing
efficient parallelization. Using other materials properties to structure the data would present
similar challenges as no property is evenly distributed across all systems.

The \underline{A}FLOW \underline{U}nique \underline{ID}entifier~(\AUID) system eliminates this
computational inefficiency. It consists of the \texttt{aflow} namespace declaration, followed by
a quasi-random 64-bit~(8-byte) hash, for example:

\begin{center}
\texttt{aflow:d3aa24a7307877b5}.
\end{center}
This identifier has the capacity for $2^{64} \approx 10^{19}$ entries and can be further extended
by using different namespaces. The \AUIDs\ are generated by hashing
the alphabetically ordered output files (\OUTCAR) of the
\AFLOW\ run using a 64-bit cyclic redundancy check~(CRC64) algorithm. Since the content of these
files depends on parameters specific to the calculation, such as the options of the \VASP\ calculations
or the input and final structure, the resulting \AUID\ is globally unique to the entry. Changing the
cut-off energy, for example, would result in a different identifier. Using \OUTCAR\ files as the basis for
the algorithm also makes the \AUID\ reproducible and thus persistent, fulfilling the ``Findable''
criterion of the \FAIR\ data sharing principles~\cite{Wilkinson_FAIR_SciData_2016}.

The byte expression of the \AUID\ offers a new way to structure data where each layer is represented
by an individual byte of the \AUID. This scheme is shown in Figure~\ref{fig:database_tree}(b). While
this leads to a larger number of branches, they are smaller~($\sim 10^2$ entries) with an
even distribution of \AUIDs\ across each byte layer due to the quasi-randomness of the hash. These
balanced search trees are fast to query and index and also enable efficient parallelization.

Due to this computational efficiency, \AUIDs\ are used throughout the \AFLOW\ software and web
applications when interfacing with the \AFLOW\ database. However, using a hash-based identifier has
the major drawback that it is not human-readable as opposed to a material-based identifier such as:

\begin{center}
\texttt{ICSD\_WEB/ORCC/Ag5S4Sb1\_ICSD\_16987}.
\end{center}
The data structures presented in this section are thus complementary. The \AFLOW\
database provides both methods of access to its materials entries.
\subsection{Accessing and Searching the Database}

\subsubsection{The \AFLOW\ \RESTAPI}
\noindent
To facilitate accessibility and interoperability to the 40~TB of data generated by the high-throughput \AFLOW\ process~\cite{curtarolo:art75},
the backing store is exposed via the \AFLOW\ Data \RESTAPI~\cite{curtarolo:art75} in a hierarchical organization.
This direct exposure of our results not only grants the end user a high degree of utility via direct access,
but more importantly, guarantees data provenance that promotes reproducibility.
The hierarchy of the \AFLOW\ Data \RESTAPI\ categorizes this abundance of information into meaningful high-level
classifications, allowing for exploration of self-similar materials that are related by stoichiometric and/or
crystallographic properties.
Once a selection of materials has been determined, the full range of available properties and procedural data are retrievable.

The organizational hierarchy of both the underlying data store and the \RESTAPI\ is project-dependent, as described
in Section~\ref{database_organization}.
Each project is equivalent to one of the catalogs listed in this section,
and in the \RESTAPI\ is denoted by the project layers \texttt{ICSD\_WEB}, \texttt{LIB1\_WEB}, \texttt{LIB2\_WEB}, and \texttt{LIB3\_WEB}.
Each project layer contains multiple set layers, which correspond to specific alloy systems in the case of \texttt{LIB1\_WEB}, \texttt{LIB2\_WEB}, \texttt{LIB3\_WEB}, etc.
For instance: \href{https://aflowlib.duke.edu/AFLOWDATA/LIB2\_RAW/}{aflowlib.duke.edu/AFLOWDATA/LIB2\_RAW} exposes the set layer for
binary entries, where each set corresponds to a different binary alloy system, allowing for pairwise atomic species examination.
Within each set is the entry layer, consisting of decorated structural prototypes which provide stoichiometric and structural variation for each alloy system.
Each entry contains the calculated results for a particular structure and composition, organized as keyword-value pairs.
The calculated values of thermodynamic, mechanical, electronic, magnetic, chemical and crystallographic properties can
be directly accessed by querying a \underline{U}niform \underline{R}esource \underline{I}dentifier (\URI) of the form \texttt{<server>/<project>/<set>/<entry>/?<keyword>},
where \texttt{<server>} is \texttt{aflowlib.duke.edu/AFLOWDATA}, \texttt{<project>} is the appropriate
project layer, \texttt{<set>} is the alloy system, \texttt{<entry>} is the structural prototype, and \texttt{<keyword>}
corresponds to the materials property of interest.
For example, the formation enthalpy per atom of AlCu$_2$Mg in the Heusler structure can be accessed at
\href{https://aflowlib.duke.edu/AFLOWDATA/LIB3\_WEB/AlCu\_pvMg\_pv/T0001.AB2C/?enthalpy\_formation\_atom}
{\nolinkurl{aflowlib.duke.edu/AFLOWDATA/LIB3\_WEB/AlCu\_pvMg\_pv/T0001.AB2C/?enthalpy\_formation\_atom}},
where the project is \texttt{LIB3\_WEB}, the set is the alloy system \texttt{AlCu\_pvMg\_pv}, the entry is the prototype label
\texttt{T0001.A2BC}, and the keyword is \texttt{enthalpy\_formation\_atom}.
A full list with descriptions of the \RESTAPI\ keywords is provided in the Appendix of this work,
combining the original list from Ref. \onlinecite{curtarolo:art92}
with the additions in the appendices of Refs. \onlinecite{curtarolo:art128, curtarolo:art115}.

The information required to construct the \URI\ for each entry in the database is available through the
\texttt{AURL} (\AFLOW\ \underline{U}niform \underline{R}esource \underline{L}ocator) keyword,
which is returned by default for each result in an \AFLUX\ search.
The \texttt{AURL} for each entry is similar in form to the first part of the \URI: \texttt{<server>/<project>/<set>/<entry>},
except that the \texttt{server} part has the form \texttt{aflowlib.duke.edu:AFLOWDATA}.
By replacing the ``:'' with a ``/'', the base \URI\ for the corresponding entry can be constructed.
The \texttt{AURL} is particularly useful when trying to retrieve data not directly accessible through \AFLUX\, such as calculation input or output files.
For example, the file with the relaxed structure for a particular entry, \texttt{CONTCAR.relax},
can be downloaded at the \URI\ \texttt{<server>/<project>/<set>/<entry>/CONTCAR.relax},
which can be constructed from the \texttt{AURL}.
\texttt{AURL}s can also be used to systematically access all of the entries in a particular \texttt{set}.

The ability to explore related entries predicated on a multitude of properties leads directly to novel materials discovery and use.
The \AFLOW\ Data \RESTAPI\ disseminates our methods and results, without restriction, to a
global research audience in order to promote scientific and engineering advancement.

\def\BLOCK{\texttt{BLOCK}}
\def\AND{\texttt{AND}}
\def\OR{\texttt{OR}}
\def\NOT{\texttt{NOT}}
\def\LOOSE{\texttt{LOOSE}}
\def\STRING{\texttt{STRING}}
\def\MUTE{\texttt{MUTE}}
\def\PAGING{\texttt{paging}}
\def\FORMAT{\texttt{format}}
\def\HELP{\texttt{help}}
\newcommand{\afluxcmd}[1]{\hfill\\\indent{\small {#1}}\hfill\\}

\subsubsection{\AFLUX: The \AFLOW\ search API}
\noindent
The \AFLUX\ API provides programmatic search functionality for the \AFLOW\ database~\cite{curtarolo:art128}.
Queries are submitted through a \URI\ and can thus be conducted with commonly available tools such
as \texttt{wget}, \texttt{curl}, Python's \texttt{urllib} module, or a web browser. Results are
returned in easily parsed formats, such as \JSON. This makes \AFLUX\ platform-independent and
enables integration into computational workflows written in all commonly used languages.

\AFLUX\ is a domain-specific language, i.e., no prior knowledge on the underlying database
structure or format is needed. It requires only a very small set of operators, shown in
Table~\ref{tab:aflux_operators}, which further reduces barriers to using it. An \AFLUX\ call, or
a {\it summons}, is written in the query portion of the \URL:
\begin{lstlisting}[language=aflux]
aflow.org/API/aflux/?<summons>
\end{lstlisting}
The summons is composed of a {\it matchbook} and {\it directives}:
\begin{lstlisting}[language=aflux]
aflow.org/API/aflux/?<matchbook>,<directive>
\end{lstlisting}
The matchbook consists of the search criteria whereas directives guide the form of the results.

\begin{table}
  \caption{List of \AFLUX\ operators, their symbols, and their scope.
  Inter-property operators can connect properties.
  Intra-property operators apply to the property alone.}
  \label{tab:aflux_operators}
  \begin{tabular*}{\columnwidth}{c @{\extracolsep{\fill}} cc}
    \hline
    Operator & Symbol & Scope \\
    \hline
    \BLOCK   & ( )    & inter- \& intra-property \\
    \NOT     & !      & intra-property \\
    \AND     & ,      & inter- \& intra-property \\
    \OR      & :      & inter- \& intra-property \\
    \LOOSE   & *      & intra-property \\
    \STRING  & ' '    & intra-property \\
    \MUTE    & \$     & intra-property \\
    \hline
  \end{tabular*}
\end{table}

\noindent \textbf{The \AFLUX\ matchbook.}
To search for and filter by properties, they need to be added to the matchbook. The following
summons lists the band gap~(\texttt{Egap} of each returned entry:
\begin{lstlisting}[language=aflux]
aflow.org/API/aflux/?Egap
\end{lstlisting}
Or in short:
\begin{lstlisting}[language=aflux]
<matchbook> = Egap
\end{lstlisting}
A full list of properties can be found in Appendix~A. Multiple properties can be returned by using
the \AND~(\texttt{,}) operator. For instance,
\begin{lstlisting}[language=aflux]
<matchbook> = Egap,spin_atom
\end{lstlisting}
returns both the band gap and the magnetic moment per atom of each material.
Properties can be restricted to a specific value using the \BLOCK~(parentheses) operator:
\begin{lstlisting}[language=aflux]
<matchbook> = species(Cr)
\end{lstlisting}
This summons will only return entries containing Chromium (Cr). These restrictions also work for
numerical values. The matchbook
\begin{lstlisting}[language=aflux]
<matchbook> = Egap(1)
\end{lstlisting}
returns all entries with a band gap of exactly 1 eV.

\noindent \textbf{Set logic.}
\AFLUX\ supports basic logical operators: \NOT~(\texttt{!}), \AND~(\texttt{,}), and \OR~(\texttt{:}).
\NOT\ can be used to exclude values from the results. For example, to show entries without Cr, the
matchbook should be:
\begin{lstlisting}[language=aflux]
<matchbook> = species(!Cr)
\end{lstlisting}
The \AND\ operator, previously used to retrieve multiple properties, can also be employed as
an intra-property operator. The following summons will return all entries containing Cr and O:
\begin{lstlisting}[language=aflux]
<matchbook> = species(Cr,O)
\end{lstlisting}
To get all entries that include Cr or Mn, but not necessarily both, the \OR\ operator can be used:
\begin{lstlisting}[language=aflux]
<matchbook> = species(Cr:Mn)
\end{lstlisting}
When combined with the \BLOCK\ operator, these logical queries can become arbitrarily complex.
The following matchbook will return all materials with either Cr or Mn~(or both) and at least one
chalcogen:
\begin{lstlisting}[language=aflux]
<matchbook> = species((Cr:Mn),(O:S:Se:Te))
\end{lstlisting}

\noindent\textbf{Value ranges and substrings.}
Exact matches may not always be desired, especially for numeric quantities like the band gap~(Egap).
The \LOOSE\ operator~(\texttt{*}) can be used to build inequalities. It is a positional operator
where $N*$ returns values $\ge N$ and $*N$ returns values $\le N$, for example:
\begin{lstlisting}[language=aflux]
<matchbook> = Egap(0.6*)
\end{lstlisting}
returns entries with band gaps of 0.6 eV and higher. This can be combined with the \AND\ operator to
form a tolerance range:
\begin{lstlisting}[language=aflux]
<matchbook> = Egap(0.6*,*1.5)
\end{lstlisting}
This returns entries with $0.6\,\textnormal{eV} \le \texttt{Egap} \le 1.5\,\textnormal{eV}$.
It is also possible to select results that satisfy multiple ranges by using
the \OR\ operator:
\begin{lstlisting}[language=aflux]
<matchbook> = Egap((0.5*,*0.7):(1.4*,*1.6))
\end{lstlisting}
To get a strictly greater than 1.6 eV match condition, the \NOT\ operator can be combined with the
\LOOSE\ operator:
\begin{lstlisting}[language=aflux]
<matchbook> = Egap(!*1.6)
\end{lstlisting}
Using the \LOOSE\ operator without any value will discard null results, which are returned by
default. The matchbook
\begin{lstlisting}[language=aflux]
<matchbook> = Egap(*)
\end{lstlisting}
will only return entries for which Egap has been calculated.

The \LOOSE\ operator can also be used for substring searches. The position of the \LOOSE\ operator
determines where the substring ought to be located. The matchbooks
\begin{lstlisting}[language=aflux]
<matchbook> = Pearson_symbol_relax('c'*)
<matchbook> = Pearson_symbol_relax(*'F'*)
<matchbook> = Pearson_symbol_relax(*'8')
\end{lstlisting}
find entries with substrings at the beginning, in the middle, and at the end, respectively.
Single quotes~(\STRING\ operator) are required for substring searches. They can also be used to
avoid collisions with \AFLUX\ operators as in
\begin{lstlisting}[language=aflux]
<matchbook> = auid('aflow:1dcf91da07337d8d')
\end{lstlisting}
Without \STRING, \AFLUX\ would search for \texttt{auid='aflow' \OR\ auid='1dcf91da07337d8d'},
resulting in an empty response.

\noindent\textbf{Removing redundancies.}
Properties in a search result may be ubiquitous and therefore of little use. For example, in the
matchbook
\begin{lstlisting}[language=aflux]
<matchbook> = Egap(0.6*,*1.5),nspecies(3)
\end{lstlisting}
each returned material will have three species, bloating the response with redundant
information. \AFLUX\ can suppress the output of a match criterion via the \MUTE~(\texttt{\$})
operator:
\begin{lstlisting}[language=aflux]
<matchbook> = Egap(0.6*,*1.5),
\end{lstlisting}
This allows a property to be matched, but not displayed.

\noindent\textbf{Aliases.}
Complex \AFLUX\ summons can become cumbersome to type. For example, to search for materials
containing a transition metal and a chalcogen, the matchbook needs to contain:
\begin{lstlisting}[language=aflux]
<matchbook> = species((Sc:Ti:V:...:Hg),(O:S:Se:Te))
\end{lstlisting}
where $\ldots$ contains 25 more species. \AFLUX\ contains pre-defined aliases to simplify this
summons to:
\begin{lstlisting}[language=aflux]
<matchbook> = species(TransitionMetals,Chalcogens)
\end{lstlisting}
A list of all supported aliases can be found at
\href{https://aflow.org/API/aflux/?help(aliases)}{aflow.org/API/aflux/?help(aliases)}.

\noindent\textbf{\AFLUX\ directives.}
Directives determine the size and format of an \AFLUX\ response and provide documentation.
The currently supported directives are as follows:
\begin{itemize}
    \item format,
    \item paging,
    \item help.
\end{itemize}

\noindent \textbf{The \FORMAT\ directive.}
By default, \AFLUX\ returns search results as minified \JSON. The \FORMAT\ directive can be
used to change the output format:
\begin{lstlisting}[language=aflux]
<directive> = format(
\end{lstlisting}
Supported formats are \textit{json}, \textit{aflow}, and \textit{html}. Using json formats the
response as a human-readable \JSON\ file. The aflow format returns the data in the aflowlib.out
format, i.e., as key-value pairs separated by \texttt{|}. Finally, html displays the results as an
HTML table.

\noindent \textbf{The \PAGING\ directive: response size and sorting.}
The size of the \AFLOW\ repository can lead to very large responses when the search criteria are
broad. This can overwhelm systems communicating with \AFLUX, especially when a web browser is
used. As a result, \AFLUX\ limits the number of entries it returns at once. The \PAGING\ directive
controls this response sizing and enumeration:
\begin{lstlisting}[language=aflux]
<directive> = paging(
\end{lstlisting}
This returns the $J^\textnormal{th}$ page with $K$ results per page. For example, the directive:
\begin{lstlisting}[language=aflux]
<directive> = paging(3,100)
\end{lstlisting}
can be used to retrieve the results 201 -- 300 (3\textsuperscript{rd} page with 100 entries per
page). The default values are $J = 1$ and $K = 64$. If $J$ is larger than the number of available
pages, the response will be an empty array or an empty string for json and aflow format,
respectively. This can be used as a stop condition in a while-loop for high-throughput retrieval.
Setting $J$ or $K$ to zero results in special behavior: $J = 0$ returns all results at once;
$K = 0$ returns the number of results, regardless of the value of $J$.

Paging can also be used to sort results. Positive and negative values for $J$ sort results in
ascending and descending order, respectively. This also works for $J = 0$ to return all results in
descending order:
\begin{lstlisting}[language=aflux]
<directive> = paging(-0)
\end{lstlisting}

When the output format is \JSON, using the \PAGING\ directive returns the results as a dictionary.
The \MUTE\ operator can be used to format the data as an array instead:
\begin{lstlisting}[language=aflux]
<directive> =
\end{lstlisting}
This is the default behavior when \PAGING\ is not added to the directive.

\noindent \textbf{The \HELP\ directive: \AFLUX\ documentation.}
Usage information on \AFLUX\ can be obtained using \HELP. The directive:
\begin{lstlisting}[language=aflux]
<directive> = help(general)
\end{lstlisting}
produces a general help file that serves as the documentation for \AFLUX. The same output can be
found when no argument to \HELP\ is given or when both the matchbook and the directive are empty.
Information on available properties~(see Appendix A) can be:
\begin{lstlisting}[language=aflux]
<directive> = help(properties)
\end{lstlisting}
Replacing ``properties'' with a list of keywords will limit the output to the selected
properties. All available help files are listed in the following directive:
\begin{lstlisting}[language=aflux]
<directive> = help(help)
\end{lstlisting}

\begin{figure*}
  \includegraphics[width=\textwidth]{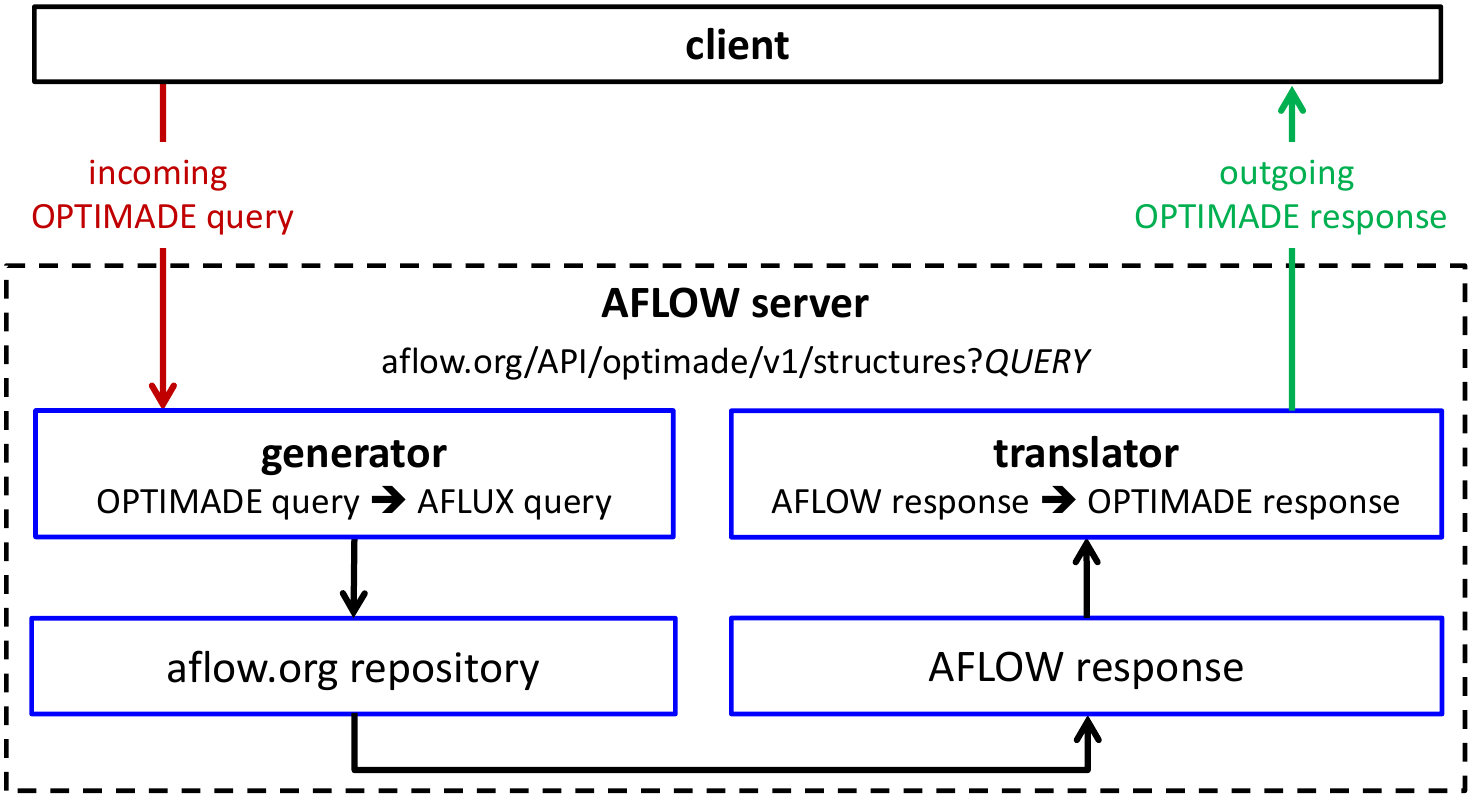}
  \caption{\textbf{Infrastructure for responding to \OPTIMADE\ queries.}
  Illustration of the workflow for processing an \OPTIMADE\ query.
  The query is translated to \AFLUX\ (via the generator) for retrieving the data
  and the \AFLOW\ response is converted to \OPTIMADE\ (via the translator) before
  returning to the client.
  }
  \label{fig:web_optimade}
\end{figure*}

\subsubsection{\AFLOWOPTIMADE\ Integration}
\noindent
To facilitate interoperability with other databases in accordance with the FAIR criteria, as well as to broaden accessibility and findability, the \OPTIMADE\ common \API~\cite{curtarolo:art178} has been implemented within the \AFLOW\ database.
\OPTIMADE\ is a common API for searching materials data across a range of repositories, including \AFLOW, NOMAD~\cite{nomadMRS}, Materials Project~\cite{APL_Mater_Jain2013}, Materials Cloud~\cite{Talirz_MatCloud_SciData_2020}, and OQMD~\cite{oqmd.org}. The full specification is available at \url{https://github.com/Materials-Consortia/OPTIMADE/blob/master/optimade.rst}.
Similar to \AFLUX, searches are encoded within the query part of URL, and can include logic operations such as \texttt{AND}, \texttt{OR} and \texttt{NOT}; searches for different element combinations using the operators \texttt{HAS ALL}, \texttt{HAS ANY} and \texttt{HAS ONLY}; substring comparisons using \texttt{CONTAINS}, \texttt{STARTS WITH} and \texttt{ENDS WITH};  and searches for property values within specific ranges using \texttt{>}, \texttt{<} and \texttt{=}.
Data is returned in as a JSON object, in a standardized format to maximize interoperability.

The \OPTIMADE\ API for \AFLOW\ can be accessed at \href{https://aflow.org/API/optimade}{aflow.org/API/optimade}.
Available queryable endpoints include ones for \texttt{structures} and \texttt{calculations}.
The server takes the \OPTIMADE\ query, converts it to an \AFLUX\ query for the \AFLOW\ database,
then reads the \AFLUX\ response and translates it into the standardized format, as illustrated in Figure~\ref{fig:web_optimade}.

\AFLOW\ keywords can be included in Optimade queries of the \AFLOW\ database by appending the prefix ``\_aflow\_'' to the front of the keyword.
For example, lead-free halide cubic perovskites with a band gap exceeding 3 eV can be searched for in the \AFLOW\ database using optimade with the following query:
\href{https://aflow.org/API/optimade/v1/structures?filter=elements\%20HAS\%20ANY\%20\%27F\%27,\%27Cl\%27,\%27Br\%27,\%27I\%27\%20AND\%20NOT\%20elements\%20HAS\%20\%27Pb\%27\%20AND\%20_aflow_aflow_prototype_label_relax\%20CONTAINS\%20AB3C_cP5_221_a_c_b}{aflow.org/API/optimade/v1/structures?filter=elements HAS ANY "F","Cl","Br","I" AND NOT elements HAS "Pb" AND \_aflow\_aflow\_prototype\_label\_relax CONTAINS AB3C\_cP5\_221\_a\_c\_b}.
Full lists of the keywords available to use with \OPTIMADE\ to query \AFLOW\ are available at the \texttt{info} endpoints \href{https://aflow.org/API/optimade/v1/info/structures}{aflow.org/API/optimade/v1/info/structures} and \href{https://aflow.org/API/optimade/v1/info/calculations}{aflow.org/API/optimade/v1/info/calculations}.

\subsubsection{The \AFLOW\ Search GUI}
\begin{figure*}
  \includegraphics[width=\textwidth]{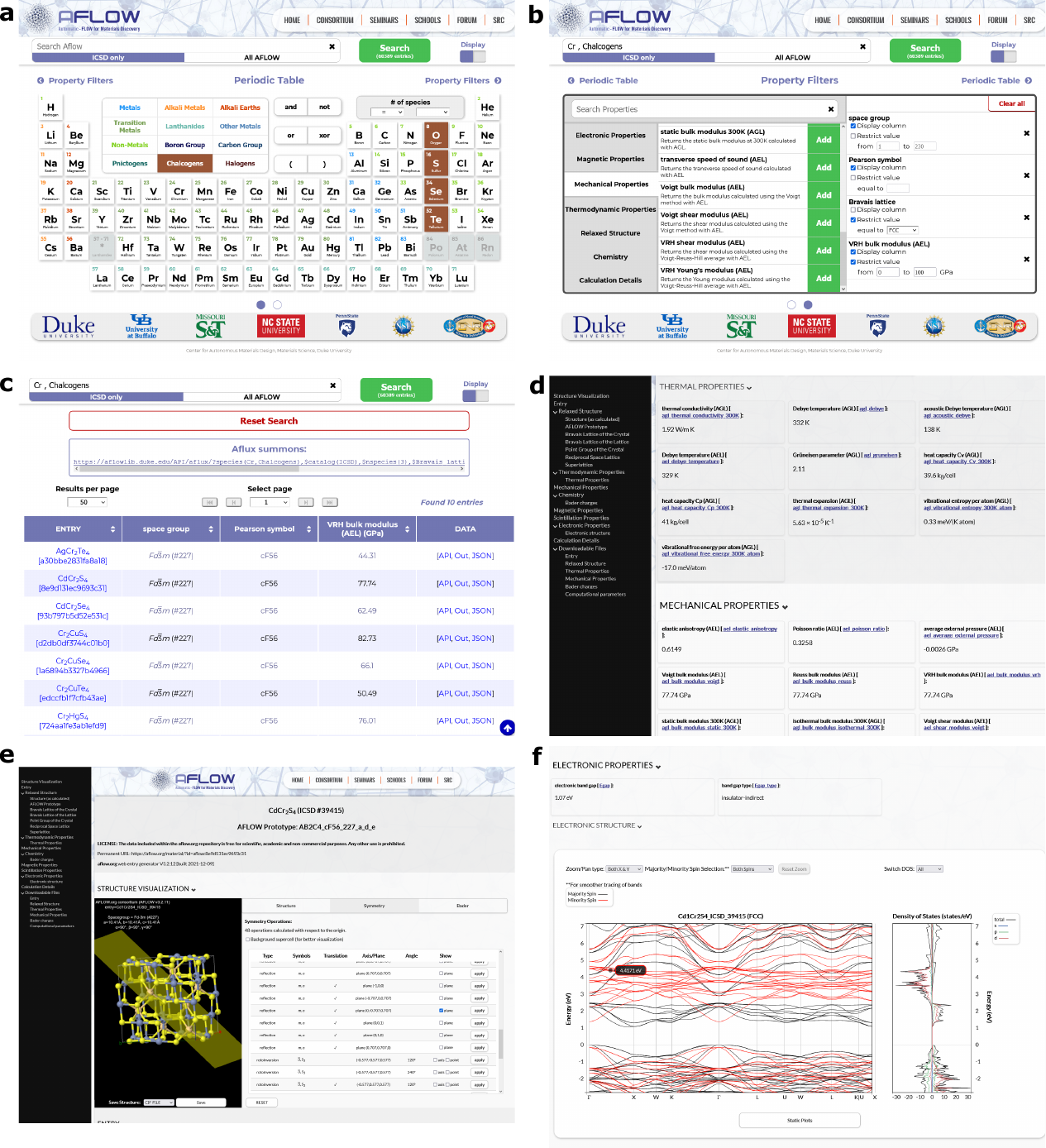}
  \caption{\textbf{The \AFLOW\ Search page and materials entry page.} (\textbf{a}) Element selection,
  (\textbf{b}) property filter, and (\textbf{c}) search results of the search page. (\textbf{d})
  Thermomechanical properties, (\textbf{e}) structure visualization and symmetry section, and
  (\textbf{f}) interactive electronic structure of the entry page for spinel CdCr$_2$S$_4$.}
  \label{fig:search_entry_page}
\end{figure*}

\noindent
Programmatic access to data is essential for integrating the \AFLOW\ data into workflows, but is
not suitable for casual searches or for users unfamiliar with coding or the \AFLUX\ language.
To fulfill these usage requirements, \AFLOWorg\ provides a search application, which can be
be accessed at \href{https://aflow.org/search/}{aflow.org/search/}. It consists of four
components: the search bar, an element filter, a properties filter, and the search results.
The search bar, at the top of the page~(see Figures~\ref{fig:search_entry_page}(a) and (b)), contains a
text field for the elements of interest, the button to conduct a search, and the ``Display'' slider.
The text field also contains buttons that can limit the search to the ICSD catalog or extend it to
the entire \AFLOW\ database. The ``Display'' slider toggles between having the element and property
filters on separate slides and to have them both displayed on a single page, depending on the
preferences of the user.

The element filter is shown in Figure~\ref{fig:search_entry_page}(a) and takes shape of the periodic
table. Elements can be added to the query by clicking on its chemical symbol or by directly typing
them into the search bar. Groups of elements can also be selected using the element group selector.
Hovering over each button reveals which elements belong to each group, as demonstrated for the
chalcogens in the figure. Next to the element group selectors is a set of logical operators.
Elements or element groups can be excluded using the \texttt{NOT} button or combined using
\texttt{AND}, \texttt{OR}, or \texttt{XOR}~(exclusive or). \texttt{AND} is automatically added when
clicking on two elements successively. Parentheses can be used to make these combinations
arbitrarily complex. Using the \AFLUX~\cite{curtarolo:art128} syntax, these logic statements can
also be directly entered into the search bar. However, using the buttons can prevent syntax errors
because they only accept valid operators, e.g., by disallowing using \texttt{OR} immediately after
\texttt{AND}. In either case, the syntax is validated before submitting a search. Lastly, the
number of species can be set with ``\# of species'' selector. Both ranges or an exact number are
possible.

The property selectors can be displayed by clicking on the arrow next to the ``Property Filters''
label or by flipping the ``Display'' slider and is shown in Figure~\ref{fig:search_entry_page}.
Available properties are organized into different categories, such as electronic, mechanical, or
symmetry properties of the relaxed structure. A property can be added to the search query by
clicking the ``Add'' button. The text field above the categories and
descriptions serves as a search field to provide a convenient way to add a property to the list
of filters. Selections can be removed via the ``$\times$'' button at the right-hand side of its
card. In the example shown in Figure~\ref{fig:search_entry_page}(b), ``Bravais lattice'' and
``VRH bulk modulus~(AEL)'' were added to the search query~(``space group'' and ``Pearson symbols''
are added by default). The figure also shows that the values for these two properties were set to
restricted to retrieve only face-centered cubic compounds with a VRH bulk modulus of 0~--~100\,\GPa.
Additionally, the ``Bravais lattice'' property has ``Display column'' unchecked, which will result
in it not being displayed in the results table. Since the Bravais lattice will be the same value
for every material, this will help focus the search output.

Figure~\ref{fig:search_entry_page}(c) shows the results of this query. The driver of the search is
\AFLUX~\cite{curtarolo:art128} --- its summons is shown in the field below the ``Reset Search''
button. The results table is divided into pages and always contains the \texttt{ENTRY} and
\texttt{DATA} columns. \texttt{ENTRY} consists of the empirical formula and the \AUID\ of the entry.
The link points to the material's entry page, which will be discussed later. The \texttt{DATA}
column contains links to the RestAPI page of the entry and its \texttt{aflowlib.out} and
\texttt{alowlib.json} files. The remaining columns contain the requested properties for which
``Display column'' is switched on --- column visibility can be toggled after the search as well.
The tables can be sorted by each visible column except \texttt{DATA} and array properties.

The entry page contains the calculated properties and downloadable files for the material. As for
the search page, properties are organized into categories to facilitate navigation. Two example
sections --- thermal properties and mechanical properties --- are
shown in Figure~\ref{fig:search_entry_page}(d). Each property has its own card containing its value and
unit. The property cards also contain links to their RestAPI entries~\cite{curtarolo:art92}. The
``Downloadable Files'' section at the bottom of the page is structured the same way and contains
various input and output files such as \VASP\ inputs, Bader charges, and outputs of the \AFLOW\
modules used for the entry. The input files and calculation parameter sections fulfill the
\FAIR\ ``Reusable'' criterion and help reproduce the results calculated by \AFLOW.

There are also two interactive applets on this page: one for a crystal structure visualization and
another one for electronic structures. Figure~\ref{fig:search_entry_page}(e) shows the structure viewer
tool. It is available for both the conventional and primitive cell of the material and is based on
JSmol~\cite{Jmol,Jmol_Hanson}. Apart from basic structure viewing options, it offers visualizations for
the symmetry elements of the structure calculated by \AFLOWSYM~\cite{curtarolo:art135}. The
``Symmetry'' tab shows all available operations and displays their types, symbols in the Schoenflies
and Hermann-Mauguin notation, whether they contain translation, their symmetry plane or axis in
fractional coordinates, and their rotation angles. The axes and planes, as well as the inversion
points~(for inversions and rotoinversions) can be displayed, as shown in
Figure~\ref{fig:search_entry_page}(e). The ``Apply'' buttons start animations that visualize the
selected symmetry operation --- a supercell of the structure can be displayed as a visual aid.
Where available, the applet can also display Bader isosurfaces for each atom in the primitive cell
for different cut-off values.

The other interactive applet shows the electronic structure of the material~(see
Figure~\ref{fig:search_entry_page}(f)). Bands of both spin channels can be traced with a mouse,
showing the energy of the band at every step. Zooming in and dragging along the $k$-path axis is
possible as well. The lines of the \underline{d}ensity \underline{o}f \underline{s}tates~(\DOS)
can also be traced, but the value of the \DOS\ is shown instead of the energy. By default, the total
\DOS\ and the projections along the orbitals are shown, but the projected \DOS\ of each
symmetrically-inequivalent atom is accessible using the ``Switch DOS'' drop-down menu. Static high-quality images of
both the total and projected \DOS\ are available through the button underneath the applet.

\section{The Prototype Encyclopedia}

\noindent
Beginning researchers in computational materials science often find it
difficult to determine exactly which structure is to be studied.  Take
as an example a study of the behavior of the ``high-temperature''
(as classified in 1969~\cite{Willens_SSC_7_837_1969})
superconductor Nb$_{3}$Al.
Performing computations for this material requires knowledge of its crystal
structure, including both the unit cell dimensions and
the position of each atom in that cell,
but novices in the field may have difficulty in finding this
information in the literature.
The original Nb$_{3}$Al article
refers to the $\beta$-W structure and hints that the structure is
similar to that of V$_{3}$Sn and Nb$_{3}$Sn.
Searching further, our hypothetical researcher might find references
to Cr$_{3}$Si as the ``prototype'' of this structure, or it might
simply be referred to as a mysterious ``A15'' superconductor.  These
sources assume that the structure is so well known that it is not
necessary to specify the atomic positions or the space group of the
structure, and might not even mention that the crystal is cubic.  This
problem persists across the literature,
obfuscating the relevant crystal structure information needed to perform computational studies.

To address this challenge, the \AFLOW\ Encyclopedia of Crystallographic Prototypes
(or Prototype Encyclopedia for brevity) serves as an extensive catalog of
crystalline prototypes hosted at
\href{https://aflow.org/prototype-encyclopedia}{aflow.org/prototype-encyclopedia/}.
Based on the ``Crystal Lattice Structures'' website started
by researchers at the U.S. Naval Research Laboratory's (NRL) Complex Systems Theory Branch
(now the Center for Computational Materials Science), the Curtarolo Group at Duke University
has continued and extended this effort via the Prototype Encyclopedia.
A web page for each structure can be found online, detailing
its space group, lattice constants, and atomic (Wyckoff) positions.
To assist new researchers and students, each page expanded this basic information by
explicitly describing the position of each atom in the structure's
primitive cell given the Wyckoff positions of its site.
It also lists compounds with the same structure as the prototype, and
included many of the common {\em Strukturbericht}
symbols~\cite{Ewald_Struk_I_1931}.
A simple search of the website would then reveal the structure of Nb$_{3}$Al~\cite{CLS_Wayback_2010}.
In addition to online search capabilities, users can generate
geometry files based on these structures with underlying \AFLOW\ prototyping functionality.

There are currently three parts to the \AFLOW\ Prototype Encyclopedia.
Each release resulted in a major update to the website --- adding a page for each new prototype ---
and is accompanied with an article in {\em Computational Materials Science}.
A summary of each part is as follows:
\begin{itemize}
\item Part 1~\cite{curtarolo:art121}: comprised of 288 structures found on the
  original ``Crystal Lattice Structures'' page.
  It also includes a discussion of topics in crystallography (found both online and in the article),
  namely three-dimensional systems, including crystal systems, lattice types,
  symmetry operations, and Wyckoff positions.
\item Part 2~\cite{curtarolo:art145}: added 302 new structures, including at least
  one structure from each of the 230 three-dimensional space groups.
  Additional topics in crystallography are discussed, including
  enantiomorphic space groups, the Wigner-Seitz description of
  primitive cells in a periodic system, and descriptions of the seventeen two-dimensional plane groups.
\item Part 3~\cite{aflowANRL3}: added 510 new structures,
  including the structure of each inorganic {\em Strukturbericht} prototype found in the original
  German publications from 1937 through 1943 and later additions to
  made by
  Smithells~\cite{Smithells_Metals_1949,Smithells_Metals_II_1955} and
  Pearson~\cite{Pearson_NRC_1958}.
  The article concluded with a list of
  the {\em Strukturbericht} symbols and the first complete index of the
  inorganic {\em Strukturbericht} prototypes.
\end{itemize}
To date, there are 1,100 structure prototypes in the Prototype Encyclopedia, and it is
expected to continue to grow.

\begin{figure*}
  \includegraphics[width=\textwidth]{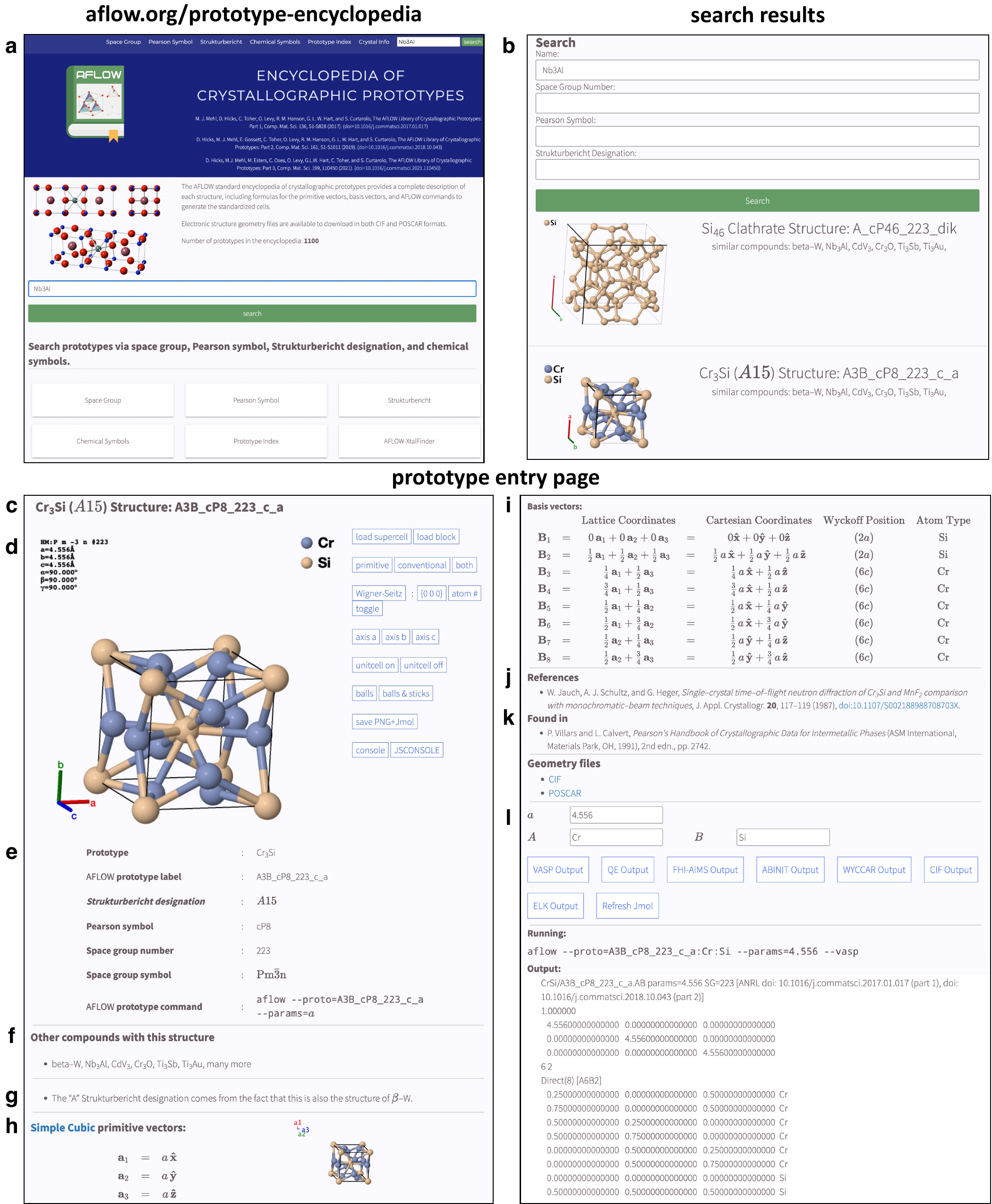}
  \caption{\label{fig:a15} \textbf{The \AFLOW\ Prototype Encyclopedia web page.}
    ({\bf a})~The main web page with a different methods to find specific prototypes,
    including a search bar.
    ({\bf b})~An example response page when searching for ``Nb3Al''.
    The prototype page for
    Cr$_{3}$Al, the ``A15'' prototype of Nb$_{3}$Al described in the
    text can be found at
    \href{https://www.aflow.org/prototype-encyclopedia/A3B\_cP8\_223\_c\_a.html}{aflow.org/prototype-encyclopedia/A3B\_cP8\_223\_c\_a.html}.
    Each page contains the following:
    ({\bf c})~title,
    ({\bf d})~Jmol viewer,
    ({\bf e})~prototype designations,
    ({\bf f})~a list of compounds/elements exhibiting this structure,
    ({\bf g})~notable comments,
    ({\bf h})~primitive vectors,
    ({\bf i})~basis vectors,
    ({\bf j})~references,
    ({\bf k})~additional references where the structure was found, and
    ({\bf l})~an automatic prototype generator.
}
\end{figure*}

\noindent\textbf{Prototype page content.}
Our hypothetical new user can now easily search for the mysterious
Nb$_{3}$Al or ``A15'' structure.  Once on the Library's home page~(Figure~\ref{fig:a15}(a)), the
user enters ``Nb3Al'' in the search box, revealing the Cr$_{3}$Si
(A15) prototype~(Figure~\ref{fig:a15}(b)).  Clicking on the link reveals the page shown in
Figures~\ref{fig:a15}(c-l). This page describes the structure in detail:
\begin{itemize}
\item The title includes the \AFLOW\ label for the structure.  The label
  also serves as the basis for the web page URL (Figure~\ref{fig:a15}(c)).
\item A rendering of the structure, using Jmol~\cite{Jmol,Jmol_Hanson}.
  Users can rotate and zoom in on the structure, as shown in
  Figure~\ref{fig:a15}(d).  They may also change the window to see a
  supercell of the structure.  Right-clicking on the image brings up
  the standard Jmol menu, and the user can even access the Jmol
  console for more options.
\item The chemical formula of the prototype of the structure (Figure~\ref{fig:a15}(e)).
\item The \AFLOW\ prototype label (Figure~\ref{fig:a15}(e)).
\item The {\em Strukturbericht} designation, if any (Figure~\ref{fig:a15}(e)).
\item The Pearson symbol~\cite{Pearson_NRC_1967}, showing the crystal
  class and the number of atoms in the unit cell (Figure~\ref{fig:a15}(e)).
\item The space group symbol (Figure~\ref{fig:a15}(e)).
\item The format of the \AFLOW\ prototype command which was used the
  generate the structure (Figure~\ref{fig:a15}(e)).
\item A list of known compounds which have this structure, if any.
  This list is updated when similar compounds are found in the
  literature (Figure~\ref{fig:a15}(f)).
\item Any comments about the structure (Figure~\ref{fig:a15}(g)).  This may include links to
  other structures associated with the prototype or related to the
  current crystal structure.
\item The vectors defining the primitive unit cell of the structure (Figure~\ref{fig:a15}(h)).
\item The basis vectors: the position of each atom in the cell is
  described in terms of the primitive vectors and the Cartesian
  coordinates, as a function of the lattice parameters and the Wyckoff
  positions (Figure~\ref{fig:a15}(i)).  The atoms are grouped by Wyckoff position and atomic
  species.
\item The reference to the structure in the literature, if any,
  including a link to the publication, if available (Figure~\ref{fig:a15}(j)).
\item If the above reference was found in another publication or website, that reference and
  its associated link are given (Figure~\ref{fig:a15}(k)).
\end{itemize}

Finally, each page contains a link to the \AFLOW\ prototype generator
for this structure (Figure~\ref{fig:a15}(l)).
The user can
change the chemistry of
the prototype, and enter new lattice constants and Wyckoff coordinates
appropriate to that structure.  Once entered, this information can be
used to update the page showing the new structure.
Information about the structure can be obtained
in a variety of formats,
including popular electronic structure programs and the standard
\underline{C}rystallographic \underline{I}nformation \underline{F}ile (\CIF).

\noindent\textbf{Searching structures by crystallographic descriptors.} The Prototype Encyclopedia website includes resources
which serve as a useful reference for beginning students in
crystallography, adapted from published
articles~\cite{curtarolo:art121,curtarolo:art145,aflowANRL3}.

All periodic systems in three-dimensional space belong to one of seven
crystal systems, each with the same holohedry (rotational symmetry).
Different possible translational symmetries in each class lead to the
fourteen Bravais lattices.  This information is summarized under the
``Crystal Info'' tab on the home page.  This tab provides a brief
description of each
crystal class, the Bravais lattices included
in the class, and the space groups associated with each Bravais
lattice.  Although each lattice can be described by a large (infinite)
variety of
primitive vectors, \AFLOW\ makes a standard choice for
each lattice~\cite{curtarolo:art58}, which we show here.  So, for
example, if a structure is known to have a body-centered orthorhombic
lattice, clicking the ``Orthorhombic'' tag under ``Crystal Info'' and
scrolling down to ``Body-Centered Orthorhombic'' will show:
\begin{itemize}
\item the standard primitive vectors of the lattice,
\item an image of the primitive unit cell and the associated
  conventional unit cell,
\item and list the nine space groups which are associated with this
  lattice.
\end{itemize}

Each crystal structure is associated with a Pearson Symbol, which
summarizes the crystal system, the Bravais lattice, and the number of
atoms in the conventional unit cell of the structure.  The ``Pearson
Symbol'' tab on the home page is an index of all the crystal
structures by symbol.  If the body-centered orthorhombic system
described above
has eight atoms in the conventional cell, then the
user finds and selects the ``Orthorhombic Structures'' link on the
Pearson Symbol page.  The resulting page lists all of the orthorhombic
structures currently in the Prototype Encyclopedia.
Searching for ``oI8'' (orthorhombic class ``o'', body-centered lattice
``I'', and eight atoms for unit cell) leads to the three crystal
structures with this Pearson symbol found in the current database.

The ``Space Group'' tab indexes the structure by space group.  To
continue with the above example, the oI8 structured ferroelectric
NaNO$_{2}$ is in space group $Imm2$~$\#44$.  This structure can be
found under the Orthorhombic Space Group link of the Space Group page,
along with the other four structures currently in this space group
that are currently in the database.

Similarly, the {\em Strukturbericht} tab lists all of the structures
which have been given a {\em Strukturbericht} designation.  The
NaNO$_{2}$ structure described above is labeled $F5_{5}$, and can be
found by clicking on the
``{\em Strukturbericht} Type F''
link on this web
page and scrolling down to this label.

The ``Prototype Index'' tab lists every structure in the
Encyclopedia.  This index can be sorted by prototype formula, the number of
atomic species, the number of atoms in the primitive cell, the Pearson
Symbol, {\em Strukturbericht} designation (if any), \AFLOW\ prototype
label, space group label, space group number, and structure name (if
available).  The default formula chemical ordering on this page is
alphabetical by element, so we can find or sample structure by
ordering by prototype structure and scrolling down to NNaO$_{2}$.

\noindent\textbf{Generic search functionality.} The prototypes within the Encyclopedia
can be searched online via different criteria.
In the navigation bar, the users can type the chemical formula, mineral, or common name to search for
specific materials/prototypes.
This function also searches through the ``similar elements/compounds'' to consider
atom decorations beyond the ``prototype material'' for a given prototype.
Matching prototypes are returned on a separate page, with hyperlinks to navigate the corresponding entries.
By selecting ``Search'' in either the navigation bar or the main web page, users will be redirected to
a separate webpage, enabling searches via
``Name'' (same functionality as with the navigation bar),
``Space Group'',
``Pearson Symbol'', and/or
``{\em Strukturbericht} Designation''.
To further narrow down results, users can type in each search field to perform compound queries.

\noindent\textbf{Comparing to the Encyclopedia with \AFLOWXTALFINDER.}
In the ``AFLOW-XtalFinder`` tab, users can upload structures to determine if they
match to one of the prototypes in the Encyclopedia.
Leveraging routines in the \AFLOWXTALFINDER\ module~\cite{curtarolo:art170}, this functionality automatically
determines
\textbf{i.} the input structure's ideal prototype designation
(i.e., label and degrees of freedom),
\textbf{ii.} structurally similar prototypes (isoconfigurational) in the Encyclopedia, and
\textbf{iii.} symmetrically similar prototypes (isopointal) in the Encyclopedia.
Uploaded geometries can be in a variety of common file formats (auto-detected), namely,
\VASP~\cite{vasp_JPCM_1994,kresse_vasp_1,vasp},
\QUANTUMESPRESSO~\cite{quantum_espresso_2009},
\FHIAIMS~\cite{blum:fhi-aims},
\ABINIT~\cite{gonze:abinit},
\WYCCAR~\cite{curtarolo:art135},
\CIF, and
\ELK~\cite{elk}.
If an uploaded structure does not match to an existing \AFLOW\ prototype,
a message will appear on the web page.
Users are encouraged to fill out the displayed submission form
providing references/citations to the structure,
any notable comments about the structure, and
user information (optional).
Submission of new structures will be considered for possible inclusion
into future iterations of the Prototype Encyclopedia.

\noindent\textbf{Future outlook.} The Prototype Encyclopedia continues to be expanded,
with updates to the database posted periodically.  Over 600 new
structures are already prepared for future release.  We invite users
to submit new structures for inclusion in the database.  See {\em The
\AFLOW\ Prototype Encyclopedia: Part 3},~\cite{aflowANRL3} and \AFLOWXTALFINDER\ at
\href{https://aflow.org/prototype-encyclopedia/xtal-finder.html}{aflow.org/prototype-encyclopedia/xtal-finder.html}
for more information.

\section{Web Applications}

\begin{figure*}
  \includegraphics[width=\textwidth]{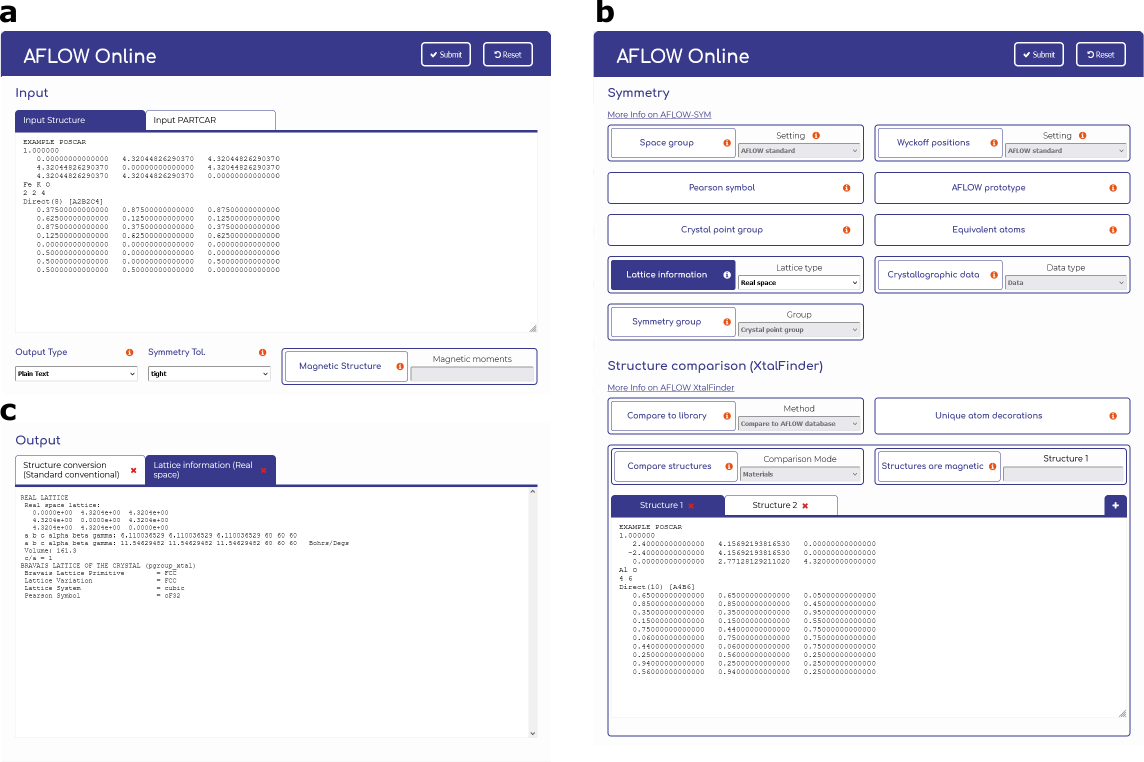}
  \caption{\textbf{The \AFLOWONLINE\ interface.} (\textbf{a}) The structure input section, (\textbf{b})
  a representative feature section, (\textbf{c}) the corresponding output section.}
  \label{fig:aflow_online}
\end{figure*}

\subsection{\AFLOWONLINE}
\noindent
Autonomous computational frameworks such as \AFLOW\ are powerful tools with a rich diversity of
features. However, the size of the codebase presents a large barrier to entry to users who wish
to use only specific functionality, only use these functions occasionally, or are not experienced
with using a command line. A graphical user interface, ideally available online, is more suitable
for these use cases. \AFLOWONLINE\ is such a platform and implements many commonly-used features
of the \AFLOW\ software. It can be accessed at
\href{https://aflow.org/aflow-online/}{aflow.org/aflow-online/}.

The interface consists of a structure input section, a section containing select \AFLOW\ features,
and an output section.
The structure input section~(Figure~\ref{fig:aflow_online}(a)) contains the structure that is used to
perform the \AFLOW\ calculation. It supports all formats that the \AFLOW\ software supports, which
includes \VASP~\cite{vasp_JPCM_1994,kresse_vasp_1,vasp},
\QUANTUMESPRESSO~\cite{quantum_espresso_2009}, \ABINIT~\cite{gonze:abinit}, \ELK~\cite{elk},
\FHIAIMS~\cite{blum:fhi-aims}, and Crystallographic Information Files~(\CIF).
In a separate tab, a PARTCAR file for the features of \AFLOW's
\underline{P}artial \underline{OCC}upation module~(\AFLOWPOCC)
can be entered~\cite{curtarolo:art110}. Magnetic moments can be added to the structure by entering
the spins as a comma-separated list into the ``Magnetic moments'' field after activating the
``Magnetic Structure'' button. There are two other settings in this section: the
``Output Type'' drop-down menu can toggle between free text and \JSON\ output~(where available), and
``Symmetry Tol.'' sets the symmetry tolerance to either ``tight'' or ``loose'', as defined in the
\AFLOWSYM\ module~\cite{curtarolo:art135}.

Below the structure input section are all \AFLOW\ modules that are available via \AFLOWONLINE.
They are categorized based on the type of tasks they perform. Figure~\ref{fig:aflow_online}(b) shows two
example categories: symmetry and structure comparison~\cite{curtarolo:art135,curtarolo:art170}.
The symmetry section contains all features provided by \AFLOWSYM: it calculates space groups,
Wyckoff positions, Pearson symbols, prototype labels, symmetry groups, information on both direct
and reciprocal lattice, and equivalency between atoms.

The structure comparison section is the interface to \AFLOWXTALFINDER~\cite{curtarolo:art170}.
It provides the options to compare the input structure to the \AFLOW\ database and the prototype
encyclopedia, as well as determining unique atom decorations, in a simple button-drop-down setup.
The application for comparing multiple structures is different: the button and drop down select
the comparison mode~(structure and material), but the structure is not taken from the input
structure section. Instead, it has its own section where multiple inputs can be added in separate
tabs. Magnetic moments can also be added for each structure using the ``Structures are magnetic''
button nearby.

To submit a calculation, the button for the feature of interest needs to be clicked --- tooltips are
available to help users select the appropriate functionality. Some options require additional input,
e.g., through a drop-down menu. In the example shown in Figure~\ref{fig:aflow_online}(b)~(``Lattice
information''), users can choose between real space, reciprocal space, and a superlattice with equal
decorations. Pressing the ``Submit'' button will start the calculation and the screen is scrolled to
the output section where the results will be displayed once available.
Figure~\ref{fig:aflow_online}(c) shows such a submission with the real space lattice
information calculated by \AFLOW\ for KFeO$_2$. Output is stored for each submitted calculation
and can be accessed by clicking on the corresponding tab --- in this case, the prior calculation
is a conversion of the input structure to the conventional unit cell. Output can be
removed using the ``$\times$'' buttons.

\AFLOWONLINE\ also provides tools to prepare \textit{ab-initio} calculations. Its structure file
manipulation tools allow conversion between different structure file formats, coordinate
systems~(Cartesian and fractional), and unit cell formats~(primitive, \AFLOW\ standard
primitive~\cite{curtarolo:art65}, standard conventional, Niggli, and Minkowksi). It also has modules
for calculating the dimensions of a k-point mesh based on k-points per reciprocal atom or a
maximum distance in reciprocal space, and to output the standard k-point path for band structure
calculations based on the \AFLOW\ standard~\cite{curtarolo:art58}. The format for the k-points are
in \VASP's \KPOINTS\ file format. Disordered materials can also be set up using
\AFLOWPOCC~\cite{curtarolo:art110}. Based on an input \PARTCAR, the ideal supercell size can be
determined, and the list of representative structures can be generated in \POSCAR\ format.
The latter can be converted into the desired format using the structure file manipulation tools.

Finally, the \AFLOW\ \underline{C}oordination \underline{C}orrected \underline{E}nthalpies~(\AFLOWCCE)
module~\cite{curtarolo:art150,curtarolo:art172} is available as well. It can calculate the
corrections to the formation enthalpy for ionic materials calculated from density functional theory
and supports the \PBE, \LDA, and \SCAN\ functionals. It requires knowledge of the oxidation states
in the material, which \AFLOW\ can determine automatically. However, users also have the option to
provide these numbers directly. The determined oxidation states and the coordination
numbers of the cations can also be output separately through the appropriate buttons.

\AFLOWONLINE\ is thus a user-friendly tool that provides a variety of features implemented in
\AFLOW\ for structural analysis, to support \textit{ab-initio} calculations, and to correct energies
from density functional theory. It is designed to be used in low-throughput settings without
installing the entire \AFLOW\ framework and without requiring a command line. Thus, it is
particularly suited for new users and for users with little or no scripting experience.

\begin{figure*}
  \includegraphics[width=\textwidth]{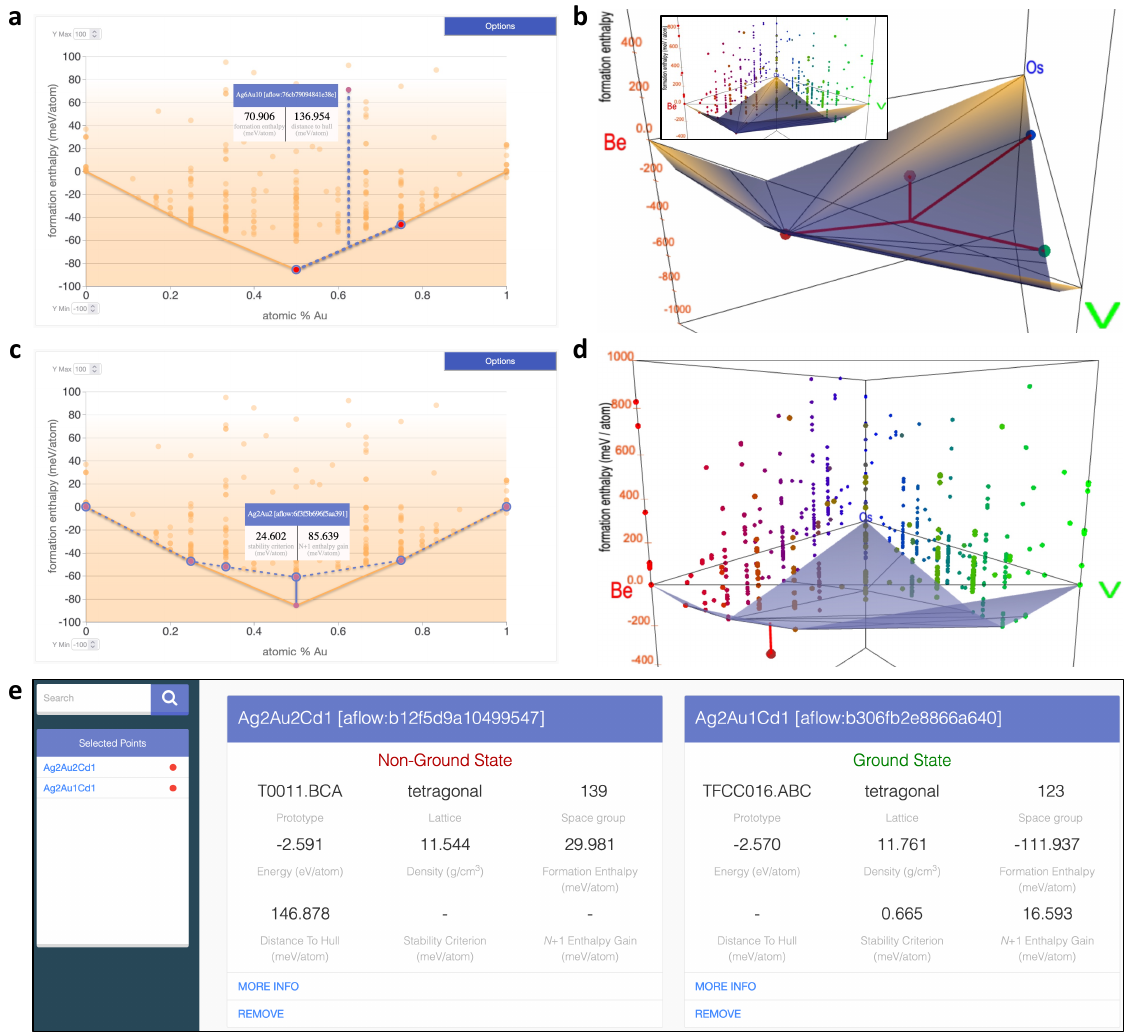}
  \caption{\textbf{Distance to the hull and stability criterion visualizations online.}
  Example visualizations are provided of ({\bf a} and {\bf b})
  the distance to the hull (and corresponding decompositions)
  and ({\bf c} and {\bf d}) stability criterion
  for binary (AgAu) and ternary (BeOsV) convex hulls.
  An inset of the original BeOsV hull is included in ({\bf b}) for reference.
  ({\bf e})~A snapshot of the information component populated
  with non-ground state and ground state entries.
  }
  \label{fig:web_chull_1}
\end{figure*}

\begin{figure*}
  \includegraphics[width=\textwidth]{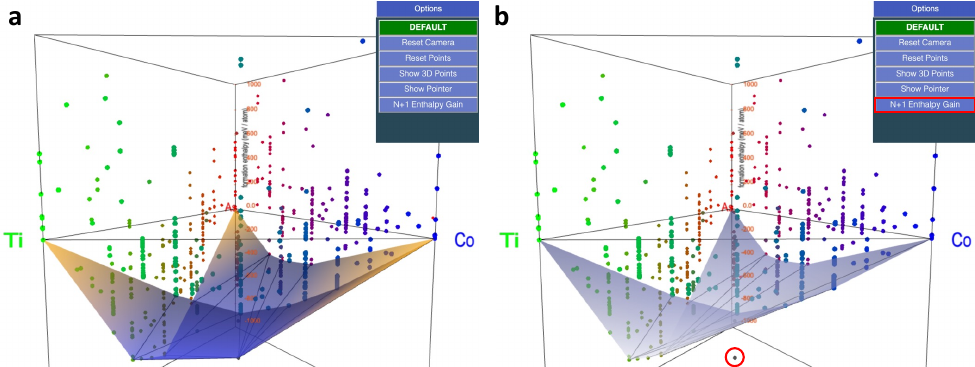}
  \caption{\textbf{Visualization of the $N+1$ enthalpy gain descriptor.}
  The original AsCoTi convex hull is plotted in ({\bf a}) and
  the pseudo-hull composed only of unaries and binaries is shown in blue in ({\bf b}).
  A stable ternary --- not included in the pseudo-hull construction --- is highlighted in red.
  }
  \label{fig:web_chull_2}
\end{figure*}

\subsection{\AFLOWCHULLONLINE}
\noindent
The \AFLOWCHULL\ web application (\href{https://aflow.org/aflow_chull}{aflow.org/aflow-chull}) was developed to provide an enhanced, command-line-free platform
for the \AFLOWCHULL\ module and was introduced in Ref.~\onlinecite{curtarolo:art144}.
The application offers an interactive visualization of
binary and ternary convex hulls,
enabling the selection and investigation of entries of interest.
The ternary convex hull allows mouse-enabled pan and zoom, and has been a useful instructional tool
for understanding the stability analysis and grasping higher dimensional hulls with multi-component systems.
Hulls can be selected from a responsive periodic table interface
that reacts to the input based on the data availability.
The number of entries
is used as an indicator of the
convergence of the analysis~\cite{curtarolo:art144,curtarolo:art20,curtarolo:art87,curtarolo:art54}:
\textcolor{pranab_green}{{\bf green}} (fully reliable, $N_{\mathrm{entries}} \geq 200$),
\textcolor{orange}{{\bf orange}} (potentially reliable, $100 \leq N_{\mathrm{entries}} < 200$),
\textcolor{pranab_red}{{\bf red}} (unreliable, $N_{\mathrm{entries}} < 100$), and
\textcolor{gray}{\bf gray} (unavailable, $N_{\mathrm{entries}} =0$).

To accelerate loading speeds for repeated queries, a caching layer has been added.
It saves the results of each request and loads them up automatically.
The caching layer is cleared periodically and
fresh calculations employing the most up-to-date database are triggered with a new request.
Longer wait times for non-cached queries should be expected.
Once the results are retrieved,
the application loads the visualization viewport,
prompting a redirect to the \URL\ endpoint of the selected hull, e.g., {\sf /hull/BeOsV}.
While the \URL\ is ubiquitous, it is dependent on the cached state at the time the query is submitted.

Users can select and highlight points on both binary and ternary hulls.
When a point is selected, its name will appear within the sidebar.
If the selected point is above the convex hull (meta-stable),
the hull visualization will highlight the distance to the hull (vertical line)
and tie-line/tie-surface directly below formed by stable phases (Figures~\ref{fig:web_chull_1}(a) and (b)),
corresponding to the decomposition reaction in the thermodynamic limit.
For example, in Figure~\ref{fig:web_chull_1}(a),
the reduced decomposition reaction is
\[
\text{Ag}_{0.375}\text{Au}_{0.625} \xrightarrow[\text{}]{-137 \atop \text{meV/atom}} \frac{1}{2}~\text{Ag}_{0.5}\text{Au}_{0.5} + \frac{1}{2}~\text{Ag}_{0.25}\text{Au}_{0.75},
\]
where the distance to the hull is the energy gained
from phase separation.
If the selected point is on the convex hull (stable),
the hull visualization will show the stability criterion~\cite{curtarolo:art144,curtarolo:art109}
(Figures~\ref{fig:web_chull_1}(c) and (d)):
the distance of the point from the pseudo-hull constructed without it,
illustrated by the dotted lines and blue surfaces on binary and ternary hulls, respectively.
This distance quantifies the effect of the phase on the convex hull, as well as its susceptibility to
destabilization by a new phase that has yet to be explored.
The stability criterion helped find two new magnetic phases, the first ever discovered with computation~\cite{curtarolo:art109}.
Clicking on the name of one of the selected points on the sidebar will bring up
the information component (Figure~\ref{fig:web_chull_1}(e)),
providing a selection of properties, such as the distance to the hull and stability criterion,
with a link to the \AFLOWorg\ entry page (``More Info'') offering the full set of
properties for the entry.

New to the stability analysis is the calculation of the $N+1$ enthalpy gain~\cite{curtarolo:art152}:
the distance to the hull for an $N$-compound (binaries are 2-compounds) from the
pseudo-hull constructed with $\{1,\ldots,N-1\}$-compounds.
The descriptor was used to establish the role of disorder in stabilizing multi-component systems~\cite{curtarolo:art152}.
It was shown that, with an increasing number of species,
the average enthalpy gained rapidly diminishes with respect to the configurational entropy gained.
The descriptor corresponds to the formation enthalpy for binaries,
and for ternaries becomes the distance from the pseudo-hull constructed with
the full set of unaries ($A$, $B$, and $C$) and binaries ($A$-$B$, $B$-$C$, and $A$-$C$).
An example visualization of the analysis is provided in Figure~\ref{fig:web_chull_2},
where the full AsCoTi convex hull and its $\{1,2\}$-pseudo-hull
are shown in Figures~\ref{fig:web_chull_2}(a) and (b), respectively.
A stable ternary, not included in the pseudo-hull construction, is highlighted in red.
The pseudo-hull can be visualized from the Options menu.

\begin{figure*}
  \includegraphics[width=\textwidth]{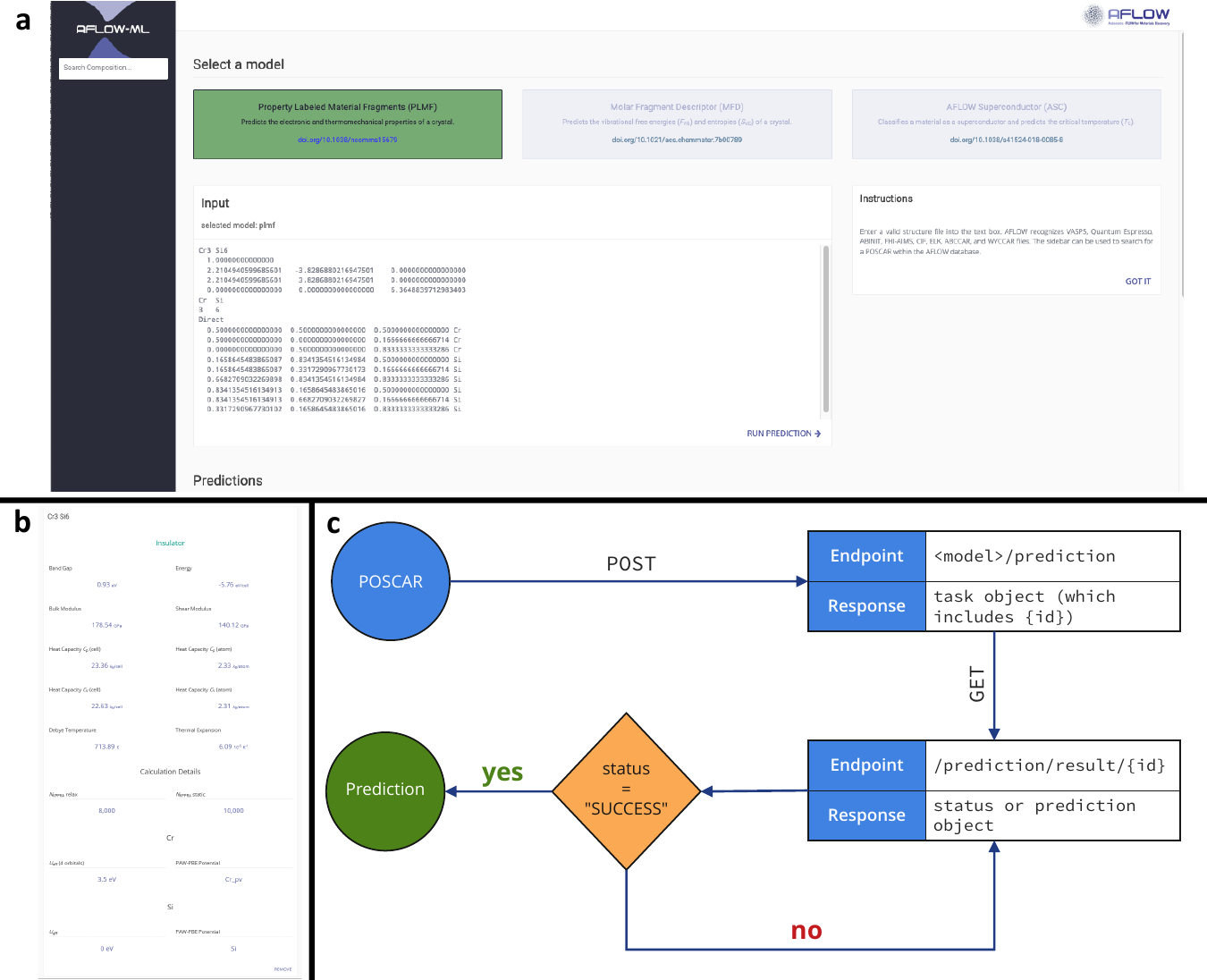}
  \caption{\textbf{\AFLOWML.} (\textbf{a}) The \AFLOWMLONLINE\ web interface, with three different models available;
(\textbf{b}) card with prediction results for  \AFLOWMLONLINE;
(\textbf{c}) the \AFLOWML\ \API\ enables programmatic access to the models.}
  \label{fig:aflow_ml}
\end{figure*}

\subsection{\AFLOWMLONLINE}
\noindent
The \AFLOW\ \underline{m}achine \underline{l}earning (\AFLOWML) online application provides a user interface to leverage machine-learning models trained on \AFLOW\ data, as shown in Figure~\ref{fig:aflow_ml}.
The application can be accessed through a web browser at \href{https://aflow.org/aflow-ml}{aflow.org/aflow-ml}, or through the \AFLOWML\ API described below.
The application takes in structural and/or compositional information as an input and outputs predictions,
including electronic, thermomechanical, and superconducting properties.
This application provides an accessible medium to retrieve machine learning predictions without the need to install specialized software libraries or machine learning packages.

Currently, \AFLOWML\ supports three different machine-learning models:
{\bf i.} The \underline{p}roperty-\underline{l}abeled \underline{m}aterials \underline{f}ragments model~\cite{curtarolo:art124}, \texttt{plmf}, has been trained using data from the \AFLOWorg\ repository, and predicts properties such as the electronic band gap, specific heat capacities, bulk and shear moduli, Debye temperature, and coefficient of thermal expansion.
It is based on structural information, using a Voronoi tesellation to identify bonded or ``connected'' atoms to form a crystal graph, and thus requires a structure file describing the material.
Atoms in the graph are labeled with their corresponding elemental properties, and pieces of the graph form the property-labeled materials fragments that form the feature vector for the model.
{\bf ii.} The \underline{m}olar \underline{f}raction \underline{d}escriptor model~\cite{curtarolo:art129}, \texttt{mfd}, predicts vibrational properties such as vibrational free energy and entropy, and is based only on the chemical composition of the material.
{\bf iii.} The \underline{A}FLOW \underline{S}uper\underline{c}onductor model~\cite{curtarolo:art137}, \texttt{asc}, predicts the superconducting critical temperature of materials based on their chemical composition, and accepts the chemical formula as input.

Using the online application involves, first, selecting the required model from the three listed and then, posting either a structure file (in the case of the \texttt{plmf} and \texttt{mfd} models) or a chemical formula (for \texttt{asc}) into the ``Input'' box, as illustrated in Figure~\ref{fig:aflow_ml}(a).
Structures can be in any format that \AFLOW\ can read,
including \VASP\ \POSCAR~\cite{vasp_JPCM_1994,kresse_vasp_1,vasp},
\QUANTUMESPRESSO~\cite{quantum_espresso_2009}, \ABINIT~\cite{gonze:abinit}, \ELK~\cite{elk},
\FHIAIMS~\cite{blum:fhi-aims}, and \underline{C}rystallographic \underline{I}nformation \underline{F}iles~(\CIF)~\cite{Hall_CIF_1991}.
Additionally, structures within the \AFLOWorg\ repository can be imported via the sidebar.
The structure is then submitted and the model is run using the ``RUN PREDICTION'' button.
When the prediction is complete (which usually takes several seconds, depending on the material, model, and the demand on the server), a card will be displayed containing the predicted property values, as shown in Figure~\ref{fig:aflow_ml}(b).

The \AFLOWML\ \API~\cite{aflowmlapi} offers programmatic access to the \AFLOWML\ online application, and provides a simplified abstraction that facilitates leveraging powerful machine-learning models.
This distills the prediction process down to its essence: from a structure file, return a prediction.
Using the \API\ is a two-step process as illustrated in Figure~\ref{fig:aflow_ml}(c): first a structure file is posted (i.e. uploaded) to the endpoint \texttt{<server>/<model>/prediction} on the \texttt{aflow.org} server using standard HTTP libraries or dedicated programs such \texttt{curl} or \texttt{wget}, where \texttt{<server>} is \texttt{aflow.org/API/aflow-ml}, and \texttt{<model>} specifies the machine-learning model to use in the prediction (current options: \texttt{plmf} and \texttt{mfd}).
For example, the full endpoint for uploading for the \texttt{plmf} model would be \texttt{aflow.org/API/aflow-ml/plmf/prediction}.
When a prediction is submitted, a \JSON\ response object is returned that includes a task \texttt{id}.
Second, the results of the prediction need to be retrieved from the \texttt{/prediction/result/} endpoint
on the \texttt{aflow.org} server by appending the task \texttt{id} to the end of the URL,
i.e. \texttt{/prediction/result/\{id\}/}.
This endpoint monitors the prediction task and responds with a \JSON\ object that details its status.
When complete, the endpoint responds with the results of the prediction, represented as a \JSON\ object containing a key-value pair for each predicted property.

The \AFLOWML\ \API\ can also be accessed using the \AFLOWML\ Python client, which is available for download at \href{https://aflow.org/src/aflow-ml/}{aflow.org/src/aflow-ml/}.
Upon installation, the Python client can be accessed through the command-line interface using the command \texttt{aflow-ml}.
Predictions tasks can be submitted by specifying the structure file name and the machine-learning model.
For example, a structure in the file \texttt{test.poscar} can be submitted to the \texttt{plmf} machine-learning model
using the command \texttt{aflowml test.poscar -{}-model=plmf}, where the model is specified using the flags
\texttt{-{}-model} or \texttt{-m}.
The results can be saved to a file usng the flag \texttt{-s} or \texttt{-{}-save}, and the output filename can be specified using the flag \texttt{-{}-outfile}
(the default filename is \texttt{prediction.txt}).
The output format can be set using the \texttt{-{}-output} flag, while the output can be limited to specific predicted fields using the \texttt{-{}-fields} flag.
Verbose mode can be activated using the flags \texttt{-v} or \texttt{-{}-verbose}.
\section{Education and Outreach}

\noindent
In addition to the rational data dissemination discussed above, the \AFLOW\ consortium
is engaged in global education and outreach activities to train the next cadre of
materials researchers.

\noindent {\bf \AFLOW\ Schools.} We offer tutorials, virtual and in-person,
  teaching the structure of our repositories and the strategies to retrieve
  data to a diverse audience.
  Particular emphasis is put on the discussion of the features
  of the \AFLOW\ C++ code and python interfaces, so that they can be
  easily adapted into workflows.
  These schools cover the \AFLOW\ code, \DFT, symmetry determination,
  prototypes, convex hulls, partial occupation, coordination corrected
  enthalpies, elastic and Gibbs library extensions, database
  integration, machine learning within \AFLOW, prototype crystal
  finder, and harmonic and quasiharmonic phonon calculations.
  Videos, slides presentations, and other educational
  materials are available online at
  \href{https://aflow.org/aflow-school/}{aflow.org/aflow-school/}.

\noindent {\bf \AFLOW\ Seminars.} We organize biweekly free virtual
  scientific seminars in the field of materials science and materials
  physics. Speakers are experts in experimental, theoretical or computational techniques,
  covering a variety of sub-fields.
  These seminars attract a global audience from over 50 countries on five continents,
  providing access to scientific talks that would otherwise require considerable resources
  to attend.
  More information can be found online at
   \href{https://aflow.org/seminars/}{aflow.org/seminars/}.

\appendix
\section{\AFLOW\ keywords}
\noindent
The \AFLOW\ database contains a large variety of materials properties that can be queried through
the appropriate keyword using the \RESTAPI\ or \AFLUX.
\AFLOW\ goes beyond data and provides rich metadata in multiple ways:
\textbf{i.}~through the keywords, which contain identifiers for the methods and modules
used to compute the property~(e.g., ael\_bulk\_modulus\_reuss describes the bulk modulus
calculated with the Reuss method using the \AFLOWAEL\ module);
\textbf{ii.}~by serving detailed information on how the data was generated with each entry,
such as input parameters needed to reproduce results~(e.g., cut-off energies and
pseudopotentials), software versions, data sources, and more;
and \textbf{iii.}~by providing metadata for the keywords themselves --- including references to
the methods used --- that can be accessed at
\href{https://aflow.org/API/aapi-schema/}{aflow.org/API/aapi-schema/}. This fulfills the
``Reusable'' criterion of the FAIR principles, which requires describing the context of the data
so that users can assess its suitability for their particular purpose.

This appendix contains all available keywords at the time of this publication.
It provides a description for each keyword along with references
to the methods used to calculate the property, units (where applicable), the data type of the
property, and an example value.
The list is also available online at \href{https://aflow.org/documentation/}{aflow.org/documentation/}
where it is continuously updated.

\noindent
\fcolorbox{black}[gray]{1}{
\begin{minipage}{\columnwidth}
    \texttt{ael\_applied\_pressure} \strut \hspace{\fill} \cite{curtarolo:art115}
    \\[-2mm]
    \rule{\columnwidth}{1pt}
    Returns the applied pressure for the AEL calculations.\\[-2mm]
    \rule{\columnwidth}{0.5pt}
    \begin{minipage}{0.4\columnwidth}\textbf{Unit}: GPa\end{minipage}\textbf{Type:} number\\
    \textbf{Example:} \seqsplit{ael\_applied\_pressure=0.0}
    \\\vspace{-2mm}
\end{minipage}
}

\noindent
\fcolorbox{black}[gray]{1}{
\begin{minipage}{\columnwidth}
    \texttt{ael\_average\_external\_pressure} \strut \hspace{\fill} \cite{curtarolo:art115}
    \\[-2mm]
    \rule{\columnwidth}{1pt}
    Returns the average external pressure for the AEL calculations.\\[-2mm]
    \rule{\columnwidth}{0.5pt}
    \begin{minipage}{0.4\columnwidth}\textbf{Unit}: GPa\end{minipage}\textbf{Type:} number\\
    \textbf{Example:} \seqsplit{ael\_average\_external\_pressure=0.079875}
    \\\vspace{-2mm}
\end{minipage}
}

\noindent
\fcolorbox{black}[gray]{1}{
\begin{minipage}{\columnwidth}
    \texttt{ael\_bulk\_modulus\_reuss} \strut \hspace{\fill} \cite{curtarolo:art115,curtarolo:art128}
    \\[-2mm]
    \rule{\columnwidth}{1pt}
    Returns the bulk modulus calculated, using the Reuss method, by AEL.\\[-2mm]
    \rule{\columnwidth}{0.5pt}
    \begin{minipage}{0.4\columnwidth}\textbf{Unit}: GPa\end{minipage}\textbf{Type:} number\\
    \textbf{Example:} \seqsplit{ael\_bulk\_modulus\_reuss=105.315}
    \\\vspace{-2mm}
\end{minipage}
}

\noindent
\fcolorbox{black}[gray]{1}{
\begin{minipage}{\columnwidth}
    \texttt{ael\_bulk\_modulus\_voigt} \strut \hspace{\fill} \cite{curtarolo:art115,curtarolo:art128}
    \\[-2mm]
    \rule{\columnwidth}{1pt}
    Returns the bulk modulus calculated, using the Voigt method, by AEL.\\[-2mm]
    \rule{\columnwidth}{0.5pt}
    \begin{minipage}{0.4\columnwidth}\textbf{Unit}: GPa\end{minipage}\textbf{Type:} number\\
    \textbf{Example:} \seqsplit{ael\_bulk\_modulus\_voigt=105.315}
    \\\vspace{-2mm}
\end{minipage}
}

\noindent
\fcolorbox{black}[gray]{1}{
\begin{minipage}{\columnwidth}
    \texttt{ael\_bulk\_modulus\_vrh} \strut \hspace{\fill} \cite{curtarolo:art115,curtarolo:art128}
    \\[-2mm]
    \rule{\columnwidth}{1pt}
    Returns the bulk modulus calculated, using the Voigt-Reuss-Hill average, by AEL.\\[-2mm]
    \rule{\columnwidth}{0.5pt}
    \begin{minipage}{0.4\columnwidth}\textbf{Unit}: GPa\end{minipage}\textbf{Type:} number\\
    \textbf{Example:} \seqsplit{ael\_bulk\_modulus\_vrh=105.315}
    \\\vspace{-2mm}
\end{minipage}
}

\noindent
\fcolorbox{black}[gray]{1}{
\begin{minipage}{\columnwidth}
    \texttt{ael\_compliance\_tensor} \strut \hspace{\fill} \cite{curtarolo:art115}
    \\[-2mm]
    \rule{\columnwidth}{1pt}
    Returns the compliance tensor calculated by AEL.\\[-2mm]
    \rule{\columnwidth}{0.5pt}
    \begin{minipage}{0.4\columnwidth}\textbf{Unit}: GPa$^{-1}$\end{minipage}\textbf{Type:} numbers\\
    \textbf{Example:} \seqsplit{74.2667,49.5167,49.5167,0,0,0;49.5167,74.2667,49.5167,0,0,0;49.5167,49.5167,74.2667,0,0,0;0,0,0,27.2833,0,0;0,0,0,0,27.2833,0;0,0,0,0,0,27.2833}
    \\\vspace{-2mm}
\end{minipage}
}

\noindent
\fcolorbox{black}[gray]{1}{
\begin{minipage}{\columnwidth}
    \texttt{ael\_debye\_temperature} \strut \hspace{\fill} \cite{curtarolo:art115}
    \\[-2mm]
    \rule{\columnwidth}{1pt}
    Returns the Debye temperature calculated by AEL.\\[-2mm]
    \rule{\columnwidth}{0.5pt}
    \begin{minipage}{0.4\columnwidth}\textbf{Unit}: K\end{minipage}\textbf{Type:} number\\
    \textbf{Example:} \seqsplit{ael\_debye\_temperature=163.348}
    \\\vspace{-2mm}
\end{minipage}
}

\noindent
\fcolorbox{black}[gray]{1}{
\begin{minipage}{\columnwidth}
    \texttt{ael\_elastic\_anisotropy} \strut \hspace{\fill} \cite{curtarolo:art115,curtarolo:art128}
    \\[-2mm]
    \rule{\columnwidth}{1pt}
    Returns the elastic anisotropy calculated by AEL.\\[-2mm]
    \rule{\columnwidth}{0.5pt}
    \begin{minipage}{0.4\columnwidth}\textbf{Unit}: none\end{minipage}\textbf{Type:} number\\
    \textbf{Example:} \seqsplit{ael\_elastic\_anisotropy=0.0008165}
    \\\vspace{-2mm}
\end{minipage}
}

\noindent
\fcolorbox{black}[gray]{1}{
\begin{minipage}{\columnwidth}
    \texttt{ael\_poisson\_ratio} \strut \hspace{\fill} \cite{curtarolo:art115,curtarolo:art128}
    \\[-2mm]
    \rule{\columnwidth}{1pt}
    Returns the isotropic Poisson ratio calculated by AEL.\\[-2mm]
    \rule{\columnwidth}{0.5pt}
    \begin{minipage}{0.4\columnwidth}\textbf{Unit}: none\end{minipage}\textbf{Type:} number\\
    \textbf{Example:} \seqsplit{ael\_poisson\_ratio=0.216}
    \\\vspace{-2mm}
\end{minipage}
}

\noindent
\fcolorbox{black}[gray]{1}{
\begin{minipage}{\columnwidth}
    \texttt{ael\_pughs\_modulus\_ratio} \strut \hspace{\fill} \cite{curtarolo:art115}
    \\[-2mm]
    \rule{\columnwidth}{1pt}
    Returns the Pugh's modulus ratio calculated by AEL.\\[-2mm]
    \rule{\columnwidth}{0.5pt}
    \begin{minipage}{0.4\columnwidth}\textbf{Unit}: none\end{minipage}\textbf{Type:} number\\
    \textbf{Example:} \seqsplit{ael\_pughs\_modulus\_ratio=0.332265}
    \\\vspace{-2mm}
\end{minipage}
}

\noindent
\fcolorbox{black}[gray]{1}{
\begin{minipage}{\columnwidth}
    \texttt{ael\_shear\_modulus\_reuss} \strut \hspace{\fill} \cite{curtarolo:art115,curtarolo:art128}
    \\[-2mm]
    \rule{\columnwidth}{1pt}
    Returns the shear modulus calculated, using the Reuss method, by AEL.\\[-2mm]
    \rule{\columnwidth}{0.5pt}
    \begin{minipage}{0.4\columnwidth}\textbf{Unit}: GPa\end{minipage}\textbf{Type:} number\\
    \textbf{Example:} \seqsplit{ael\_shear\_modulus\_reuss=73.787}
    \\\vspace{-2mm}
\end{minipage}
}

\noindent
\fcolorbox{black}[gray]{1}{
\begin{minipage}{\columnwidth}
    \texttt{ael\_shear\_modulus\_voigt} \strut \hspace{\fill} \cite{curtarolo:art115,curtarolo:art128}
    \\[-2mm]
    \rule{\columnwidth}{1pt}
    Returns the shear modulus calculated, using the Voigt method, by AEL.\\[-2mm]
    \rule{\columnwidth}{0.5pt}
    \begin{minipage}{0.4\columnwidth}\textbf{Unit}: GPa\end{minipage}\textbf{Type:} number\\
    \textbf{Example:} \seqsplit{ael\_shear\_modulus\_voigt=73.799}
    \\\vspace{-2mm}
\end{minipage}
}

\noindent
\fcolorbox{black}[gray]{1}{
\begin{minipage}{\columnwidth}
    \texttt{ael\_shear\_modulus\_vrh} \strut \hspace{\fill} \cite{curtarolo:art115,curtarolo:art128}
    \\[-2mm]
    \rule{\columnwidth}{1pt}
    Returns the shear modulus calculated, using the Voigt-Reuss-Hill average, by AEL.\\[-2mm]
    \rule{\columnwidth}{0.5pt}
    \begin{minipage}{0.4\columnwidth}\textbf{Unit}: GPa\end{minipage}\textbf{Type:} number\\
    \textbf{Example:} \seqsplit{ael\_shear\_modulus\_vrh=73.793}
    \\\vspace{-2mm}
\end{minipage}
}

\noindent
\fcolorbox{black}[gray]{1}{
\begin{minipage}{\columnwidth}
    \texttt{ael\_speed\_sound\_average} \strut \hspace{\fill} \cite{curtarolo:art115,curtarolo:art128}
    \\[-2mm]
    \rule{\columnwidth}{1pt}
    Returns the average speed of sound calculated by AEL.\\[-2mm]
    \rule{\columnwidth}{0.5pt}
    \begin{minipage}{0.4\columnwidth}\textbf{Unit}: m/s\end{minipage}\textbf{Type:} number\\
    \textbf{Example:} \seqsplit{ael\_speed\_sound\_average=1540.09}
    \\\vspace{-2mm}
\end{minipage}
}

\noindent
\fcolorbox{black}[gray]{1}{
\begin{minipage}{\columnwidth}
    \texttt{ael\_speed\_sound\_longitudinal} \strut \hspace{\fill} \cite{curtarolo:art115,curtarolo:art128}
    \\[-2mm]
    \rule{\columnwidth}{1pt}
    Returns the longitudinal speed of sound calculated by AEL.\\[-2mm]
    \rule{\columnwidth}{0.5pt}
    \begin{minipage}{0.4\columnwidth}\textbf{Unit}: m/s\end{minipage}\textbf{Type:} number\\
    \textbf{Example:} \seqsplit{ael\_speed\_sound\_longitudinal=2147.05}
    \\\vspace{-2mm}
\end{minipage}
}

\noindent
\fcolorbox{black}[gray]{1}{
\begin{minipage}{\columnwidth}
    \texttt{ael\_speed\_sound\_transverse} \strut \hspace{\fill} \cite{curtarolo:art115,curtarolo:art128}
    \\[-2mm]
    \rule{\columnwidth}{1pt}
    Returns the transverse speed of sound calculated by AEL.\\[-2mm]
    \rule{\columnwidth}{0.5pt}
    \begin{minipage}{0.4\columnwidth}\textbf{Unit}: m/s\end{minipage}\textbf{Type:} number\\
    \textbf{Example:} \seqsplit{ael\_speed\_sound\_transverse=1405.57}
    \\\vspace{-2mm}
\end{minipage}
}

\noindent
\fcolorbox{black}[gray]{1}{
\begin{minipage}{\columnwidth}
    \texttt{ael\_stiffness\_tensor} \strut \hspace{\fill} \cite{curtarolo:art115}
    \\[-2mm]
    \rule{\columnwidth}{1pt}
    Returns the stiffness tensor calculated by AEL.\\[-2mm]
    \rule{\columnwidth}{0.5pt}
    \begin{minipage}{0.4\columnwidth}\textbf{Unit}: GPa\end{minipage}\textbf{Type:} numbers\\
    \textbf{Example:} \seqsplit{0.0288595,-0.0115446,-0.0115446,0,0,0;-0.0115446,0.0288595,-0.0115446,0,0,0;-0.0115446,-0.0115446,0.0288595,0,0,0;0,0,0,0.0366524,0,0;0,0,0,0,0.0366524,0;0,0,0,0,0,0.0366524}
    \\\vspace{-2mm}
\end{minipage}
}

\noindent
\fcolorbox{black}[gray]{1}{
\begin{minipage}{\columnwidth}
    \texttt{ael\_youngs\_modulus\_vrh} \strut \hspace{\fill} \cite{curtarolo:art115}
    \\[-2mm]
    \rule{\columnwidth}{1pt}
    Returns the Young modulus calculated, using the Voigt-Reuss-Hill average, by AEL.\\[-2mm]
    \rule{\columnwidth}{0.5pt}
    \begin{minipage}{0.4\columnwidth}\textbf{Unit}: GPa\end{minipage}\textbf{Type:} number\\
    \textbf{Example:} \seqsplit{ael\_youngs\_modulus\_vrh=31.6223}
    \\\vspace{-2mm}
\end{minipage}
}

\noindent
\fcolorbox{black}[gray]{1}{
\begin{minipage}{\columnwidth}
    \texttt{aflow\_prototype\_label\_orig} \strut \hspace{\fill} \cite{curtarolo:art170}
    \\[-2mm]
    \rule{\columnwidth}{1pt}
    Returns the AFLOW prototype label for the unrelaxed structure.\\[-2mm]
    \rule{\columnwidth}{0.5pt}
    \begin{minipage}{0.4\columnwidth}\textbf{Unit}: none\end{minipage}\textbf{Type:} string\\
    \textbf{Example:} \seqsplit{aflow\_prototype\_label\_orig=A2BC4\_cF56\_227\_c\_b\_e}
    \\\vspace{-2mm}
\end{minipage}
}

\noindent
\fcolorbox{black}[gray]{1}{
\begin{minipage}{\columnwidth}
    \texttt{aflow\_prototype\_label\_relax} \strut \hspace{\fill} \cite{curtarolo:art170}
    \\[-2mm]
    \rule{\columnwidth}{1pt}
    Returns the AFLOW prototype label for the relaxed structure.\\[-2mm]
    \rule{\columnwidth}{0.5pt}
    \begin{minipage}{0.4\columnwidth}\textbf{Unit}: none\end{minipage}\textbf{Type:} string\\
    \textbf{Example:} \seqsplit{aflow\_prototype\_label\_relax=A2BC4\_cF56\_227\_c\_b\_e}
    \\\vspace{-2mm}
\end{minipage}
}

\noindent
\fcolorbox{black}[gray]{1}{
\begin{minipage}{\columnwidth}
    \texttt{aflow\_prototype\_params\_list\_orig} \strut \hspace{\fill} \cite{curtarolo:art170}
    \\[-2mm]
    \rule{\columnwidth}{1pt}
    Returns the AFLOW prototype parameter labels for the unrelaxed structure.\\[-2mm]
    \rule{\columnwidth}{0.5pt}
    \begin{minipage}{0.4\columnwidth}\textbf{Unit}: none\end{minipage}\textbf{Type:} strings\\
    \textbf{Example:} \seqsplit{aflow\_prototype\_params\_list\_orig=a,x3}
    \\\vspace{-2mm}
\end{minipage}
}

\noindent
\fcolorbox{black}[gray]{1}{
\begin{minipage}{\columnwidth}
    \texttt{aflow\_prototype\_params\_list\_relax} \strut \hspace{\fill} \cite{curtarolo:art170}
    \\[-2mm]
    \rule{\columnwidth}{1pt}
    Returns the AFLOW prototype parameter labels for the relaxed structure.\\[-2mm]
    \rule{\columnwidth}{0.5pt}
    \begin{minipage}{0.4\columnwidth}\textbf{Unit}: none\end{minipage}\textbf{Type:} strings\\
    \textbf{Example:} \seqsplit{aflow\_prototype\_params\_list\_relax=a,x3}
    \\\vspace{-2mm}
\end{minipage}
}

\noindent
\fcolorbox{black}[gray]{1}{
\begin{minipage}{\columnwidth}
    \texttt{aflow\_prototype\_params\_values\_orig} \strut \hspace{\fill} \cite{curtarolo:art170}
    \\[-2mm]
    \rule{\columnwidth}{1pt}
    Returns the AFLOW prototype parameter values for the unrelaxed structure.\\[-2mm]
    \rule{\columnwidth}{0.5pt}
    \begin{minipage}{0.4\columnwidth}\textbf{Unit}: none\end{minipage}\textbf{Type:} numbers\\
    \textbf{Example:} \seqsplit{aflow\_prototype\_params\_values\_orig=10.336,0.757}
    \\\vspace{-2mm}
\end{minipage}
}

\noindent
\fcolorbox{black}[gray]{1}{
\begin{minipage}{\columnwidth}
    \texttt{aflow\_prototype\_params\_values\_relax} \strut \hspace{\fill} \cite{curtarolo:art170}
    \\[-2mm]
    \rule{\columnwidth}{1pt}
    Returns the AFLOW prototype parameter values for the relaxed structure.\\[-2mm]
    \rule{\columnwidth}{0.5pt}
    \begin{minipage}{0.4\columnwidth}\textbf{Unit}: none\end{minipage}\textbf{Type:} numbers\\
    \textbf{Example:} \seqsplit{aflow\_prototype\_params\_values\_relax=10.336,0.757}
    \\\vspace{-2mm}
\end{minipage}
}

\noindent
\fcolorbox{black}[gray]{1}{
\begin{minipage}{\columnwidth}
    \texttt{aflow\_version} \strut \hspace{\fill} \cite{curtarolo:art75,curtarolo:art92}
    \\[-2mm]
    \rule{\columnwidth}{1pt}
    Returns the version number of AFLOW used to perform the calculation.\\[-2mm]
    \rule{\columnwidth}{0.5pt}
    \begin{minipage}{0.4\columnwidth}\textbf{Unit}: none\end{minipage}\textbf{Type:} string\\
    \textbf{Example:} \seqsplit{aflow\_version=aflow30641}
    \\\vspace{-2mm}
\end{minipage}
}

\noindent
\fcolorbox{black}[gray]{1}{
\begin{minipage}{\columnwidth}
    \texttt{aflowlib\_date} \strut \hspace{\fill} \cite{curtarolo:art92}
    \\[-2mm]
    \rule{\columnwidth}{1pt}
    Returns the date when the AFLOW post-processor generated the entry in the library.\\[-2mm]
    \rule{\columnwidth}{0.5pt}
    \begin{minipage}{0.4\columnwidth}\textbf{Unit}: none\end{minipage}\textbf{Type:} string\\
    \textbf{Example:} \seqsplit{aflowlib\_date=20140204\_13:10:39\_GMT-5}
    \\\vspace{-2mm}
\end{minipage}
}

\noindent
\fcolorbox{black}[gray]{1}{
\begin{minipage}{\columnwidth}
    \texttt{aflowlib\_version} \strut \hspace{\fill} \cite{curtarolo:art92}
    \\[-2mm]
    \rule{\columnwidth}{1pt}
    Returns the version of the AFLOW post-processor which generated the entry in the library.\\[-2mm]
    \rule{\columnwidth}{0.5pt}
    \begin{minipage}{0.4\columnwidth}\textbf{Unit}: none\end{minipage}\textbf{Type:} string\\
    \textbf{Example:} \seqsplit{aflowlib\_version=3.1.103}
    \\\vspace{-2mm}
\end{minipage}
}

\noindent
\fcolorbox{black}[gray]{1}{
\begin{minipage}{\columnwidth}
    \texttt{agl\_acoustic\_debye} \strut \hspace{\fill} \cite{curtarolo:art96,curtarolo:art115,curtarolo:art128}
    \\[-2mm]
    \rule{\columnwidth}{1pt}
    Returns the acoustic Debye temperature calculated by AGL.\\[-2mm]
    \rule{\columnwidth}{0.5pt}
    \begin{minipage}{0.4\columnwidth}\textbf{Unit}: K\end{minipage}\textbf{Type:} number\\
    \textbf{Example:} \seqsplit{agl\_acoustic\_debye=492}
    \\\vspace{-2mm}
\end{minipage}
}

\noindent
\fcolorbox{black}[gray]{1}{
\begin{minipage}{\columnwidth}
    \texttt{agl\_bulk\_modulus\_isothermal\_300K} \strut \hspace{\fill} \cite{curtarolo:art96,curtarolo:art115,curtarolo:art128}
    \\[-2mm]
    \rule{\columnwidth}{1pt}
    Returns the isothermal bulk modulus calculated by AGL at 300 K.\\[-2mm]
    \rule{\columnwidth}{0.5pt}
    \begin{minipage}{0.4\columnwidth}\textbf{Unit}: GPa\end{minipage}\textbf{Type:} number\\
    \textbf{Example:} \seqsplit{agl\_bulk\_modulus\_isothermal\_300K=96.6}
    \\\vspace{-2mm}
\end{minipage}
}

\noindent
\fcolorbox{black}[gray]{1}{
\begin{minipage}{\columnwidth}
    \texttt{agl\_bulk\_modulus\_static\_300K} \strut \hspace{\fill} \cite{curtarolo:art96,curtarolo:art115,curtarolo:art128}
    \\[-2mm]
    \rule{\columnwidth}{1pt}
    Returns the static bulk modulus calculated by AGL at 300 K.\\[-2mm]
    \rule{\columnwidth}{0.5pt}
    \begin{minipage}{0.4\columnwidth}\textbf{Unit}: GPa\end{minipage}\textbf{Type:} number\\
    \textbf{Example:} \seqsplit{agl\_bulk\_modulus\_static\_300K=99.6}
    \\\vspace{-2mm}
\end{minipage}
}

\noindent
\fcolorbox{black}[gray]{1}{
\begin{minipage}{\columnwidth}
    \texttt{agl\_debye} \strut \hspace{\fill} \cite{curtarolo:art96,curtarolo:art115,curtarolo:art128}
    \\[-2mm]
    \rule{\columnwidth}{1pt}
    Returns the Debye temperature calculated by AGL.\\[-2mm]
    \rule{\columnwidth}{0.5pt}
    \begin{minipage}{0.4\columnwidth}\textbf{Unit}: K\end{minipage}\textbf{Type:} number\\
    \textbf{Example:} \seqsplit{agl\_debye=620}
    \\\vspace{-2mm}
\end{minipage}
}

\noindent
\fcolorbox{black}[gray]{1}{
\begin{minipage}{\columnwidth}
    \texttt{agl\_gruneisen} \strut \hspace{\fill} \cite{curtarolo:art96,curtarolo:art115,curtarolo:art128}
    \\[-2mm]
    \rule{\columnwidth}{1pt}
    Returns the Gr\"{u}neisen parameter calculated by AGL.\\[-2mm]
    \rule{\columnwidth}{0.5pt}
    \begin{minipage}{0.4\columnwidth}\textbf{Unit}: none\end{minipage}\textbf{Type:} number\\
    \textbf{Example:} \seqsplit{agl\_gruneisen=2.06}
    \\\vspace{-2mm}
\end{minipage}
}

\noindent
\fcolorbox{black}[gray]{1}{
\begin{minipage}{\columnwidth}
    \texttt{agl\_heat\_capacity\_Cp\_300K} \strut \hspace{\fill} \cite{curtarolo:art96,curtarolo:art115,curtarolo:art128}
    \\[-2mm]
    \rule{\columnwidth}{1pt}
    Returns the heat capacity per cell, at constant pressure, calculated by AGL at 300 K.\\[-2mm]
    \rule{\columnwidth}{0.5pt}
    \begin{minipage}{0.4\columnwidth}\textbf{Unit}: $k_\textnormal{B}$/cell\end{minipage}\textbf{Type:} number\\
    \textbf{Example:} \seqsplit{agl\_heat\_capacity\_Cp\_300K=5.502}
    \\\vspace{-2mm}
\end{minipage}
}

\noindent
\fcolorbox{black}[gray]{1}{
\begin{minipage}{\columnwidth}
    \texttt{agl\_heat\_capacity\_Cv\_300K} \strut \hspace{\fill} \cite{curtarolo:art96,curtarolo:art115,curtarolo:art128}
    \\[-2mm]
    \rule{\columnwidth}{1pt}
    Returns the heat capacity per cell, at constant volume, calculated by AGL at 300 K.\\[-2mm]
    \rule{\columnwidth}{0.5pt}
    \begin{minipage}{0.4\columnwidth}\textbf{Unit}: $k_\textnormal{B}$/cell\end{minipage}\textbf{Type:} number\\
    \textbf{Example:} \seqsplit{agl\_heat\_capacity\_Cv\_300K=4.901}
    \\\vspace{-2mm}
\end{minipage}
}

\noindent
\fcolorbox{black}[gray]{1}{
\begin{minipage}{\columnwidth}
    \texttt{agl\_poisson\_ratio\_source} \strut \hspace{\fill} \cite{curtarolo:art96,curtarolo:art115,curtarolo:art128}
    \\[-2mm]
    \rule{\columnwidth}{1pt}
    Returns the source of the Poisson ratio used for AGL calculations.\\[-2mm]
    \rule{\columnwidth}{0.5pt}
    \begin{minipage}{0.4\columnwidth}\textbf{Unit}: none\end{minipage}\textbf{Type:} string\\
    \textbf{Example:} \seqsplit{agl\_poisson\_ratio\_source=ael\_poisson\_ratio\_0.350433}
    \\\vspace{-2mm}
\end{minipage}
}

\noindent
\fcolorbox{black}[gray]{1}{
\begin{minipage}{\columnwidth}
    \texttt{agl\_thermal\_conductivity\_300K} \strut \hspace{\fill} \cite{curtarolo:art96,curtarolo:art115,curtarolo:art128}
    \\[-2mm]
    \rule{\columnwidth}{1pt}
    Returns the thermal conductivity calculated by AGL at 300 K.\\[-2mm]
    \rule{\columnwidth}{0.5pt}
    \begin{minipage}{0.4\columnwidth}\textbf{Unit}: W m$^{-1}$ K$^{-1}$\end{minipage}\textbf{Type:} number\\
    \textbf{Example:} \seqsplit{agl\_thermal\_conductivity\_300K=24.41}
    \\\vspace{-2mm}
\end{minipage}
}

\noindent
\fcolorbox{black}[gray]{1}{
\begin{minipage}{\columnwidth}
    \texttt{agl\_thermal\_expansion\_300K} \strut \hspace{\fill} \cite{curtarolo:art96,curtarolo:art115,curtarolo:art128}
    \\[-2mm]
    \rule{\columnwidth}{1pt}
    Returns the thermal expansion coefficient calculated by AGL at 300 K.\\[-2mm]
    \rule{\columnwidth}{0.5pt}
    \begin{minipage}{0.4\columnwidth}\textbf{Unit}: K$^{-1}$\end{minipage}\textbf{Type:} number\\
    \textbf{Example:} \seqsplit{agl\_thermal\_expansion\_300K=4.997e-05}
    \\\vspace{-2mm}
\end{minipage}
}

\noindent
\fcolorbox{black}[gray]{1}{
\begin{minipage}{\columnwidth}
    \texttt{agl\_vibrational\_entropy\_300K\_atom} \strut \hspace{\fill} \cite{curtarolo:art96,curtarolo:art115}
    \\[-2mm]
    \rule{\columnwidth}{1pt}
    Returns the vibrational entropy per atom calculated by AGL at 300 K.\\[-2mm]
    \rule{\columnwidth}{0.5pt}
    \begin{minipage}{0.4\columnwidth}\textbf{Unit}: meV/(K atom)\end{minipage}\textbf{Type:} number\\
    \textbf{Example:} \seqsplit{agl\_vibrational\_entropy\_300K\_atom=0.454761}
    \\\vspace{-2mm}
\end{minipage}
}

\noindent
\fcolorbox{black}[gray]{1}{
\begin{minipage}{\columnwidth}
    \texttt{agl\_vibrational\_entropy\_300K\_cell} \strut \hspace{\fill} \cite{curtarolo:art96,curtarolo:art115}
    \\[-2mm]
    \rule{\columnwidth}{1pt}
    Returns the vibrational entropy per cell calculated by AGL at 300 K.\\[-2mm]
    \rule{\columnwidth}{0.5pt}
    \begin{minipage}{0.4\columnwidth}\textbf{Unit}: meV/(K cell)\end{minipage}\textbf{Type:} number\\
    \textbf{Example:} \seqsplit{agl\_vibrational\_entropy\_300K\_cell=9.09521}
    \\\vspace{-2mm}
\end{minipage}
}

\noindent
\fcolorbox{black}[gray]{1}{
\begin{minipage}{\columnwidth}
    \texttt{agl\_vibrational\_free\_energy\_300K\_atom} \strut \hspace{\fill} \cite{curtarolo:art96,curtarolo:art115}
    \\[-2mm]
    \rule{\columnwidth}{1pt}
    Returns the vibrational free energy per atom calculated by AGL at 300 K.\\[-2mm]
    \rule{\columnwidth}{0.5pt}
    \begin{minipage}{0.4\columnwidth}\textbf{Unit}: meV/atom\end{minipage}\textbf{Type:} number\\
    \textbf{Example:} \seqsplit{agl\_vibrational\_free\_energy\_300K\_atom=-57.1897}
    \\\vspace{-2mm}
\end{minipage}
}

\noindent
\fcolorbox{black}[gray]{1}{
\begin{minipage}{\columnwidth}
    \texttt{agl\_vibrational\_free\_energy\_300K\_cell} \strut \hspace{\fill} \cite{curtarolo:art96,curtarolo:art115}
    \\[-2mm]
    \rule{\columnwidth}{1pt}
    Returns the vibrational free energy per cell calculated by AGL at 300 K.\\[-2mm]
    \rule{\columnwidth}{0.5pt}
    \begin{minipage}{0.4\columnwidth}\textbf{Unit}: meV/cell\end{minipage}\textbf{Type:} number\\
    \textbf{Example:} \seqsplit{agl\_vibrational\_free\_energy\_300K\_cell=-1143.79}
    \\\vspace{-2mm}
\end{minipage}
}

\noindent
\fcolorbox{black}[gray]{1}{
\begin{minipage}{\columnwidth}
    \texttt{auid} \strut \hspace{\fill} \cite{curtarolo:art92}
    \\[-2mm]
    \rule{\columnwidth}{1pt}
    Returns the AFLOWLIB unique identifier (AUID) for the entry.\\[-2mm]
    \rule{\columnwidth}{0.5pt}
    \begin{minipage}{0.4\columnwidth}\textbf{Unit}: none\end{minipage}\textbf{Type:} string\\
    \textbf{Example:} \seqsplit{auid=aflow:e9c6d914c4b8d9ca}
    \\\vspace{-2mm}
\end{minipage}
}

\noindent
\fcolorbox{black}[gray]{1}{
\begin{minipage}{\columnwidth}
    \texttt{aurl} \strut \hspace{\fill} \cite{curtarolo:art92}
    \\[-2mm]
    \rule{\columnwidth}{1pt}
    Returns the AFLOWLIB uniform resource locator (AURL) for the entry.\\[-2mm]
    \rule{\columnwidth}{0.5pt}
    \begin{minipage}{0.4\columnwidth}\textbf{Unit}: none\end{minipage}\textbf{Type:} string\\
    \textbf{Example:} \seqsplit{aurl=aflowlib.duke.edu:AFLOWDATA/LIB3\_RAW/Bi\_dRh\_pvTi\_sv/T0003.ABC:LDAU2}
    \\\vspace{-2mm}
\end{minipage}
}

\noindent
\fcolorbox{black}[gray]{1}{
\begin{minipage}{\columnwidth}
    \texttt{bader\_atomic\_volumes} \strut \hspace{\fill} \cite{aflow_bader}
    \\[-2mm]
    \rule{\columnwidth}{1pt}
    Returns the volume of each atom calculated by the Atoms in Molecules (AIM) Bader analysis.\\[-2mm]
    \rule{\columnwidth}{0.5pt}
    \begin{minipage}{0.4\columnwidth}\textbf{Unit}: {\AA}$^{3}$\end{minipage}\textbf{Type:} numbers\\
    \textbf{Example:} \seqsplit{bader\_atomic\_volumes=15.235,12.581,13.009}
    \\\vspace{-2mm}
\end{minipage}
}

\noindent
\fcolorbox{black}[gray]{1}{
\begin{minipage}{\columnwidth}
    \texttt{bader\_net\_charges} \strut \hspace{\fill} \cite{aflow_bader}
    \\[-2mm]
    \rule{\columnwidth}{1pt}
    Returns the partial charge of each atom calculated by the Atoms in Molecules (AIM) Bader analysis.\\[-2mm]
    \rule{\columnwidth}{0.5pt}
    \begin{minipage}{0.4\columnwidth}\textbf{Unit}: e$^{-}$\end{minipage}\textbf{Type:} numbers\\
    \textbf{Example:} \seqsplit{bader\_net\_charges=0.125,0.125,-0.25}
    \\\vspace{-2mm}
\end{minipage}
}

\noindent
\fcolorbox{black}[gray]{1}{
\begin{minipage}{\columnwidth}
    \texttt{Bravais\_lattice\_lattice\_system} \strut \hspace{\fill} \cite{curtarolo:art75,curtarolo:art135}
    \\[-2mm]
    \rule{\columnwidth}{1pt}
    Returns the Bravais lattice of the lattice system for the relaxed structure.\\[-2mm]
    \rule{\columnwidth}{0.5pt}
    \begin{minipage}{0.4\columnwidth}\textbf{Unit}: none\end{minipage}\textbf{Type:} string\\
    \textbf{Example:} \seqsplit{Bravais\_lattice\_lattice\_system=cubic}
    \\\vspace{-2mm}
\end{minipage}
}

\noindent
\fcolorbox{black}[gray]{1}{
\begin{minipage}{\columnwidth}
    \texttt{Bravais\_lattice\_lattice\_system\_orig} \strut \hspace{\fill} \cite{curtarolo:art75,curtarolo:art135}
    \\[-2mm]
    \rule{\columnwidth}{1pt}
    Returns the Bravais lattice of the lattice system for the unrelaxed structure.\\[-2mm]
    \rule{\columnwidth}{0.5pt}
    \begin{minipage}{0.4\columnwidth}\textbf{Unit}: none\end{minipage}\textbf{Type:} string\\
    \textbf{Example:} \seqsplit{Bravais\_lattice\_lattice\_system=cubic}
    \\\vspace{-2mm}
\end{minipage}
}

\noindent
\fcolorbox{black}[gray]{1}{
\begin{minipage}{\columnwidth}
    \texttt{Bravais\_lattice\_lattice\_type} \strut \hspace{\fill} \cite{curtarolo:art75,curtarolo:art135}
    \\[-2mm]
    \rule{\columnwidth}{1pt}
    Returns the lattice centering type for the relaxed structure.\\[-2mm]
    \rule{\columnwidth}{0.5pt}
    \begin{minipage}{0.4\columnwidth}\textbf{Unit}: none\end{minipage}\textbf{Type:} string\\
    \textbf{Example:} \seqsplit{Bravais\_lattice\_lattice\_type=BCC}
    \\\vspace{-2mm}
\end{minipage}
}

\noindent
\fcolorbox{black}[gray]{1}{
\begin{minipage}{\columnwidth}
    \texttt{Bravais\_lattice\_lattice\_type\_orig} \strut \hspace{\fill} \cite{curtarolo:art75,curtarolo:art135}
    \\[-2mm]
    \rule{\columnwidth}{1pt}
    Returns the lattice centering type for the unrelaxed structure.\\[-2mm]
    \rule{\columnwidth}{0.5pt}
    \begin{minipage}{0.4\columnwidth}\textbf{Unit}: none\end{minipage}\textbf{Type:} string\\
    \textbf{Example:} \seqsplit{Bravais\_lattice\_lattice\_type\_orig=BCC}
    \\\vspace{-2mm}
\end{minipage}
}

\noindent
\fcolorbox{black}[gray]{1}{
\begin{minipage}{\columnwidth}
    \texttt{Bravais\_lattice\_lattice\_variation\_type} \strut \hspace{\fill} \cite{curtarolo:art75,curtarolo:art135}
    \\[-2mm]
    \rule{\columnwidth}{1pt}
    Returns the Bravais lattice variation of the lattice system for the relaxed structure.\\[-2mm]
    \rule{\columnwidth}{0.5pt}
    \begin{minipage}{0.4\columnwidth}\textbf{Unit}: none\end{minipage}\textbf{Type:} string\\
    \textbf{Example:} \seqsplit{Bravais\_lattice\_lattice\_variation\_type=BCC}
    \\\vspace{-2mm}
\end{minipage}
}

\noindent
\fcolorbox{black}[gray]{1}{
\begin{minipage}{\columnwidth}
    \texttt{Bravais\_lattice\_lattice\_variation\_type\_orig} \strut \hspace{\fill} \cite{curtarolo:art75,curtarolo:art135}
    \\[-2mm]
    \rule{\columnwidth}{1pt}
    Returns the Bravais lattice variation of the lattice system for the unrelaxed structure.\\[-2mm]
    \rule{\columnwidth}{0.5pt}
    \begin{minipage}{0.4\columnwidth}\textbf{Unit}: none\end{minipage}\textbf{Type:} string\\
    \textbf{Example:} \seqsplit{Bravais\_lattice\_lattice\_variation\_type\_orig=BCC}
    \\\vspace{-2mm}
\end{minipage}
}

\noindent
\fcolorbox{black}[gray]{1}{
\begin{minipage}{\columnwidth}
    \texttt{Bravais\_lattice\_orig} \strut \hspace{\fill} \cite{curtarolo:art58,curtarolo:art75,curtarolo:art92,curtarolo:art135}
    \\[-2mm]
    \rule{\columnwidth}{1pt}
    Returns the Bravais lattice of the crystal for the unrelaxed structure.\\[-2mm]
    \rule{\columnwidth}{0.5pt}
    \begin{minipage}{0.4\columnwidth}\textbf{Unit}: none\end{minipage}\textbf{Type:} string\\
    \textbf{Example:} \seqsplit{Bravais\_lattice\_orig=MCLC}
    \\\vspace{-2mm}
\end{minipage}
}

\noindent
\fcolorbox{black}[gray]{1}{
\begin{minipage}{\columnwidth}
    \texttt{Bravais\_lattice\_relax} \strut \hspace{\fill} \cite{curtarolo:art58,curtarolo:art75,curtarolo:art92,curtarolo:art135}
    \\[-2mm]
    \rule{\columnwidth}{1pt}
    Returns the Bravais lattice of the crystal for the relaxed structure.\\[-2mm]
    \rule{\columnwidth}{0.5pt}
    \begin{minipage}{0.4\columnwidth}\textbf{Unit}: none\end{minipage}\textbf{Type:} string\\
    \textbf{Example:} \seqsplit{Bravais\_lattice\_relax=MCLC}
    \\\vspace{-2mm}
\end{minipage}
}

\noindent
\fcolorbox{black}[gray]{1}{
\begin{minipage}{\columnwidth}
    \texttt{Bravais\_superlattice\_lattice\_system} \strut \hspace{\fill} \cite{curtarolo:art75,curtarolo:art135}
    \\[-2mm]
    \rule{\columnwidth}{1pt}
    Returns the Bravais superlattice of the lattice system for the relaxed structure.\\[-2mm]
    \rule{\columnwidth}{0.5pt}
    \begin{minipage}{0.4\columnwidth}\textbf{Unit}: none\end{minipage}\textbf{Type:} string\\
    \textbf{Example:} \seqsplit{Bravais\_superlattice\_lattice\_system=cubic}
    \\\vspace{-2mm}
\end{minipage}
}

\noindent
\fcolorbox{black}[gray]{1}{
\begin{minipage}{\columnwidth}
    \texttt{Bravais\_superlattice\_lattice\_system\_orig} \strut \hspace{\fill} \cite{curtarolo:art75,curtarolo:art135}
    \\[-2mm]
    \rule{\columnwidth}{1pt}
    Returns the Bravais superlattice of the lattice system for the unrelaxed structure.\\[-2mm]
    \rule{\columnwidth}{0.5pt}
    \begin{minipage}{0.4\columnwidth}\textbf{Unit}: none\end{minipage}\textbf{Type:} string\\
    \textbf{Example:} \seqsplit{Bravais\_superlattice\_lattice\_system\_orig=cubic}
    \\\vspace{-2mm}
\end{minipage}
}

\noindent
\fcolorbox{black}[gray]{1}{
\begin{minipage}{\columnwidth}
    \texttt{Bravais\_superlattice\_lattice\_type} \strut \hspace{\fill} \cite{curtarolo:art75,curtarolo:art135}
    \\[-2mm]
    \rule{\columnwidth}{1pt}
    Returns the Bravais superlattice centering type for the relaxed structure.\\[-2mm]
    \rule{\columnwidth}{0.5pt}
    \begin{minipage}{0.4\columnwidth}\textbf{Unit}: none\end{minipage}\textbf{Type:} string\\
    \textbf{Example:} \seqsplit{Bravais\_superlattice\_lattice\_type=BCC}
    \\\vspace{-2mm}
\end{minipage}
}

\noindent
\fcolorbox{black}[gray]{1}{
\begin{minipage}{\columnwidth}
    \texttt{Bravais\_superlattice\_lattice\_type\_orig} \strut \hspace{\fill} \cite{curtarolo:art75,curtarolo:art135}
    \\[-2mm]
    \rule{\columnwidth}{1pt}
    Returns the Bravais superlattice centering type for the unrelaxed structure.\\[-2mm]
    \rule{\columnwidth}{0.5pt}
    \begin{minipage}{0.4\columnwidth}\textbf{Unit}: none\end{minipage}\textbf{Type:} string\\
    \textbf{Example:} \seqsplit{Bravais\_superlattice\_lattice\_type\_orig=BCC}
    \\\vspace{-2mm}
\end{minipage}
}

\noindent
\fcolorbox{black}[gray]{1}{
\begin{minipage}{\columnwidth}
    \texttt{Bravais\_superlattice\_lattice\_variation\_type} \strut \hspace{\fill} \cite{curtarolo:art75,curtarolo:art135}
    \\[-2mm]
    \rule{\columnwidth}{1pt}
    Returns the Bravais superlattice variation of the lattice system for the relaxed structure.\\[-2mm]
    \rule{\columnwidth}{0.5pt}
    \begin{minipage}{0.4\columnwidth}\textbf{Unit}: none\end{minipage}\textbf{Type:} string\\
    \textbf{Example:} \seqsplit{Bravais\_superlattice\_lattice\_variation\_type=BCC}
    \\\vspace{-2mm}
\end{minipage}
}

\noindent
\fcolorbox{black}[gray]{1}{
\begin{minipage}{\columnwidth}
    \texttt{Bravais\_superlattice\_lattice\_variation\_type\_orig} \strut \hspace{\fill} \cite{curtarolo:art75,curtarolo:art135}
    \\[-2mm]
    \rule{\columnwidth}{1pt}
    Returns the Bravais superlattice variation of the lattice system for the unrelaxed structure.\\[-2mm]
    \rule{\columnwidth}{0.5pt}
    \begin{minipage}{0.4\columnwidth}\textbf{Unit}: none\end{minipage}\textbf{Type:} string\\
    \textbf{Example:} \seqsplit{Bravais\_superlattice\_lattice\_variation\_type\_orig=BCC}
    \\\vspace{-2mm}
\end{minipage}
}

\noindent
\fcolorbox{black}[gray]{1}{
\begin{minipage}{\columnwidth}
    \texttt{calculation\_cores} \strut \hspace{\fill} \cite{curtarolo:art92}
    \\[-2mm]
    \rule{\columnwidth}{1pt}
    Returns the number of CPUs used by the calculation.\\[-2mm]
    \rule{\columnwidth}{0.5pt}
    \begin{minipage}{0.4\columnwidth}\textbf{Unit}: none\end{minipage}\textbf{Type:} number\\
    \textbf{Example:} \seqsplit{calculation\_cores=32}
    \\\vspace{-2mm}
\end{minipage}
}

\noindent
\fcolorbox{black}[gray]{1}{
\begin{minipage}{\columnwidth}
    \texttt{calculation\_memory} \strut \hspace{\fill} \cite{curtarolo:art92}
    \\[-2mm]
    \rule{\columnwidth}{1pt}
    Returns the maximum RAM used by the calculation.\\[-2mm]
    \rule{\columnwidth}{0.5pt}
    \begin{minipage}{0.4\columnwidth}\textbf{Unit}: MB\end{minipage}\textbf{Type:} number\\
    \textbf{Example:} \seqsplit{calculation\_memory=32}
    \\\vspace{-2mm}
\end{minipage}
}

\noindent
\fcolorbox{black}[gray]{1}{
\begin{minipage}{\columnwidth}
    \texttt{calculation\_time} \strut \hspace{\fill} \cite{curtarolo:art92}
    \\[-2mm]
    \rule{\columnwidth}{1pt}
    Returns the total time taken by the calculation.\\[-2mm]
    \rule{\columnwidth}{0.5pt}
    \begin{minipage}{0.4\columnwidth}\textbf{Unit}: seconds\end{minipage}\textbf{Type:} number\\
    \textbf{Example:} \seqsplit{calculation\_time=32}
    \\\vspace{-2mm}
\end{minipage}
}

\noindent
\fcolorbox{black}[gray]{1}{
\begin{minipage}{\columnwidth}
    \texttt{catalog} \strut \hspace{\fill} \cite{curtarolo:art75}
    \\[-2mm]
    \rule{\columnwidth}{1pt}
    Returns the database name for the calculation.\\[-2mm]
    \rule{\columnwidth}{0.5pt}
    \begin{minipage}{0.4\columnwidth}\textbf{Unit}: none\end{minipage}\textbf{Type:} string\\
    \textbf{Example:} \seqsplit{catalog=icsd}
    \\\vspace{-2mm}
\end{minipage}
}

\noindent
\fcolorbox{black}[gray]{1}{
\begin{minipage}{\columnwidth}
    \texttt{code} \strut \hspace{\fill} \cite{curtarolo:art92}
    \\[-2mm]
    \rule{\columnwidth}{1pt}
    Returns the software name and version used to perform the calculation.\\[-2mm]
    \rule{\columnwidth}{0.5pt}
    \begin{minipage}{0.4\columnwidth}\textbf{Unit}: none\end{minipage}\textbf{Type:} string\\
    \textbf{Example:} \seqsplit{code=vasp.4.6.35}
    \\\vspace{-2mm}
\end{minipage}
}

\noindent
\fcolorbox{black}[gray]{1}{
\begin{minipage}{\columnwidth}
    \texttt{composition} \strut \hspace{\fill} \cite{curtarolo:art92}
    \\[-2mm]
    \rule{\columnwidth}{1pt}
    Returns the number of atoms per type in the simulation cell.\\[-2mm]
    \rule{\columnwidth}{0.5pt}
    \begin{minipage}{0.4\columnwidth}\textbf{Unit}: none\end{minipage}\textbf{Type:} numbers\\
    \textbf{Example:} \seqsplit{composition=2,6,6}
    \\\vspace{-2mm}
\end{minipage}
}

\noindent
\fcolorbox{black}[gray]{1}{
\begin{minipage}{\columnwidth}
    \texttt{compound} \strut \hspace{\fill} \cite{curtarolo:art92}
    \\[-2mm]
    \rule{\columnwidth}{1pt}
    Returns the chemical formula of the structure.\\[-2mm]
    \rule{\columnwidth}{0.5pt}
    \begin{minipage}{0.4\columnwidth}\textbf{Unit}: none\end{minipage}\textbf{Type:} string\\
    \textbf{Example:} \seqsplit{compound=Co2Er6Si6}
    \\\vspace{-2mm}
\end{minipage}
}

\noindent
\fcolorbox{black}[gray]{1}{
\begin{minipage}{\columnwidth}
    \texttt{crystal\_class} \strut \hspace{\fill} \cite{curtarolo:art135}
    \\[-2mm]
    \rule{\columnwidth}{1pt}
    Returns the crystal class for the relaxed structure.\\[-2mm]
    \rule{\columnwidth}{0.5pt}
    \begin{minipage}{0.4\columnwidth}\textbf{Unit}: none\end{minipage}\textbf{Type:} string\\
    \textbf{Example:} \seqsplit{crystal\_class=tetrahedral}
    \\\vspace{-2mm}
\end{minipage}
}

\noindent
\fcolorbox{black}[gray]{1}{
\begin{minipage}{\columnwidth}
    \texttt{crystal\_class\_orig} \strut \hspace{\fill} \cite{curtarolo:art135}
    \\[-2mm]
    \rule{\columnwidth}{1pt}
    Returns the crystal class for the unrelaxed structure.\\[-2mm]
    \rule{\columnwidth}{0.5pt}
    \begin{minipage}{0.4\columnwidth}\textbf{Unit}: none\end{minipage}\textbf{Type:} string\\
    \textbf{Example:} \seqsplit{crystal\_class\_orig=tetrahedral}
    \\\vspace{-2mm}
\end{minipage}
}

\noindent
\fcolorbox{black}[gray]{1}{
\begin{minipage}{\columnwidth}
    \texttt{crystal\_family} \strut \hspace{\fill} \cite{curtarolo:art135}
    \\[-2mm]
    \rule{\columnwidth}{1pt}
    Returns the crystal family for the relaxed structure.\\[-2mm]
    \rule{\columnwidth}{0.5pt}
    \begin{minipage}{0.4\columnwidth}\textbf{Unit}: none\end{minipage}\textbf{Type:} string\\
    \textbf{Example:} \seqsplit{crystal\_family=cubic}
    \\\vspace{-2mm}
\end{minipage}
}

\noindent
\fcolorbox{black}[gray]{1}{
\begin{minipage}{\columnwidth}
    \texttt{crystal\_family\_orig} \strut \hspace{\fill} \cite{curtarolo:art135}
    \\[-2mm]
    \rule{\columnwidth}{1pt}
    Returns the crystal family for the unrelaxed structure.\\[-2mm]
    \rule{\columnwidth}{0.5pt}
    \begin{minipage}{0.4\columnwidth}\textbf{Unit}: none\end{minipage}\textbf{Type:} string\\
    \textbf{Example:} \seqsplit{crystal\_family\_orig=cubic}
    \\\vspace{-2mm}
\end{minipage}
}

\noindent
\fcolorbox{black}[gray]{1}{
\begin{minipage}{\columnwidth}
    \texttt{crystal\_system} \strut \hspace{\fill} \cite{curtarolo:art135}
    \\[-2mm]
    \rule{\columnwidth}{1pt}
    Returns the crystal system for the relaxed structure.\\[-2mm]
    \rule{\columnwidth}{0.5pt}
    \begin{minipage}{0.4\columnwidth}\textbf{Unit}: none\end{minipage}\textbf{Type:} string\\
    \textbf{Example:} \seqsplit{crystal\_system=cubic}
    \\\vspace{-2mm}
\end{minipage}
}

\noindent
\fcolorbox{black}[gray]{1}{
\begin{minipage}{\columnwidth}
    \texttt{crystal\_system\_orig} \strut \hspace{\fill} \cite{curtarolo:art135}
    \\[-2mm]
    \rule{\columnwidth}{1pt}
    Returns the crystal system for the unrelaxed structure.\\[-2mm]
    \rule{\columnwidth}{0.5pt}
    \begin{minipage}{0.4\columnwidth}\textbf{Unit}: none\end{minipage}\textbf{Type:} string\\
    \textbf{Example:} \seqsplit{crystal\_system\_orig=cubic}
    \\\vspace{-2mm}
\end{minipage}
}

\noindent
\fcolorbox{black}[gray]{1}{
\begin{minipage}{\columnwidth}
    \texttt{data\_api} \strut \hspace{\fill} \cite{curtarolo:art92}
    \\[-2mm]
    \rule{\columnwidth}{1pt}
    Returns the REST API version for the entry.\\[-2mm]
    \rule{\columnwidth}{0.5pt}
    \begin{minipage}{0.4\columnwidth}\textbf{Unit}: none\end{minipage}\textbf{Type:} string\\
    \textbf{Example:} \seqsplit{data\_api=aapi1.0}
    \\\vspace{-2mm}
\end{minipage}
}

\noindent
\fcolorbox{black}[gray]{1}{
\begin{minipage}{\columnwidth}
    \texttt{data\_source} \strut \hspace{\fill} \cite{curtarolo:art92}
    \\[-2mm]
    \rule{\columnwidth}{1pt}
    Returns the data source for the entry.\\[-2mm]
    \rule{\columnwidth}{0.5pt}
    \begin{minipage}{0.4\columnwidth}\textbf{Unit}: none\end{minipage}\textbf{Type:} string\\
    \textbf{Example:} \seqsplit{data\_source=aflowlib}
    \\\vspace{-2mm}
\end{minipage}
}

\noindent
\fcolorbox{black}[gray]{1}{
\begin{minipage}{\columnwidth}
    \texttt{delta\_electronic\_energy\_convergence} \strut \hspace{\fill} \cite{curtarolo:art104,curtarolo:art128}
    \\[-2mm]
    \rule{\columnwidth}{1pt}
    Returns the change in total energy from the last step of the self-consistent field (SCF) iteration.\\[-2mm]
    \rule{\columnwidth}{0.5pt}
    \begin{minipage}{0.4\columnwidth}\textbf{Unit}: eV\end{minipage}\textbf{Type:} number\\
    \textbf{Example:} \seqsplit{delta\_electronic\_energy\_convergence=6.09588e-05}
    \\\vspace{-2mm}
\end{minipage}
}

\noindent
\fcolorbox{black}[gray]{1}{
\begin{minipage}{\columnwidth}
    \texttt{delta\_electronic\_energy\_threshold} \strut \hspace{\fill} \cite{curtarolo:art104,curtarolo:art128}
    \\[-2mm]
    \rule{\columnwidth}{1pt}
    Returns the threshold for the self-consistent field (SCF) convergence.\\[-2mm]
    \rule{\columnwidth}{0.5pt}
    \begin{minipage}{0.4\columnwidth}\textbf{Unit}: eV\end{minipage}\textbf{Type:} number\\
    \textbf{Example:} \seqsplit{delta\_electronic\_energy\_threshold=0.0001}
    \\\vspace{-2mm}
\end{minipage}
}

\noindent
\fcolorbox{black}[gray]{1}{
\begin{minipage}{\columnwidth}
    \texttt{density} \strut \hspace{\fill} \cite{curtarolo:art92}
    \\[-2mm]
    \rule{\columnwidth}{1pt}
    Returns the mass density of the unit cell.\\[-2mm]
    \rule{\columnwidth}{0.5pt}
    \begin{minipage}{0.4\columnwidth}\textbf{Unit}: g/cm$^{3}$\end{minipage}\textbf{Type:} number\\
    \textbf{Example:} \seqsplit{density=7.76665}
    \\\vspace{-2mm}
\end{minipage}
}

\noindent
\fcolorbox{black}[gray]{1}{
\begin{minipage}{\columnwidth}
    \texttt{dft\_type} \strut \hspace{\fill} \cite{curtarolo:art92,curtarolo:art104}
    \\[-2mm]
    \rule{\columnwidth}{1pt}
    Returns the level of theory used in the calculation, i.e., pseudopotential type, exchange-correlation functional used, and use of GW.\\[-2mm]
    \rule{\columnwidth}{0.5pt}
    \begin{minipage}{0.4\columnwidth}\textbf{Unit}: none\end{minipage}\textbf{Type:} strings\\
    \textbf{Example:} \seqsplit{dft\_type=PAW\_PBE}
    \\\vspace{-2mm}
\end{minipage}
}

\noindent
\fcolorbox{black}[gray]{1}{
\begin{minipage}{\columnwidth}
    \texttt{eentropy\_atom} \strut \hspace{\fill} \cite{curtarolo:art92}
    \\[-2mm]
    \rule{\columnwidth}{1pt}
    Returns the electronic entropy per atom used to converge the calculation.\\[-2mm]
    \rule{\columnwidth}{0.5pt}
    \begin{minipage}{0.4\columnwidth}\textbf{Unit}: eV/atom\end{minipage}\textbf{Type:} number\\
    \textbf{Example:} \seqsplit{eentropy\_atom=0.0011}
    \\\vspace{-2mm}
\end{minipage}
}

\noindent
\fcolorbox{black}[gray]{1}{
\begin{minipage}{\columnwidth}
    \texttt{eentropy\_cell} \strut \hspace{\fill} \cite{curtarolo:art92}
    \\[-2mm]
    \rule{\columnwidth}{1pt}
    Returns the electronic entropy per cell used to converge the calculation.\\[-2mm]
    \rule{\columnwidth}{0.5pt}
    \begin{minipage}{0.4\columnwidth}\textbf{Unit}: eV/cell\end{minipage}\textbf{Type:} number\\
    \textbf{Example:} \seqsplit{eentropy\_cell=0.0011}
    \\\vspace{-2mm}
\end{minipage}
}

\noindent
\fcolorbox{black}[gray]{1}{
\begin{minipage}{\columnwidth}
    \texttt{Egap} \strut \hspace{\fill} \cite{curtarolo:art58,curtarolo:art75,curtarolo:art92,curtarolo:art104}
    \\[-2mm]
    \rule{\columnwidth}{1pt}
    Returns the electronic band gap.\\[-2mm]
    \rule{\columnwidth}{0.5pt}
    \begin{minipage}{0.4\columnwidth}\textbf{Unit}: eV\end{minipage}\textbf{Type:} number\\
    \textbf{Example:} \seqsplit{Egap=2.5}
    \\\vspace{-2mm}
\end{minipage}
}

\noindent
\fcolorbox{black}[gray]{1}{
\begin{minipage}{\columnwidth}
    \texttt{Egap\_type} \strut \hspace{\fill} \cite{curtarolo:art58,curtarolo:art75,curtarolo:art92,curtarolo:art104}
    \\[-2mm]
    \rule{\columnwidth}{1pt}
    Returns the electronic band gap type.\\[-2mm]
    \rule{\columnwidth}{0.5pt}
    \begin{minipage}{0.4\columnwidth}\textbf{Unit}: none\end{minipage}\textbf{Type:} string\\
    \textbf{Example:} \seqsplit{Egap\_type=insulator\_direct}
    \\\vspace{-2mm}
\end{minipage}
}

\noindent
\fcolorbox{black}[gray]{1}{
\begin{minipage}{\columnwidth}
    \texttt{energy\_atom} \strut \hspace{\fill} \cite{curtarolo:art75,curtarolo:art92}
    \\[-2mm]
    \rule{\columnwidth}{1pt}
    Returns the total ab-initio energy per atom.\\[-2mm]
    \rule{\columnwidth}{0.5pt}
    \begin{minipage}{0.4\columnwidth}\textbf{Unit}: eV/atom\end{minipage}\textbf{Type:} number\\
    \textbf{Example:} \seqsplit{energy\_atom=-82.1656}
    \\\vspace{-2mm}
\end{minipage}
}

\noindent
\fcolorbox{black}[gray]{1}{
\begin{minipage}{\columnwidth}
    \texttt{energy\_cell} \strut \hspace{\fill} \cite{curtarolo:art75,curtarolo:art92}
    \\[-2mm]
    \rule{\columnwidth}{1pt}
    Returns the total ab-initio energy per cell.\\[-2mm]
    \rule{\columnwidth}{0.5pt}
    \begin{minipage}{0.4\columnwidth}\textbf{Unit}: eV/cell\end{minipage}\textbf{Type:} number\\
    \textbf{Example:} \seqsplit{energy\_cell=-82.1656}
    \\\vspace{-2mm}
\end{minipage}
}

\noindent
\fcolorbox{black}[gray]{1}{
\begin{minipage}{\columnwidth}
    \texttt{energy\_cutoff} \strut \hspace{\fill} \cite{curtarolo:art92,curtarolo:art104}
    \\[-2mm]
    \rule{\columnwidth}{1pt}
    Return the plane-wave energy cut-off used for the calculation.\\[-2mm]
    \rule{\columnwidth}{0.5pt}
    \begin{minipage}{0.4\columnwidth}\textbf{Unit}: eV\end{minipage}\textbf{Type:} number\\
    \textbf{Example:} \seqsplit{energy\_cutoff=384.1,384.1,384.1}
    \\\vspace{-2mm}
\end{minipage}
}

\noindent
\fcolorbox{black}[gray]{1}{
\begin{minipage}{\columnwidth}
    \texttt{enthalpy\_atom} \strut \hspace{\fill} \cite{curtarolo:art92}
    \\[-2mm]
    \rule{\columnwidth}{1pt}
    Returns the enthalpy per atom.\\[-2mm]
    \rule{\columnwidth}{0.5pt}
    \begin{minipage}{0.4\columnwidth}\textbf{Unit}: eV/atom\end{minipage}\textbf{Type:} number\\
    \textbf{Example:} \seqsplit{enthalpy\_atom=-82.1656}
    \\\vspace{-2mm}
\end{minipage}
}

\noindent
\fcolorbox{black}[gray]{1}{
\begin{minipage}{\columnwidth}
    \texttt{enthalpy\_cell} \strut \hspace{\fill} \cite{curtarolo:art92}
    \\[-2mm]
    \rule{\columnwidth}{1pt}
    Returns the enthalpy per cell.\\[-2mm]
    \rule{\columnwidth}{0.5pt}
    \begin{minipage}{0.4\columnwidth}\textbf{Unit}: eV/cell\end{minipage}\textbf{Type:} number\\
    \textbf{Example:} \seqsplit{enthalpy\_cell=-82.1656}
    \\\vspace{-2mm}
\end{minipage}
}

\noindent
\fcolorbox{black}[gray]{1}{
\begin{minipage}{\columnwidth}
    \texttt{enthalpy\_formation\_atom} \strut \hspace{\fill} \cite{curtarolo:art75,curtarolo:art92}
    \\[-2mm]
    \rule{\columnwidth}{1pt}
    Returns the formation enthalpy per atom.\\[-2mm]
    \rule{\columnwidth}{0.5pt}
    \begin{minipage}{0.4\columnwidth}\textbf{Unit}: eV/atom\end{minipage}\textbf{Type:} number\\
    \textbf{Example:} \seqsplit{enthalpy\_formation\_atom=-33.1587}
    \\\vspace{-2mm}
\end{minipage}
}

\noindent
\fcolorbox{black}[gray]{1}{
\begin{minipage}{\columnwidth}
    \texttt{enthalpy\_formation\_cell} \strut \hspace{\fill} \cite{curtarolo:art75,curtarolo:art92}
    \\[-2mm]
    \rule{\columnwidth}{1pt}
    Returns the formation enthalpy per cell.\\[-2mm]
    \rule{\columnwidth}{0.5pt}
    \begin{minipage}{0.4\columnwidth}\textbf{Unit}: eV/cell\end{minipage}\textbf{Type:} number\\
    \textbf{Example:} \seqsplit{enthalpy\_formation\_cell=-33.1587}
    \\\vspace{-2mm}
\end{minipage}
}

\noindent
\fcolorbox{black}[gray]{1}{
\begin{minipage}{\columnwidth}
    \texttt{entropic\_temperature} \strut \hspace{\fill} \cite{curtarolo:art81,curtarolo:art87}
    \\[-2mm]
    \rule{\columnwidth}{1pt}
    Returns the entropic temperature.\\[-2mm]
    \rule{\columnwidth}{0.5pt}
    \begin{minipage}{0.4\columnwidth}\textbf{Unit}: K\end{minipage}\textbf{Type:} number\\
    \textbf{Example:} \seqsplit{entropic\_temperature=1072.1}
    \\\vspace{-2mm}
\end{minipage}
}

\noindent
\fcolorbox{black}[gray]{1}{
\begin{minipage}{\columnwidth}
    \texttt{files} \strut \hspace{\fill} \cite{curtarolo:art92}
    \\[-2mm]
    \rule{\columnwidth}{1pt}
    Returns the input and output files used in the simulation.\\[-2mm]
    \rule{\columnwidth}{0.5pt}
    \begin{minipage}{0.4\columnwidth}\textbf{Unit}: none\end{minipage}\textbf{Type:} strings\\
    \textbf{Example:} \seqsplit{files=Bi\_dRh\_pv.33.cif,Bi\_dRh\_pv.33.png,CONTCAR.relax,CONTCAR.relax1,}
    \\\vspace{-2mm}
\end{minipage}
}

\noindent
\fcolorbox{black}[gray]{1}{
\begin{minipage}{\columnwidth}
    \texttt{forces} \strut \hspace{\fill} \cite{curtarolo:art92}
    \\[-2mm]
    \rule{\columnwidth}{1pt}
    Returns the forces on the atoms for the relaxed structure.\\[-2mm]
    \rule{\columnwidth}{0.5pt}
    \begin{minipage}{0.4\columnwidth}\textbf{Unit}: eV/{\AA}\end{minipage}\textbf{Type:} numbers\\
    \textbf{Example:} \seqsplit{forces=0,-0.023928,0.000197;0,0.023928,-0.000197;...}
    \\\vspace{-2mm}
\end{minipage}
}

\noindent
\fcolorbox{black}[gray]{1}{
\begin{minipage}{\columnwidth}
    \texttt{geometry} \strut \hspace{\fill} \cite{curtarolo:art58,curtarolo:art75,curtarolo:art92,curtarolo:art135}
    \\[-2mm]
    \rule{\columnwidth}{1pt}
    Returns the lattice parameters of the relaxed simulation cell.\\[-2mm]
    \rule{\columnwidth}{0.5pt}
    \begin{minipage}{0.4\columnwidth}\textbf{Unit}: none\end{minipage}\textbf{Type:} numbers\\
    \textbf{Example:} \seqsplit{geometry=18.82,18.82,18.82,32.41,32.41,32.41}
    \\\vspace{-2mm}
\end{minipage}
}

\noindent
\fcolorbox{black}[gray]{1}{
\begin{minipage}{\columnwidth}
    \texttt{geometry\_orig} \strut \hspace{\fill} \cite{curtarolo:art58,curtarolo:art75,curtarolo:art92,curtarolo:art135}
    \\[-2mm]
    \rule{\columnwidth}{1pt}
    Returns the lattice parameters of the unrelaxed simulation cell.\\[-2mm]
    \rule{\columnwidth}{0.5pt}
    \begin{minipage}{0.4\columnwidth}\textbf{Unit}: none\end{minipage}\textbf{Type:} numbers\\
    \textbf{Example:} \seqsplit{geometry\_orig=18.82,18.82,18.82,32.41,32.41,32.41}
    \\\vspace{-2mm}
\end{minipage}
}

\noindent
\fcolorbox{black}[gray]{1}{
\begin{minipage}{\columnwidth}
    \texttt{kpoints\_bands\_nkpts} \strut \hspace{\fill} \cite{curtarolo:art58,curtarolo:art92,curtarolo:art104,curtarolo:art128}
    \\[-2mm]
    \rule{\columnwidth}{1pt}
    Returns the number of points, between the high-symmetry k-points, used for the band structure calculation.\\[-2mm]
    \rule{\columnwidth}{0.5pt}
    \begin{minipage}{0.4\columnwidth}\textbf{Unit}: none\end{minipage}\textbf{Type:} number\\
    \textbf{Example:} \seqsplit{kpoints\_bands\_nkpts=20}
    \\\vspace{-2mm}
\end{minipage}
}

\noindent
\fcolorbox{black}[gray]{1}{
\begin{minipage}{\columnwidth}
    \texttt{kpoints\_bands\_path} \strut \hspace{\fill} \cite{curtarolo:art58,curtarolo:art92,curtarolo:art104,curtarolo:art128}
    \\[-2mm]
    \rule{\columnwidth}{1pt}
    Returns the high-symmetry k-point path used for the band structure calculation.\\[-2mm]
    \rule{\columnwidth}{0.5pt}
    \begin{minipage}{0.4\columnwidth}\textbf{Unit}: none\end{minipage}\textbf{Type:} strings\\
    \textbf{Example:} \seqsplit{kpoints\_bands\_path=G-X-W-K-G-L-U-W-K-K-U-X}
    \\\vspace{-2mm}
\end{minipage}
}

\noindent
\fcolorbox{black}[gray]{1}{
\begin{minipage}{\columnwidth}
    \texttt{kpoints\_relax} \strut \hspace{\fill} \cite{curtarolo:art58,curtarolo:art92,curtarolo:art104,curtarolo:art128}
    \\[-2mm]
    \rule{\columnwidth}{1pt}
    Returns the k-point grid used for the structural relaxation calculation.\\[-2mm]
    \rule{\columnwidth}{0.5pt}
    \begin{minipage}{0.4\columnwidth}\textbf{Unit}: none\end{minipage}\textbf{Type:} numbers\\
    \textbf{Example:} \seqsplit{kpoints\_relax=14,14,14}
    \\\vspace{-2mm}
\end{minipage}
}

\noindent
\fcolorbox{black}[gray]{1}{
\begin{minipage}{\columnwidth}
    \texttt{kpoints\_static} \strut \hspace{\fill} \cite{curtarolo:art58,curtarolo:art92,curtarolo:art104,curtarolo:art128}
    \\[-2mm]
    \rule{\columnwidth}{1pt}
    Returns the k-point grid used for the static calculation.\\[-2mm]
    \rule{\columnwidth}{0.5pt}
    \begin{minipage}{0.4\columnwidth}\textbf{Unit}: none\end{minipage}\textbf{Type:} numbers\\
    \textbf{Example:} \seqsplit{kpoints\_static=14,14,14}
    \\\vspace{-2mm}
\end{minipage}
}

\noindent
\fcolorbox{black}[gray]{1}{
\begin{minipage}{\columnwidth}
    \texttt{lattice\_system\_orig} \strut \hspace{\fill} \cite{curtarolo:art58,curtarolo:art75,curtarolo:art92,curtarolo:art135}
    \\[-2mm]
    \rule{\columnwidth}{1pt}
    Returns the lattice system for the unrelaxed structure.\\[-2mm]
    \rule{\columnwidth}{0.5pt}
    \begin{minipage}{0.4\columnwidth}\textbf{Unit}: none\end{minipage}\textbf{Type:} string\\
    \textbf{Example:} \seqsplit{lattice\_system\_orig=rhombohedral}
    \\\vspace{-2mm}
\end{minipage}
}

\noindent
\fcolorbox{black}[gray]{1}{
\begin{minipage}{\columnwidth}
    \texttt{lattice\_system\_relax} \strut \hspace{\fill} \cite{curtarolo:art58,curtarolo:art75,curtarolo:art92,curtarolo:art135}
    \\[-2mm]
    \rule{\columnwidth}{1pt}
    Returns the lattice system for the relaxed structure.\\[-2mm]
    \rule{\columnwidth}{0.5pt}
    \begin{minipage}{0.4\columnwidth}\textbf{Unit}: none\end{minipage}\textbf{Type:} string\\
    \textbf{Example:} \seqsplit{lattice\_system\_relax=rhombohedral}
    \\\vspace{-2mm}
\end{minipage}
}

\noindent
\fcolorbox{black}[gray]{1}{
\begin{minipage}{\columnwidth}
    \texttt{lattice\_variation\_orig} \strut \hspace{\fill} \cite{curtarolo:art58,curtarolo:art75,curtarolo:art92,curtarolo:art135}
    \\[-2mm]
    \rule{\columnwidth}{1pt}
    Returns the lattice variation for the unrelaxed structure.\\[-2mm]
    \rule{\columnwidth}{0.5pt}
    \begin{minipage}{0.4\columnwidth}\textbf{Unit}: none\end{minipage}\textbf{Type:} string\\
    \textbf{Example:} \seqsplit{lattice\_variation\_orig=rhombohedral}
    \\\vspace{-2mm}
\end{minipage}
}

\noindent
\fcolorbox{black}[gray]{1}{
\begin{minipage}{\columnwidth}
    \texttt{lattice\_variation\_relax} \strut \hspace{\fill} \cite{curtarolo:art58,curtarolo:art75,curtarolo:art92,curtarolo:art135}
    \\[-2mm]
    \rule{\columnwidth}{1pt}
    Returns the lattice variation for the relaxed structure.\\[-2mm]
    \rule{\columnwidth}{0.5pt}
    \begin{minipage}{0.4\columnwidth}\textbf{Unit}: none\end{minipage}\textbf{Type:} string\\
    \textbf{Example:} \seqsplit{lattice\_variation\_relax=rhombohedral}
    \\\vspace{-2mm}
\end{minipage}
}

\noindent
\fcolorbox{black}[gray]{1}{
\begin{minipage}{\columnwidth}
    \texttt{ldau\_j} \strut \hspace{\fill} \cite{curtarolo:art92,curtarolo:art104}
    \\[-2mm]
    \rule{\columnwidth}{1pt}
    Returns the J parameters of the DFT+U calculation.\\[-2mm]
    \rule{\columnwidth}{0.5pt}
    \begin{minipage}{0.4\columnwidth}\textbf{Unit}: eV\end{minipage}\textbf{Type:} numbers\\
    \textbf{Example:} \seqsplit{ldau\_j=0,0,0}
    \\\vspace{-2mm}
\end{minipage}
}

\noindent
\fcolorbox{black}[gray]{1}{
\begin{minipage}{\columnwidth}
    \texttt{ldau\_l} \strut \hspace{\fill} \cite{curtarolo:art92,curtarolo:art104}
    \\[-2mm]
    \rule{\columnwidth}{1pt}
    Returns the orbitals of the DFT+U calculation.\\[-2mm]
    \rule{\columnwidth}{0.5pt}
    \begin{minipage}{0.4\columnwidth}\textbf{Unit}: none\end{minipage}\textbf{Type:} numbers\\
    \textbf{Example:} \seqsplit{ldau\_l=2,0,0}
    \\\vspace{-2mm}
\end{minipage}
}

\noindent
\fcolorbox{black}[gray]{1}{
\begin{minipage}{\columnwidth}
    \texttt{ldau\_type} \strut \hspace{\fill} \cite{curtarolo:art92,curtarolo:art104}
    \\[-2mm]
    \rule{\columnwidth}{1pt}
    Returns the type of DFT+U calculation performed.\\[-2mm]
    \rule{\columnwidth}{0.5pt}
    \begin{minipage}{0.4\columnwidth}\textbf{Unit}: none\end{minipage}\textbf{Type:} number\\
    \textbf{Example:} \seqsplit{ldau\_type=2}
    \\\vspace{-2mm}
\end{minipage}
}

\noindent
\fcolorbox{black}[gray]{1}{
\begin{minipage}{\columnwidth}
    \texttt{ldau\_u} \strut \hspace{\fill} \cite{curtarolo:art92,curtarolo:art104}
    \\[-2mm]
    \rule{\columnwidth}{1pt}
    Returns the U parameters of the DFT+U calculation.\\[-2mm]
    \rule{\columnwidth}{0.5pt}
    \begin{minipage}{0.4\columnwidth}\textbf{Unit}: eV\end{minipage}\textbf{Type:} numbers\\
    \textbf{Example:} \seqsplit{ldau\_u=5,0,0}
    \\\vspace{-2mm}
\end{minipage}
}

\noindent
\fcolorbox{black}[gray]{1}{
\begin{minipage}{\columnwidth}
    \texttt{loop} \strut \hspace{\fill} \cite{curtarolo:art92}
    \\[-2mm]
    \rule{\columnwidth}{1pt}
    Returns information about the type of post-processing that was performed.\\[-2mm]
    \rule{\columnwidth}{0.5pt}
    \begin{minipage}{0.4\columnwidth}\textbf{Unit}: none\end{minipage}\textbf{Type:} strings\\
    \textbf{Example:} \seqsplit{loop=thermodynamics,bands,magnetic}
    \\\vspace{-2mm}
\end{minipage}
}

\noindent
\fcolorbox{black}[gray]{1}{
\begin{minipage}{\columnwidth}
    \texttt{natoms} \strut \hspace{\fill} \cite{curtarolo:art92}
    \\[-2mm]
    \rule{\columnwidth}{1pt}
    Returns the number of atoms in the simulation cell.\\[-2mm]
    \rule{\columnwidth}{0.5pt}
    \begin{minipage}{0.4\columnwidth}\textbf{Unit}: none\end{minipage}\textbf{Type:} number\\
    \textbf{Example:} \seqsplit{natoms=12}
    \\\vspace{-2mm}
\end{minipage}
}

\noindent
\fcolorbox{black}[gray]{1}{
\begin{minipage}{\columnwidth}
    \texttt{nbondxx} \strut \hspace{\fill} \cite{curtarolo:art92}
    \\[-2mm]
    \rule{\columnwidth}{1pt}
    Returns the nearest neighbors distances for the relaxed structure.\\[-2mm]
    \rule{\columnwidth}{0.5pt}
    \begin{minipage}{0.4\columnwidth}\textbf{Unit}: {\AA}\end{minipage}\textbf{Type:} numbers\\
    \textbf{Example:} \seqsplit{nbondxx=1.2599,1.0911,1.0911,1.7818,1.2599,1.7818}
    \\\vspace{-2mm}
\end{minipage}
}

\noindent
\fcolorbox{black}[gray]{1}{
\begin{minipage}{\columnwidth}
    \texttt{node\_CPU\_Cores} \strut \hspace{\fill} \cite{curtarolo:art92}
    \\[-2mm]
    \rule{\columnwidth}{1pt}
    Returns information about the number of CPUs on the node/cluster where the calculation was performed.\\[-2mm]
    \rule{\columnwidth}{0.5pt}
    \begin{minipage}{0.4\columnwidth}\textbf{Unit}: none\end{minipage}\textbf{Type:} number\\
    \textbf{Example:} \seqsplit{node\_CPU\_Cores=12}
    \\\vspace{-2mm}
\end{minipage}
}

\noindent
\fcolorbox{black}[gray]{1}{
\begin{minipage}{\columnwidth}
    \texttt{node\_CPU\_MHz} \strut \hspace{\fill} \cite{curtarolo:art92}
    \\[-2mm]
    \rule{\columnwidth}{1pt}
    Returns information about the speed of CPUs on the node/cluster where the calculation was performed.\\[-2mm]
    \rule{\columnwidth}{0.5pt}
    \begin{minipage}{0.4\columnwidth}\textbf{Unit}: MHz\end{minipage}\textbf{Type:} number\\
    \textbf{Example:} \seqsplit{node\_CPU\_MHz=12}
    \\\vspace{-2mm}
\end{minipage}
}

\noindent
\fcolorbox{black}[gray]{1}{
\begin{minipage}{\columnwidth}
    \texttt{node\_CPU\_Model} \strut \hspace{\fill} \cite{curtarolo:art92}
    \\[-2mm]
    \rule{\columnwidth}{1pt}
    Returns information about the model of CPUs on the node/cluster where the calculation was performed.\\[-2mm]
    \rule{\columnwidth}{0.5pt}
    \begin{minipage}{0.4\columnwidth}\textbf{Unit}: none\end{minipage}\textbf{Type:} string\\
    \textbf{Example:} \seqsplit{node\_CPU\_Model=12}
    \\\vspace{-2mm}
\end{minipage}
}

\noindent
\fcolorbox{black}[gray]{1}{
\begin{minipage}{\columnwidth}
    \texttt{node\_RAM\_GB} \strut \hspace{\fill} \cite{curtarolo:art92}
    \\[-2mm]
    \rule{\columnwidth}{1pt}
    Returns information about the RAM on the node/cluster where the calculation was performed.\\[-2mm]
    \rule{\columnwidth}{0.5pt}
    \begin{minipage}{0.4\columnwidth}\textbf{Unit}: Gb\end{minipage}\textbf{Type:} number\\
    \textbf{Example:} \seqsplit{node\_RAM\_GB=12}
    \\\vspace{-2mm}
\end{minipage}
}

\noindent
\fcolorbox{black}[gray]{1}{
\begin{minipage}{\columnwidth}
    \texttt{nspecies} \strut \hspace{\fill} \cite{curtarolo:art75,curtarolo:art92}
    \\[-2mm]
    \rule{\columnwidth}{1pt}
    Returns the number of unique species in the structure.\\[-2mm]
    \rule{\columnwidth}{0.5pt}
    \begin{minipage}{0.4\columnwidth}\textbf{Unit}: none\end{minipage}\textbf{Type:} number\\
    \textbf{Example:} \seqsplit{nspecies=3}
    \\\vspace{-2mm}
\end{minipage}
}

\noindent
\fcolorbox{black}[gray]{1}{
\begin{minipage}{\columnwidth}
    \texttt{Pearson\_symbol\_orig} \strut \hspace{\fill} \cite{curtarolo:art58,curtarolo:art75,curtarolo:art92,curtarolo:art135}
    \\[-2mm]
    \rule{\columnwidth}{1pt}
    Returns the Pearson symbol for the unrelaxed structure.\\[-2mm]
    \rule{\columnwidth}{0.5pt}
    \begin{minipage}{0.4\columnwidth}\textbf{Unit}: none\end{minipage}\textbf{Type:} string\\
    \textbf{Example:} \seqsplit{Pearson\_symbol\_orig=mS32}
    \\\vspace{-2mm}
\end{minipage}
}

\noindent
\fcolorbox{black}[gray]{1}{
\begin{minipage}{\columnwidth}
    \texttt{Pearson\_symbol\_relax} \strut \hspace{\fill} \cite{curtarolo:art58,curtarolo:art75,curtarolo:art92,curtarolo:art135}
    \\[-2mm]
    \rule{\columnwidth}{1pt}
    Returns the Pearson symbol for the relaxed structure.\\[-2mm]
    \rule{\columnwidth}{0.5pt}
    \begin{minipage}{0.4\columnwidth}\textbf{Unit}: none\end{minipage}\textbf{Type:} string\\
    \textbf{Example:} \seqsplit{Pearson\_symbol\_relax=mS32}
    \\\vspace{-2mm}
\end{minipage}
}

\noindent
\fcolorbox{black}[gray]{1}{
\begin{minipage}{\columnwidth}
    \texttt{Pearson\_symbol\_superlattice} \strut \hspace{\fill} \cite{curtarolo:art75,curtarolo:art135}
    \\[-2mm]
    \rule{\columnwidth}{1pt}
    Returns the Pearson symbol of the superlattice for the relaxed structure.\\[-2mm]
    \rule{\columnwidth}{0.5pt}
    \begin{minipage}{0.4\columnwidth}\textbf{Unit}: none\end{minipage}\textbf{Type:} string\\
    \textbf{Example:} \seqsplit{Pearson\_symbol\_superlattice=cI52}
    \\\vspace{-2mm}
\end{minipage}
}

\noindent
\fcolorbox{black}[gray]{1}{
\begin{minipage}{\columnwidth}
    \texttt{Pearson\_symbol\_superlattice\_orig} \strut \hspace{\fill} \cite{curtarolo:art75,curtarolo:art135}
    \\[-2mm]
    \rule{\columnwidth}{1pt}
    Returns the Pearson symbol of the superlattice for the unrelaxed structure.\\[-2mm]
    \rule{\columnwidth}{0.5pt}
    \begin{minipage}{0.4\columnwidth}\textbf{Unit}: none\end{minipage}\textbf{Type:} string\\
    \textbf{Example:} \seqsplit{Pearson\_symbol\_superlattice\_orig=cI52}
    \\\vspace{-2mm}
\end{minipage}
}

\noindent
\fcolorbox{black}[gray]{1}{
\begin{minipage}{\columnwidth}
    \texttt{point\_group\_Hermann\_Mauguin} \strut \hspace{\fill} \cite{curtarolo:art75,curtarolo:art135}
    \\[-2mm]
    \rule{\columnwidth}{1pt}
    Returns the point group, in Hermann-Mauguin notation, for the relaxed structure.\\[-2mm]
    \rule{\columnwidth}{0.5pt}
    \begin{minipage}{0.4\columnwidth}\textbf{Unit}: none\end{minipage}\textbf{Type:} string\\
    \textbf{Example:} \seqsplit{point\_group\_Hermann\_Mauguin=-43m}
    \\\vspace{-2mm}
\end{minipage}
}

\noindent
\fcolorbox{black}[gray]{1}{
\begin{minipage}{\columnwidth}
    \texttt{point\_group\_Hermann\_Mauguin\_orig} \strut \hspace{\fill} \cite{curtarolo:art75,curtarolo:art135}
    \\[-2mm]
    \rule{\columnwidth}{1pt}
    Returns the point group, in Hermann-Mauguin notation, for the unrelaxed structure.\\[-2mm]
    \rule{\columnwidth}{0.5pt}
    \begin{minipage}{0.4\columnwidth}\textbf{Unit}: none\end{minipage}\textbf{Type:} string\\
    \textbf{Example:} \seqsplit{point\_group\_Hermann\_Mauguin\_orig=-43m}
    \\\vspace{-2mm}
\end{minipage}
}

\noindent
\fcolorbox{black}[gray]{1}{
\begin{minipage}{\columnwidth}
    \texttt{point\_group\_orbifold} \strut \hspace{\fill} \cite{curtarolo:art75,curtarolo:art135}
    \\[-2mm]
    \rule{\columnwidth}{1pt}
    Returns the point group orbifold for the relaxed structure.\\[-2mm]
    \rule{\columnwidth}{0.5pt}
    \begin{minipage}{0.4\columnwidth}\textbf{Unit}: none\end{minipage}\textbf{Type:} string\\
    \textbf{Example:} \seqsplit{point\_group\_orbifold=*332}
    \\\vspace{-2mm}
\end{minipage}
}

\noindent
\fcolorbox{black}[gray]{1}{
\begin{minipage}{\columnwidth}
    \texttt{point\_group\_orbifold\_orig} \strut \hspace{\fill} \cite{curtarolo:art75,curtarolo:art135}
    \\[-2mm]
    \rule{\columnwidth}{1pt}
    Returns the point group orbifold for the unrelaxed structure.\\[-2mm]
    \rule{\columnwidth}{0.5pt}
    \begin{minipage}{0.4\columnwidth}\textbf{Unit}: none\end{minipage}\textbf{Type:} string\\
    \textbf{Example:} \seqsplit{point\_group\_orbifold\_orig=*332}
    \\\vspace{-2mm}
\end{minipage}
}

\noindent
\fcolorbox{black}[gray]{1}{
\begin{minipage}{\columnwidth}
    \texttt{point\_group\_order} \strut \hspace{\fill} \cite{curtarolo:art75,curtarolo:art135}
    \\[-2mm]
    \rule{\columnwidth}{1pt}
    Returns the point group order for the relaxed structure.\\[-2mm]
    \rule{\columnwidth}{0.5pt}
    \begin{minipage}{0.4\columnwidth}\textbf{Unit}: none\end{minipage}\textbf{Type:} number\\
    \textbf{Example:} \seqsplit{point\_group\_order=24}
    \\\vspace{-2mm}
\end{minipage}
}

\noindent
\fcolorbox{black}[gray]{1}{
\begin{minipage}{\columnwidth}
    \texttt{point\_group\_order\_orig} \strut \hspace{\fill} \cite{curtarolo:art75,curtarolo:art135}
    \\[-2mm]
    \rule{\columnwidth}{1pt}
    Returns the point group order for the unrelaxed structure.\\[-2mm]
    \rule{\columnwidth}{0.5pt}
    \begin{minipage}{0.4\columnwidth}\textbf{Unit}: none\end{minipage}\textbf{Type:} number\\
    \textbf{Example:} \seqsplit{point\_group\_order\_orig=24}
    \\\vspace{-2mm}
\end{minipage}
}

\noindent
\fcolorbox{black}[gray]{1}{
\begin{minipage}{\columnwidth}
    \texttt{point\_group\_Schoenflies} \strut \hspace{\fill} \cite{curtarolo:art75,curtarolo:art135}
    \\[-2mm]
    \rule{\columnwidth}{1pt}
    Returns the point group, in Schoenflies notation, for the relaxed structure.\\[-2mm]
    \rule{\columnwidth}{0.5pt}
    \begin{minipage}{0.4\columnwidth}\textbf{Unit}: none\end{minipage}\textbf{Type:} string\\
    \textbf{Example:} \seqsplit{point\_group\_Schoenflies=T\_d}
    \\\vspace{-2mm}
\end{minipage}
}

\noindent
\fcolorbox{black}[gray]{1}{
\begin{minipage}{\columnwidth}
    \texttt{point\_group\_Schoenflies\_orig} \strut \hspace{\fill} \cite{curtarolo:art75,curtarolo:art135}
    \\[-2mm]
    \rule{\columnwidth}{1pt}
    Returns the point group, in Schoenflies notation, for the unrelaxed structure.\\[-2mm]
    \rule{\columnwidth}{0.5pt}
    \begin{minipage}{0.4\columnwidth}\textbf{Unit}: none\end{minipage}\textbf{Type:} string\\
    \textbf{Example:} \seqsplit{point\_group\_Schoenflies\_orig=T\_d}
    \\\vspace{-2mm}
\end{minipage}
}

\noindent
\fcolorbox{black}[gray]{1}{
\begin{minipage}{\columnwidth}
    \texttt{point\_group\_structure} \strut \hspace{\fill} \cite{curtarolo:art75,curtarolo:art135}
    \\[-2mm]
    \rule{\columnwidth}{1pt}
    Returns the point group structure for the relaxed structure.\\[-2mm]
    \rule{\columnwidth}{0.5pt}
    \begin{minipage}{0.4\columnwidth}\textbf{Unit}: none\end{minipage}\textbf{Type:} string\\
    \textbf{Example:} \seqsplit{point\_group\_structure=symmetric}
    \\\vspace{-2mm}
\end{minipage}
}

\noindent
\fcolorbox{black}[gray]{1}{
\begin{minipage}{\columnwidth}
    \texttt{point\_group\_structure\_orig} \strut \hspace{\fill} \cite{curtarolo:art75,curtarolo:art135}
    \\[-2mm]
    \rule{\columnwidth}{1pt}
    Returns the point group structure for the unrelaxed structure.\\[-2mm]
    \rule{\columnwidth}{0.5pt}
    \begin{minipage}{0.4\columnwidth}\textbf{Unit}: none\end{minipage}\textbf{Type:} string\\
    \textbf{Example:} \seqsplit{point\_group\_structure\_orig=symmetric}
    \\\vspace{-2mm}
\end{minipage}
}

\noindent
\fcolorbox{black}[gray]{1}{
\begin{minipage}{\columnwidth}
    \texttt{point\_group\_type} \strut \hspace{\fill} \cite{curtarolo:art75,curtarolo:art135}
    \\[-2mm]
    \rule{\columnwidth}{1pt}
    Returns the point group type for the relaxed structure.\\[-2mm]
    \rule{\columnwidth}{0.5pt}
    \begin{minipage}{0.4\columnwidth}\textbf{Unit}: none\end{minipage}\textbf{Type:} string\\
    \textbf{Example:} \seqsplit{point\_group\_type=non-centrosymmetric,non-enantiomorphic,non-polar}
    \\\vspace{-2mm}
\end{minipage}
}

\noindent
\fcolorbox{black}[gray]{1}{
\begin{minipage}{\columnwidth}
    \texttt{point\_group\_type\_orig} \strut \hspace{\fill} \cite{curtarolo:art75,curtarolo:art135}
    \\[-2mm]
    \rule{\columnwidth}{1pt}
    Returns the point group type for the unrelaxed structure.\\[-2mm]
    \rule{\columnwidth}{0.5pt}
    \begin{minipage}{0.4\columnwidth}\textbf{Unit}: none\end{minipage}\textbf{Type:} string\\
    \textbf{Example:} \seqsplit{point\_group\_type\_orig=non-centrosymmetric,non-enantiomorphic,non-polar}
    \\\vspace{-2mm}
\end{minipage}
}

\noindent
\fcolorbox{black}[gray]{1}{
\begin{minipage}{\columnwidth}
    \texttt{positions\_cartesian} \strut \hspace{\fill} \cite{curtarolo:art92}
    \\[-2mm]
    \rule{\columnwidth}{1pt}
    Returns the Cartesian coordinates of the atoms for the relaxed structure.\\[-2mm]
    \rule{\columnwidth}{0.5pt}
    \begin{minipage}{0.4\columnwidth}\textbf{Unit}: {\AA}\end{minipage}\textbf{Type:} numbers\\
    \textbf{Example:} \seqsplit{positions\_cartesian=0,0,0;18.18438,0,2.85027;...}
    \\\vspace{-2mm}
\end{minipage}
}

\noindent
\fcolorbox{black}[gray]{1}{
\begin{minipage}{\columnwidth}
    \texttt{positions\_fractional} \strut \hspace{\fill} \cite{curtarolo:art92}
    \\[-2mm]
    \rule{\columnwidth}{1pt}
    Returns the fractional coordinates of the atoms for the relaxed structure.\\[-2mm]
    \rule{\columnwidth}{0.5pt}
    \begin{minipage}{0.4\columnwidth}\textbf{Unit}: none\end{minipage}\textbf{Type:} numbers\\
    \textbf{Example:} \seqsplit{positions\_fractional=0,0,0;0.25,0.25,0.25;...}
    \\\vspace{-2mm}
\end{minipage}
}

\noindent
\fcolorbox{black}[gray]{1}{
\begin{minipage}{\columnwidth}
    \texttt{pressure} \strut \hspace{\fill} \cite{curtarolo:art92}
    \\[-2mm]
    \rule{\columnwidth}{1pt}
    Returns the hydrostatic pressure on the simulation cell for the unrelaxed structure.\\[-2mm]
    \rule{\columnwidth}{0.5pt}
    \begin{minipage}{0.4\columnwidth}\textbf{Unit}: kbar\end{minipage}\textbf{Type:} number\\
    \textbf{Example:} \seqsplit{pressure=10.0}
    \\\vspace{-2mm}
\end{minipage}
}

\noindent
\fcolorbox{black}[gray]{1}{
\begin{minipage}{\columnwidth}
    \texttt{pressure\_residual} \strut \hspace{\fill} \cite{curtarolo:art128}
    \\[-2mm]
    \rule{\columnwidth}{1pt}
    Returns the hydrostatic pressure, corrected by the Pulay stress, on the simulation cell for the relaxed structure.\\[-2mm]
    \rule{\columnwidth}{0.5pt}
    \begin{minipage}{0.4\columnwidth}\textbf{Unit}: kbar\end{minipage}\textbf{Type:} number\\
    \textbf{Example:} \seqsplit{pressure\_residual=10.0}
    \\\vspace{-2mm}
\end{minipage}
}

\noindent
\fcolorbox{black}[gray]{1}{
\begin{minipage}{\columnwidth}
    \texttt{prototype} \strut \hspace{\fill} \cite{curtarolo:art75,curtarolo:art92}
    \\[-2mm]
    \rule{\columnwidth}{1pt}
    Returns the AFLOW prototype for the unrelaxed structure.\\[-2mm]
    \rule{\columnwidth}{0.5pt}
    \begin{minipage}{0.4\columnwidth}\textbf{Unit}: none\end{minipage}\textbf{Type:} string\\
    \textbf{Example:} \seqsplit{prototype=T0001.A2BC}
    \\\vspace{-2mm}
\end{minipage}
}

\noindent
\fcolorbox{black}[gray]{1}{
\begin{minipage}{\columnwidth}
    \texttt{Pulay\_stress} \strut \hspace{\fill} \cite{curtarolo:art104,curtarolo:art128}
    \\[-2mm]
    \rule{\columnwidth}{1pt}
    Returns the Pulay stress correcton for the calculation.\\[-2mm]
    \rule{\columnwidth}{0.5pt}
    \begin{minipage}{0.4\columnwidth}\textbf{Unit}: kbar\end{minipage}\textbf{Type:} number\\
    \textbf{Example:} \seqsplit{pulay\_stress=10.0}
    \\\vspace{-2mm}
\end{minipage}
}

\noindent
\fcolorbox{black}[gray]{1}{
\begin{minipage}{\columnwidth}
    \texttt{PV\_atom} \strut \hspace{\fill} \cite{curtarolo:art92}
    \\[-2mm]
    \rule{\columnwidth}{1pt}
    Returns the pressure multiplied by volume per atom for the relaxed structure.\\[-2mm]
    \rule{\columnwidth}{0.5pt}
    \begin{minipage}{0.4\columnwidth}\textbf{Unit}: eV/atom\end{minipage}\textbf{Type:} number\\
    \textbf{Example:} \seqsplit{PV\_atom=12.13}
    \\\vspace{-2mm}
\end{minipage}
}

\noindent
\fcolorbox{black}[gray]{1}{
\begin{minipage}{\columnwidth}
    \texttt{PV\_cell} \strut \hspace{\fill} \cite{curtarolo:art92}
    \\[-2mm]
    \rule{\columnwidth}{1pt}
    Returns the pressure multiplied by volume per atom for the relaxed structure.\\[-2mm]
    \rule{\columnwidth}{0.5pt}
    \begin{minipage}{0.4\columnwidth}\textbf{Unit}: eV/cell\end{minipage}\textbf{Type:} number\\
    \textbf{Example:} \seqsplit{PV\_cell=12.13}
    \\\vspace{-2mm}
\end{minipage}
}

\noindent
\fcolorbox{black}[gray]{1}{
\begin{minipage}{\columnwidth}
    \texttt{reciprocal\_geometry} \strut \hspace{\fill} \cite{curtarolo:art58,curtarolo:art75,curtarolo:art135}
    \\[-2mm]
    \rule{\columnwidth}{1pt}
    Returns the reciprocal lattice parameters of the relaxed simulation cell.\\[-2mm]
    \rule{\columnwidth}{0.5pt}
    \begin{minipage}{0.4\columnwidth}\textbf{Unit}: none\end{minipage}\textbf{Type:} numbers\\
    \textbf{Example:} \seqsplit{reciprocal\_geometry\_relax=0.8747,0.9747,0.8747,60.0,60.0,60.0}
    \\\vspace{-2mm}
\end{minipage}
}

\noindent
\fcolorbox{black}[gray]{1}{
\begin{minipage}{\columnwidth}
    \texttt{reciprocal\_geometry\_orig} \strut \hspace{\fill} \cite{curtarolo:art58,curtarolo:art75,curtarolo:art135}
    \\[-2mm]
    \rule{\columnwidth}{1pt}
    Returns the reciprocal lattice parameters of the unrelaxed simulation cell.\\[-2mm]
    \rule{\columnwidth}{0.5pt}
    \begin{minipage}{0.4\columnwidth}\textbf{Unit}: none\end{minipage}\textbf{Type:} numbers\\
    \textbf{Example:} \seqsplit{reciprocal\_geometry\_orig=0.8747,0.9747,0.8747,60.0,60.0,60.0}
    \\\vspace{-2mm}
\end{minipage}
}

\noindent
\fcolorbox{black}[gray]{1}{
\begin{minipage}{\columnwidth}
    \texttt{reciprocal\_lattice\_type} \strut \hspace{\fill} \cite{curtarolo:art58,curtarolo:art75,curtarolo:art135}
    \\[-2mm]
    \rule{\columnwidth}{1pt}
    Returns the reciprocal lattice centering type for the relaxed structure.\\[-2mm]
    \rule{\columnwidth}{0.5pt}
    \begin{minipage}{0.4\columnwidth}\textbf{Unit}: none\end{minipage}\textbf{Type:} string\\
    \textbf{Example:} \seqsplit{reciprocal\_lattice\_type=FCC}
    \\\vspace{-2mm}
\end{minipage}
}

\noindent
\fcolorbox{black}[gray]{1}{
\begin{minipage}{\columnwidth}
    \texttt{reciprocal\_lattice\_type\_orig} \strut \hspace{\fill} \cite{curtarolo:art58,curtarolo:art75,curtarolo:art135}
    \\[-2mm]
    \rule{\columnwidth}{1pt}
    Returns the reciprocal lattice centering type for the unrelaxed structure.\\[-2mm]
    \rule{\columnwidth}{0.5pt}
    \begin{minipage}{0.4\columnwidth}\textbf{Unit}: none\end{minipage}\textbf{Type:} string\\
    \textbf{Example:} \seqsplit{reciprocal\_lattice\_type\_orig=FCC}
    \\\vspace{-2mm}
\end{minipage}
}

\noindent
\fcolorbox{black}[gray]{1}{
\begin{minipage}{\columnwidth}
    \texttt{reciprocal\_lattice\_variation\_type} \strut \hspace{\fill} \cite{curtarolo:art58,curtarolo:art75,curtarolo:art135}
    \\[-2mm]
    \rule{\columnwidth}{1pt}
    Returns the reciprocal lattice centering type variation for the relaxed structure.\\[-2mm]
    \rule{\columnwidth}{0.5pt}
    \begin{minipage}{0.4\columnwidth}\textbf{Unit}: none\end{minipage}\textbf{Type:} string\\
    \textbf{Example:} \seqsplit{reciprocal\_lattice\_variation\_type=FCC}
    \\\vspace{-2mm}
\end{minipage}
}

\noindent
\fcolorbox{black}[gray]{1}{
\begin{minipage}{\columnwidth}
    \texttt{reciprocal\_lattice\_variation\_type\_orig} \strut \hspace{\fill} \cite{curtarolo:art58,curtarolo:art75,curtarolo:art135}
    \\[-2mm]
    \rule{\columnwidth}{1pt}
    Returns the reciprocal lattice centering type variation for the unrelaxed structure.\\[-2mm]
    \rule{\columnwidth}{0.5pt}
    \begin{minipage}{0.4\columnwidth}\textbf{Unit}: none\end{minipage}\textbf{Type:} string\\
    \textbf{Example:} \seqsplit{reciprocal\_lattice\_variation\_type\_orig=FCC}
    \\\vspace{-2mm}
\end{minipage}
}

\noindent
\fcolorbox{black}[gray]{1}{
\begin{minipage}{\columnwidth}
    \texttt{reciprocal\_volume\_cell} \strut \hspace{\fill} \cite{curtarolo:art58,curtarolo:art75,curtarolo:art135}
    \\[-2mm]
    \rule{\columnwidth}{1pt}
    Returns the volume of the reciprocal cell for the relaxed structure.\\[-2mm]
    \rule{\columnwidth}{0.5pt}
    \begin{minipage}{0.4\columnwidth}\textbf{Unit}: {\AA}$^{-3}$\end{minipage}\textbf{Type:} number\\
    \textbf{Example:} \seqsplit{reciprocal\_volume\_cell=0.4733}
    \\\vspace{-2mm}
\end{minipage}
}

\noindent
\fcolorbox{black}[gray]{1}{
\begin{minipage}{\columnwidth}
    \texttt{reciprocal\_volume\_cell\_orig} \strut \hspace{\fill} \cite{curtarolo:art58,curtarolo:art75,curtarolo:art135}
    \\[-2mm]
    \rule{\columnwidth}{1pt}
    Returns the volume of the reciprocal cell for the unrelaxed structure.\\[-2mm]
    \rule{\columnwidth}{0.5pt}
    \begin{minipage}{0.4\columnwidth}\textbf{Unit}: {\AA}$^{-3}$\end{minipage}\textbf{Type:} number\\
    \textbf{Example:} \seqsplit{reciprocal\_volume\_cell\_orig=0.4733}
    \\\vspace{-2mm}
\end{minipage}
}

\noindent
\fcolorbox{black}[gray]{1}{
\begin{minipage}{\columnwidth}
    \texttt{scintillation\_attenuation\_length} \strut \hspace{\fill} \cite{curtarolo:art64,curtarolo:art46,curtarolo:art92}
    \\[-2mm]
    \rule{\columnwidth}{1pt}
    Returns the scintillation attenuation length.\\[-2mm]
    \rule{\columnwidth}{0.5pt}
    \begin{minipage}{0.4\columnwidth}\textbf{Unit}: cm\end{minipage}\textbf{Type:} number\\
    \textbf{Example:} \seqsplit{scintillation\_attenuation\_length=2.21895}
    \\\vspace{-2mm}
\end{minipage}
}

\noindent
\fcolorbox{black}[gray]{1}{
\begin{minipage}{\columnwidth}
    \texttt{sg} \strut \hspace{\fill} \cite{curtarolo:art75,curtarolo:art92,curtarolo:art135}
    \\[-2mm]
    \rule{\columnwidth}{1pt}
    Returns the space groups for the structure, before the first relaxation step (unrelaxed), after the first relaxation step and after the last relaxation step (relaxed), using a loose tolerance.\\[-2mm]
    \rule{\columnwidth}{0.5pt}
    \begin{minipage}{0.4\columnwidth}\textbf{Unit}: none\end{minipage}\textbf{Type:} strings\\
    \textbf{Example:} \seqsplit{sg=Fm-3m\#225,Fm-3m\#225,Fm-3m\#225}
    \\\vspace{-2mm}
\end{minipage}
}

\noindent
\fcolorbox{black}[gray]{1}{
\begin{minipage}{\columnwidth}
    \texttt{sg2} \strut \hspace{\fill} \cite{curtarolo:art75,curtarolo:art92,curtarolo:art135}
    \\[-2mm]
    \rule{\columnwidth}{1pt}
    Returns the space groups for the structure, before the first relaxation step (unrelaxed), after the first relaxation step and after the last relaxation step (relaxed), using a tight (default) tolerance.\\[-2mm]
    \rule{\columnwidth}{0.5pt}
    \begin{minipage}{0.4\columnwidth}\textbf{Unit}: none\end{minipage}\textbf{Type:} strings\\
    \textbf{Example:} \seqsplit{sg2=Fm-3m\#225,Fm-3m\#225,Fm-3m\#225}
    \\\vspace{-2mm}
\end{minipage}
}

\noindent
\fcolorbox{black}[gray]{1}{
\begin{minipage}{\columnwidth}
    \texttt{spacegroup\_orig} \strut \hspace{\fill} \cite{curtarolo:art75,curtarolo:art92,curtarolo:art135}
    \\[-2mm]
    \rule{\columnwidth}{1pt}
    Returns the space group for the unrelaxed structure.\\[-2mm]
    \rule{\columnwidth}{0.5pt}
    \begin{minipage}{0.4\columnwidth}\textbf{Unit}: none\end{minipage}\textbf{Type:} number\\
    \textbf{Example:} \seqsplit{spacegroup\_orig=225}
    \\\vspace{-2mm}
\end{minipage}
}

\noindent
\fcolorbox{black}[gray]{1}{
\begin{minipage}{\columnwidth}
    \texttt{spacegroup\_relax} \strut \hspace{\fill} \cite{curtarolo:art75,curtarolo:art92,curtarolo:art135}
    \\[-2mm]
    \rule{\columnwidth}{1pt}
    Returns the space group number for the relaxed structure.\\[-2mm]
    \rule{\columnwidth}{0.5pt}
    \begin{minipage}{0.4\columnwidth}\textbf{Unit}: none\end{minipage}\textbf{Type:} number\\
    \textbf{Example:} \seqsplit{spacegroup\_relax=225}
    \\\vspace{-2mm}
\end{minipage}
}

\noindent
\fcolorbox{black}[gray]{1}{
\begin{minipage}{\columnwidth}
    \texttt{species} \strut \hspace{\fill} \cite{curtarolo:art75,curtarolo:art92}
    \\[-2mm]
    \rule{\columnwidth}{1pt}
    Returns the unique species.\\[-2mm]
    \rule{\columnwidth}{0.5pt}
    \begin{minipage}{0.4\columnwidth}\textbf{Unit}: none\end{minipage}\textbf{Type:} strings\\
    \textbf{Example:} \seqsplit{species=Y,Zn,Zr}
    \\\vspace{-2mm}
\end{minipage}
}

\noindent
\fcolorbox{black}[gray]{1}{
\begin{minipage}{\columnwidth}
    \texttt{species\_pp} \strut \hspace{\fill} \cite{curtarolo:art92,curtarolo:art104}
    \\[-2mm]
    \rule{\columnwidth}{1pt}
    Returns the pseudopotential of the species.\\[-2mm]
    \rule{\columnwidth}{0.5pt}
    \begin{minipage}{0.4\columnwidth}\textbf{Unit}: none\end{minipage}\textbf{Type:} strings\\
    \textbf{Example:} \seqsplit{species\_pp=Y,Zn,Zr}
    \\\vspace{-2mm}
\end{minipage}
}

\noindent
\fcolorbox{black}[gray]{1}{
\begin{minipage}{\columnwidth}
    \texttt{species\_pp\_version} \strut \hspace{\fill} \cite{curtarolo:art92}
    \\[-2mm]
    \rule{\columnwidth}{1pt}
    Returns the pseudopotential version of the species.\\[-2mm]
    \rule{\columnwidth}{0.5pt}
    \begin{minipage}{0.4\columnwidth}\textbf{Unit}: none\end{minipage}\textbf{Type:} strings\\
    \textbf{Example:} \seqsplit{species\_pp\_version=Y,Zn,Zr}
    \\\vspace{-2mm}
\end{minipage}
}

\noindent
\fcolorbox{black}[gray]{1}{
\begin{minipage}{\columnwidth}
    \texttt{species\_pp\_ZVAL} \strut \hspace{\fill} \cite{curtarolo:art104,aflow_bader}
    \\[-2mm]
    \rule{\columnwidth}{1pt}
    Returns the number of valence electrons of the species.\\[-2mm]
    \rule{\columnwidth}{0.5pt}
    \begin{minipage}{0.4\columnwidth}\textbf{Unit}: e$^{-}$\end{minipage}\textbf{Type:} numbers\\
    \textbf{Example:} \seqsplit{species\_pp\_ZVAL=3}
    \\\vspace{-2mm}
\end{minipage}
}

\noindent
\fcolorbox{black}[gray]{1}{
\begin{minipage}{\columnwidth}
    \texttt{spin\_atom} \strut \hspace{\fill} \cite{curtarolo:art75,curtarolo:art92}
    \\[-2mm]
    \rule{\columnwidth}{1pt}
    Returns the magnetization of the simulation cell per atom.\\[-2mm]
    \rule{\columnwidth}{0.5pt}
    \begin{minipage}{0.4\columnwidth}\textbf{Unit}: $\mu_\textnormal{B}$/atom\end{minipage}\textbf{Type:} number\\
    \textbf{Example:} \seqsplit{spin\_atom=2.16419}
    \\\vspace{-2mm}
\end{minipage}
}

\noindent
\fcolorbox{black}[gray]{1}{
\begin{minipage}{\columnwidth}
    \texttt{spin\_cell} \strut \hspace{\fill} \cite{curtarolo:art75,curtarolo:art92}
    \\[-2mm]
    \rule{\columnwidth}{1pt}
    Returns the magnetization of the simulation cell.\\[-2mm]
    \rule{\columnwidth}{0.5pt}
    \begin{minipage}{0.4\columnwidth}\textbf{Unit}: $\mu_\textnormal{B}$/cell\end{minipage}\textbf{Type:} number\\
    \textbf{Example:} \seqsplit{spin\_cell=2.16419}
    \\\vspace{-2mm}
\end{minipage}
}

\noindent
\fcolorbox{black}[gray]{1}{
\begin{minipage}{\columnwidth}
    \texttt{spinD} \strut \hspace{\fill} \cite{curtarolo:art75,curtarolo:art92}
    \\[-2mm]
    \rule{\columnwidth}{1pt}
    Returns the magnetic moment on each atom.\\[-2mm]
    \rule{\columnwidth}{0.5pt}
    \begin{minipage}{0.4\columnwidth}\textbf{Unit}: $\mu_\textnormal{B}$\end{minipage}\textbf{Type:} numbers\\
    \textbf{Example:} \seqsplit{spinD=0.236,0.236,-0.023,1.005}
    \\\vspace{-2mm}
\end{minipage}
}

\noindent
\fcolorbox{black}[gray]{1}{
\begin{minipage}{\columnwidth}
    \texttt{spinF} \strut \hspace{\fill} \cite{curtarolo:art75,curtarolo:art92}
    \\[-2mm]
    \rule{\columnwidth}{1pt}
    Returns the magnetization of the simulation cell at the Fermi energy.\\[-2mm]
    \rule{\columnwidth}{0.5pt}
    \begin{minipage}{0.4\columnwidth}\textbf{Unit}: $\mu_\textnormal{B}$/cell\end{minipage}\textbf{Type:} number\\
    \textbf{Example:} \seqsplit{spinF=0.410879}
    \\\vspace{-2mm}
\end{minipage}
}

\noindent
\fcolorbox{black}[gray]{1}{
\begin{minipage}{\columnwidth}
    \texttt{stoichiometry} \strut \hspace{\fill} \cite{curtarolo:art92}
    \\[-2mm]
    \rule{\columnwidth}{1pt}
    Returns the normalized composition of the structure.\\[-2mm]
    \rule{\columnwidth}{0.5pt}
    \begin{minipage}{0.4\columnwidth}\textbf{Unit}: none\end{minipage}\textbf{Type:} numbers\\
    \textbf{Example:} \seqsplit{stoichiometry=0.5,0.25,0.25}
    \\\vspace{-2mm}
\end{minipage}
}

\noindent
\fcolorbox{black}[gray]{1}{
\begin{minipage}{\columnwidth}
    \texttt{stress\_tensor} \strut \hspace{\fill} \cite{curtarolo:art128}
    \\[-2mm]
    \rule{\columnwidth}{1pt}
    Returns the stress tensor for the relaxed structure.\\[-2mm]
    \rule{\columnwidth}{0.5pt}
    \begin{minipage}{0.4\columnwidth}\textbf{Unit}: kbar\end{minipage}\textbf{Type:} numbers\\
    \textbf{Example:} \seqsplit{stress\_tensor=-0.96,-0,-0,-0,-0.96,-0,-0,-0,-0.96}
    \\\vspace{-2mm}
\end{minipage}
}

\noindent
\fcolorbox{black}[gray]{1}{
\begin{minipage}{\columnwidth}
    \texttt{valence\_cell\_iupac} \strut \hspace{\fill} \cite{curtarolo:art92}
    \\[-2mm]
    \rule{\columnwidth}{1pt}
    Returns the sum of the valence electrons, based on IUPAC standards, of the atoms in the simulation cell.\\[-2mm]
    \rule{\columnwidth}{0.5pt}
    \begin{minipage}{0.4\columnwidth}\textbf{Unit}: none\end{minipage}\textbf{Type:} number\\
    \textbf{Example:} \seqsplit{valence\_cell\_iupac=22}
    \\\vspace{-2mm}
\end{minipage}
}

\noindent
\fcolorbox{black}[gray]{1}{
\begin{minipage}{\columnwidth}
    \texttt{valence\_cell\_std} \strut \hspace{\fill} \cite{curtarolo:art92}
    \\[-2mm]
    \rule{\columnwidth}{1pt}
    Returns the sum of the valence electrons, based on the outermost shell(s), of the atoms in the simulation cell.\\[-2mm]
    \rule{\columnwidth}{0.5pt}
    \begin{minipage}{0.4\columnwidth}\textbf{Unit}: none\end{minipage}\textbf{Type:} number\\
    \textbf{Example:} \seqsplit{valence\_cell\_std=22}
    \\\vspace{-2mm}
\end{minipage}
}

\noindent
\fcolorbox{black}[gray]{1}{
\begin{minipage}{\columnwidth}
    \texttt{volume\_atom} \strut \hspace{\fill} \cite{curtarolo:art75,curtarolo:art92}
    \\[-2mm]
    \rule{\columnwidth}{1pt}
    Returns the volume per atom of the simulation cell for the relaxed structure.\\[-2mm]
    \rule{\columnwidth}{0.5pt}
    \begin{minipage}{0.4\columnwidth}\textbf{Unit}: {\AA}$^{3}$/atom\end{minipage}\textbf{Type:} number\\
    \textbf{Example:} \seqsplit{volume\_atom=100.984}
    \\\vspace{-2mm}
\end{minipage}
}

\noindent
\fcolorbox{black}[gray]{1}{
\begin{minipage}{\columnwidth}
    \texttt{volume\_cell} \strut \hspace{\fill} \cite{curtarolo:art75,curtarolo:art92}
    \\[-2mm]
    \rule{\columnwidth}{1pt}
    Returns the volume of the simulation cell for the relaxed structure.\\[-2mm]
    \rule{\columnwidth}{0.5pt}
    \begin{minipage}{0.4\columnwidth}\textbf{Unit}: {\AA}$^{3}$\end{minipage}\textbf{Type:} number\\
    \textbf{Example:} \seqsplit{volume\_cell=100.984}
    \\\vspace{-2mm}
\end{minipage}
}

\noindent
\fcolorbox{black}[gray]{1}{
\begin{minipage}{\columnwidth}
    \texttt{Wyckoff\_letters} \strut \hspace{\fill} \cite{curtarolo:art135}
    \\[-2mm]
    \rule{\columnwidth}{1pt}
    Returns the Wyckoff letters of each site for the relaxed structure.\\[-2mm]
    \rule{\columnwidth}{0.5pt}
    \begin{minipage}{0.4\columnwidth}\textbf{Unit}: none\end{minipage}\textbf{Type:} strings\\
    \textbf{Example:} \seqsplit{Wyckoff\_letters=g,c,a}
    \\\vspace{-2mm}
\end{minipage}
}

\noindent
\fcolorbox{black}[gray]{1}{
\begin{minipage}{\columnwidth}
    \texttt{Wyckoff\_letters\_orig} \strut \hspace{\fill} \cite{curtarolo:art135}
    \\[-2mm]
    \rule{\columnwidth}{1pt}
    Returns the Wyckoff letters of each site for the unrelaxed structure.\\[-2mm]
    \rule{\columnwidth}{0.5pt}
    \begin{minipage}{0.4\columnwidth}\textbf{Unit}: none\end{minipage}\textbf{Type:} strings\\
    \textbf{Example:} \seqsplit{Wyckoff\_letters\_orig=g,c,a}
    \\\vspace{-2mm}
\end{minipage}
}

\noindent
\fcolorbox{black}[gray]{1}{
\begin{minipage}{\columnwidth}
    \texttt{Wyckoff\_multiplicities} \strut \hspace{\fill} \cite{curtarolo:art135}
    \\[-2mm]
    \rule{\columnwidth}{1pt}
    Returns the Wyckoff multiplicity of each site for the relaxed structure.\\[-2mm]
    \rule{\columnwidth}{0.5pt}
    \begin{minipage}{0.4\columnwidth}\textbf{Unit}: none\end{minipage}\textbf{Type:} numbers\\
    \textbf{Example:} \seqsplit{Wyckoff\_multiplicities=24,8,2}
    \\\vspace{-2mm}
\end{minipage}
}

\noindent
\fcolorbox{black}[gray]{1}{
\begin{minipage}{\columnwidth}
    \texttt{Wyckoff\_multiplicities\_orig} \strut \hspace{\fill} \cite{curtarolo:art135}
    \\[-2mm]
    \rule{\columnwidth}{1pt}
    Returns the Wyckoff multiplicity of each site for the unrelaxed structure.\\[-2mm]
    \rule{\columnwidth}{0.5pt}
    \begin{minipage}{0.4\columnwidth}\textbf{Unit}: none\end{minipage}\textbf{Type:} strings\\
    \textbf{Example:} \seqsplit{Wyckoff\_multiplicities\_orig=24,8,2}
    \\\vspace{-2mm}
\end{minipage}
}

\noindent
\fcolorbox{black}[gray]{1}{
\begin{minipage}{\columnwidth}
    \texttt{Wyckoff\_site\_symmetries} \strut \hspace{\fill} \cite{curtarolo:art135}
    \\[-2mm]
    \rule{\columnwidth}{1pt}
    Returns the Wyckoff symmetry of each site for the relaxed structure.\\[-2mm]
    \rule{\columnwidth}{0.5pt}
    \begin{minipage}{0.4\columnwidth}\textbf{Unit}: none\end{minipage}\textbf{Type:} strings\\
    \textbf{Example:} \seqsplit{Wyckoff\_site\_symmetries=m..,.-3.,m-3.}
    \\\vspace{-2mm}
\end{minipage}
}

\noindent
\fcolorbox{black}[gray]{1}{
\begin{minipage}{\columnwidth}
    \texttt{Wyckoff\_site\_symmetries\_orig} \strut \hspace{\fill} \cite{curtarolo:art135}
    \\[-2mm]
    \rule{\columnwidth}{1pt}
    Returns the Wyckoff symmetry of each site for the unrelaxed structure.\\[-2mm]
    \rule{\columnwidth}{0.5pt}
    \begin{minipage}{0.4\columnwidth}\textbf{Unit}: none\end{minipage}\textbf{Type:} strings\\
    \textbf{Example:} \seqsplit{Wyckoff\_site\_symmetries\_orig=m..,.-3.,m-3.}
    \\\vspace{-2mm}
\end{minipage}
}

\clearpage

\vspace{1cm}

{\small
\noindent{\textbf{Acknowledgments.}
The authors thank
Yoav Lederer,
Ohad Levy,
Eric Gossett,
Cheryl Li,
Harry Wang,
Mana Rose,
William Schmitt, and
Stuart Ki
for fruitful discussions.
Research sponsored by DOD-ONR
All the authors acknowledge support by DOD-ONR (N00014-21-1-2132,
N00014-20-1-2525, N00014-20-1-2299) and by NSF (NRT-HDR DGE-2022040).
C.T. acknowledges support from NSF (DMR-2219788).
R.F. acknowledges support from the Alexander von Humboldt foundation under the Feodor Lynen research fellowship.
}

\newcommand{\Ozolins}{Ozoli{\c{n}}{\v{s}}}

\end{document}